\documentclass[preprint,3p,12pt]{elsarticle}
\usepackage{mathrsfs}
\usepackage{amsmath}
\usepackage{stmaryrd}
\usepackage{bbding}
\usepackage{dcolumn}
\usepackage{graphicx}
\usepackage{amsfonts}
\usepackage{amssymb}
\usepackage{psfrag}
\usepackage{wrapfig}
\usepackage{subfigure}
\usepackage{makeidx}
\usepackage{bm}
\usepackage{epsf}
\usepackage{epsfig}
\usepackage{setspace}
\usepackage{graphicx}
\usepackage{epstopdf}
\usepackage{psfrag}
\usepackage{subfigure}
\usepackage{color}
\usepackage{comment}

\begin{document}

\title{Unified gas-kinetic wave-particle methods \uppercase\expandafter{\romannumeral6}: Disperse dilute gas-particle multiphase flow}

\author[HKUST1]{Xiaojian Yang}
\ead{xyangbm@connect.ust.hk}

\author[IAPCM]{Chang Liu}
\ead{cliuaa@.ust.hk}

\author[HKUST2]{Xing Ji}
\ead{xjiad@connect.ust.hk}

\author[HKUST1]{Wei Shyy}
\ead{weishyy@ust.hk}

\author[HKUST1,HKUST2,HKUST3,HKUST4]{Kun Xu\corref{cor1}}
\ead{makxu@ust.hk}

\address[HKUST1]{Department of Mechanical and Aerospace Engineering, Hong Kong University of Science and Technology, Clear Water Bay, Kowloon, Hong Kong, China}
\address[IAPCM]{Institute of Applied Physics and Computational Mathematics, Beijing, China}
\address[HKUST2]{Department of Mathematics, Hong Kong University of Science and Technology, Clear Water Bay, Kowloon, Hong Kong, China}
\address[HKUST3]{Guangdong-Hong Kong-Macao Joint Laboratory for Data-Driven Fluid Mechanics and Engineering Applications, Hong Kong University of Science and Technology, Hong Kong, China}
\address[HKUST4]{Shenzhen Research Institute, Hong Kong University of Science and Technology, Shenzhen, China}
\cortext[cor1]{Corresponding author}

\begin{abstract}
In this paper, a unified gas-kinetic wave-particle scheme (UGKWP) for the disperse dilute gas-particle multiphase flow is proposed.
In this study, the gas phase is always in the hydrodynamic regime. However, the particle phase covers different flow regimes from particle trajectory crossing to the hydrodynamic wave interaction with the variation of local particle phase Knudsen number. The UGKWP is an appropriate method for the capturing of the multiscale transport mechanism in the particle phase through its coupled wave-particle formulation. In the regime with intensive particle collision, the evolution of solid particle will be followed by the analytic wave with quasi-equilibrium distribution; while in the rarefied regime the non-equilibrium particle phase will be captured through particle tracking and collision, which plays a decisive role in recovering particle trajectory crossing behavior.
The gas phase in the multiphase system is assumed to be stay in the continuum flow regime, and its evolution is mainly controlled by the Navier-Stokes equations with the interaction with particle phase. The gas-kinetic scheme (GKS) is employed for the simulation of gas flow.
In the highly collision regime for the particles, no particles will be sampled in UGKWP and the wave formulation for solid particle with the hydrodynamic gas phase will reduce the system to the two-fluid Eulerian model. On the other hand, in the collisionless regime for the solid particle, the free transport of solid particle will be followed in UGKWP, and coupled system will return to the Eulerian-Lagrangian formulation for the gas and particle.
The scheme will be tested for in all flow regimes, which include the non-equilibrium particle trajectory crossing,
the particle concentration under different Knudsen number, and the dispersion of particle flow with the variation of Stokes number.
A experiment of shock-induced particle bed fluidization is simulated and the results are compared with experimental measurements.
These numerical solutions validate suitability of the proposed scheme for the simulation of gas-particle multiphase flow.
\end{abstract}


\maketitle

\section{Introduction}
Gas-particle flow is a common two-phase system and it occurs in both natural phenomena and many engineering industries, such as volcanic eruption, sandstorm propagation, petroleum and chemical industry, and aeronautics and aerospace applications, etc \cite{Gasparticle-book-crowe2011multiphase, Gasparticle-momentmethod-Fox2013computational, Gasparticle-review-Ge2017discrete, Gasparticle-review-balachandar2010turbulent}.
Numerical simulations become  powerful and indispensable tools for the study of gas-particle system due to the complex physics and
difficulties in the experiments and theoretical analysis. Therefore, the development of reliable, accurate, and efficient numerical algorithm
to study the multiscale transport associated with different flow regimes is highly demanding in both scientific research and engineering application.

The flow physics of the gas-particle system is very complicated due to the particle-particle collision and particle-gas interaction.
While the gas phase is in continuum regime and modeled by the Navier-Stokes equations, the particle phase can cover a wide range of flow regime with multiscale transport mechanism \cite{Gasparticle-momentmethod-Fox2013computational, UGKS-gas-particle-liu2019unified, Gasparticle-PIC-DEM-tian2020compressible}. The simulation of  gas-particle flow has to model the gas-particle interaction and particle-particle collision. The flow physics is mainly controlled by two non-dimensional parameters, Stokes number $St$ and Knudsen number $Kn_s$ \cite{Gasparticle-momentmethod-Fox2013computational}. The Stokes number is related to the drag force on the particle exerted by the surrounding gas flow, which accounts for the momentum exchange between gas and solid particle.
The dusty flow model is an example in the continuum flow regime when the Stokes number is very small \cite{Gasparticle-dusty-saito2002numerical, Gasparticle-dusty-flow-pelanti2006high}. Besides, the heat conduction between the gas and solid particle leads to the energy exchange.

Another important parameter is the Knudsen number, characterizing the flow regime of particle phase.
At small Knudsen number, the intensive particle-particle collision drives the particle phase to a local equilibrium state and evolves as a continuum flow. Then, the Eulerian-Eulerian (EE) model is usually employed for the gas-particle system under the Eulerian framework, and the EE model is also called two fluid model (TFM). Many studies have been conducted using TFM \cite{Gasparticle-BN-model-baer1986two, Gasparticle-Abgrall-saurel1999multiphase, Gasparticle-dusty-flow-pelanti2006high, Rogueproblem-Houim2016multiphase}.
The kinetic theory-based  granular flow (KTGF) is one representative method of TFM.
Based on the analogy between the solid particle behavior in a granular flow and the molecule movement,
the kinetic theory is used to get the granular flow equations \cite{CE-expansion, Gasparticle-KTGF-jenkins1983theory, Gasparticle-KTGF-lun1984kinetic}, which is further extended to the disperse gas-particle system \cite{Gasparticle-KTGF-ding1990bubbling, Gasparticle-KTGF-yu2007numerical}. Since the particle is very large in granular flow, the particle-particle collision should be inelastic,
which distinguishes the dynamics of solid particles and molecule \cite{Gasparticle-KTGF-lun1984kinetic}.
The limitation of TFM is that it cannot describe the non-equilibrium state under the quasi-equilibrium assumption  \cite{Gasparticle-review-balachandar2010turbulent}. A representative non-equilibrium phenomenon of disperse phase is the particle trajectory crossing (PTC), which occurs in the extremely dilute flow regime, i.e., at large Knudsen number \cite{Gasparticle-momentmethod-Fox2013computational}.
The TFM fails to capture the PTC transport \cite{Gasparticle-PTC-freret2008turbulent}. The TFM gives an accurate prediction in the fluid dynamic regime for both gas and particle \cite{Gasparticle-momentmethod-Fox2013computational}.

On the other hand, when the Knudsen number is not very small, the local equilibrium assumption for disperse phase is no longer satisfied. Therefore, both the transport and collision processes have to be considered for the particle phase movement.
Under such a non-equilibrium flow regime, the Eulerian-Lagrangian (EL) model is usually adopted. In the EL model, the governing equations of gas phase are discretized in the Eulerian frame, while the particle is tracked discretely in the Lagrangian formulation.
The representative methods of EL model include direct numerical simulation (DNS) \cite{Gasparticle-DNS-gewei-liu2017meso, Gasparticle-DNS-immersed-boundary-luokun-luo2019improved}, computational fluid dynamic-discrete element method (CFD-DEM) \cite{Gasparticle-DEM-review-guo-curtis-2015discrete, Gasparticle-DNS-DEM-DFM-comparison-lu-gewei-2017assessing}, multiphase particle-in-cell (MP-PIC) \cite{Gasparticle-PIC-andrews1996multiphase, Gasparticle-PIC-snider2001incompressible, Gasparticle-PIC-rourke-2012inclusion, Gasparticle-PIC-DEM-tian2020compressible}, etc.
In order to reduce the number of simulating particles, many particles are collected as a particle parcel in the coarse graining particle method (CGPM) \cite{Gasparticle-coarse-grained-DEM-hilton2014comparison, Gasparticle-coarse-grained-EMMS-DPM-gewei-lu2016computer, Gasparticle-coarse-grained-lu2017extension}.
For the EL model, the number of solid particles or parcels used in the simulation is an important factor to determine the computation cost and accuracy. With the increase of particle collision in the near continuum flow regime,  the computational cost for tracking particles
becomes unaffordable. The hydrodynamic flow solver is preferred in the continuum flow regime. Therefore, the development of multiscale method which connects the modeling in different flow regimes smoothly is necessary.
Many studies targeting on the multiscale methods have been conducted for the gas-particle system, such as unified gas kinetic scheme (UGKS)\cite{UGKS-xu2010unified, UGKS-gas-particle-liu2019unified, UGKS-gasparticle-wangzhao-wang2019unified}, discrete unified gas kinetic scheme (DUGKS)\cite{Gasparticle-DUGKS-immersed-boundary-guozhaoli-tao2018combined}, unified gas kinetic particle method (UGKP)\cite{WP-second-zhu-unstructured-mesh-zhu2019unified, KP-gasparticle-wangzhao-wang2020unified}, method of moment (MOM) \cite{Gasparticle-MOM-Fox-desjardins2008quadrature, Gasparticle-MOM-Fox-fox2008quadrature, Gasparticle-MOM-Fox-passalacqua2010fully, Gasparticle-momentmethod-Fox2013computational}, direct simulation Monte Carlo (DSMC) \cite{Gasparticle-DSMC-bird1976molecular}, hybrid finite-volume-particle method \cite{Gasparticle-dusty-hybrid-finite-volume-particle-chertock2017hybrid}, etc.

The UGKS is discrete velocity method (DVM) for multiscale flow dynamics \cite{UGKS-xu2010unified}. The time accumulating solution of the kinetic model equation within a time step is used in UGKS for the flux construction in the finite volume method.
The time step in UGKS doesn't need to be less than the particle collision time, which makes the scheme highly efficient in continuum flow regime. Besides the neutral gas flow, the UGKS has been successfully applied to many other transport problems, such as radiation, photon transport, plasma, etc \cite{UGKS-plasma-liu2017unified, UGKS-microflow-linearized-ke-liu2020unified, UGKS-radiative-sun2015asymptotic, UGKS-photon-li2018unified, UGKS-first-decade-zhu2021first}.
In particular, an effective multiscale scheme based on the UGKS for the disperse dilute gas-particle system has been proposed \cite{UGKS-gas-particle-liu2019unified}.
In order to further improve the efficiency of the scheme in the highly rarefied regime, a particle-based UGKS has been developed,
which is named unified gas-kinetic particle (UGKP) \cite{WP-first-liu2020unified, WP-second-zhu-unstructured-mesh-zhu2019unified}.
The particles in UGKP are categorized as free transport particle and collisional particle. The collisional particle within a time step will be
eliminated and get sampled again from the equilibrium state at the beginning of next time step. Therefore, only collisionless
particle is fully tracked in UGKP.
As a further study, it is realized that the flux from the collisional particle in UGKP can be evaluated analytically.
As a result, the collisional particles will not be sampled, but evolved with analytical wave representation in the unified gas-kinetic
wave-particle (UGKWP) method \cite{WP-first-liu2020unified, WP-second-zhu-unstructured-mesh-zhu2019unified, WP-three-photon-transport-li2020unified, WP-four-liu2020unified, WP-sample-xu2021modeling, WP-five-diatomic-molecular-xu2020unified, WP-3D-chen2020three}.
In the continuum flow regime, no particle will be sampled at all and only wave part survives in UGKWP, which becomes automatically a hydrodynamic flow solver. In the highly rarefied flow regime, the flow evolution will be controlled by the particle transport alone.
In the transition regime, the UGKWP dynamically distributes the wave and particle components according to the cell's Knudsen number.
The UGKWP is inherently appropriate for simulating particle evolution in the disperse gas-particle system.
In this paper, a multiscale method will be developed for the system with the gas-kinetic scheme (GKS) for the gas phase and the UGKWP for the particle phase \cite{GKS-2001, GKS-lecture}. The GKS is a gas-kinetic theory-based Navier-Stokes solver which is basically the limiting scheme of UGKWP in the continuum flow regime \cite{CompactGKS-ji2018-structured, CompactGKS-zhao2019-8th-order, CompactGKS-ji2020-unstructured}.
For the continuum flow, the GKS has been used in the turbulence flow \cite{GKS-turbulence-CIT-Cao2019, GKS-turbulence-implicitHGKS-Cao2019}, acoustic wave \cite{GKS-acoustic-zhao2019}, multi-component flow \cite{GKS-multicomponent-xu1997, GKS-multicomponent-pan2017}, and hypersonic flow studies \cite{GKS-hypersonic-LiQibing2005}.

The paper is organized as the following. In section 2, the governing equations for the gas-particle two-phase system are introduced. Section 3 is about the construction of the UGKWP for solid particle phase and the GKS for gas phase.
The numerical experiments are conducted to validate the proposed method in section 4. Section 5 is the conclusion.

\section{Governing equation}
\subsection{Governing equation for particle phase}
The evolution of particle phase is govern by the following kinetic equation,
\begin{gather}\label{particle phase kinetic equ}
\frac{\partial f_{s}}{\partial t}
+ \nabla_x \cdot \left(\textbf{u}f_{s}\right)
+ \nabla_u \cdot \left(\textbf{a}f_{s} + \textbf{G}f_{s}\right)
= \frac{g_{s}-f_{s}}{\tau_{s}},
\end{gather}
where $\textbf{u}$ is the particle velocity, $\textbf{a}$ is the particle acceleration caused by the interactive force between particle phase and gas phase, $\textbf{G}$ is the gravitational acceleration, $\nabla_x$ is the divergence operator with respect to space, $\nabla_u$ is the divergence operator with respect to velocity, $\tau_s$ is the relaxation time for the particle phase, $f_{s}$ is the distribution function of particle phase, and $g_{s}$ is the associated equilibrium distribution, which can be written as,
\begin{gather*}
g_{s}=\epsilon_s\rho_s\left(\frac{\lambda_s}{\pi}\right)^{\frac{3}{2}}e^{-\lambda_s \left[(\textbf{u}-\textbf{U}_s)^2\right]},
\end{gather*}
where $\epsilon_s$ is the volume fraction of particle phase, $\rho_s$ is the material density of particle phase, $\lambda_s$ is the value relevant to the granular temperature $T_s$ with $\lambda_s = \frac{m_s}{2k_BT_s}$, $m_s=\rho_s \frac{4}{3}\pi\left(\frac{d_s}{2}\right)^3$ is the mass of one particle, $d_s$ is the diameter of solid particle, and $\textbf{U}_s$ is the macroscopic velocity of particle phase.

For the inelastic collision between particles, the sum of the kinetic and thermal energy for colliding particles may not be conserved.
Therefore, the collision term should satisfy the following compatibility condition,
\begin{equation}\label{particle phase compatibility condition}
\frac{1}{\tau_s} \int g_s \bm{\psi} \text{d}\Xi=
\frac{1}{\tau_s} \int f_s \bm{\psi}' \text{d}\Xi,
\end{equation}
where $\bm{\psi}=\left(1,\textbf{u},\displaystyle \frac{1}{2}\left(\textbf{u}^2+\bm{\xi}^2\right)\right)^T$ and $\bm{\psi}'=\left(1,\textbf{u},\displaystyle \frac{1}{2}\left(\textbf{u}^2+\bm{\xi}^2\right)+\frac{r^2-1}{2}\left(\textbf{u}-\textbf{U}_s\right)^2\right)^T$. The lost energy due to inelastic collision in 3D can be written as,
\begin{gather*}
Q_{loss} = \frac{\left(1-r^2\right)3p_s}{2},
\end{gather*}
where $r\in\left[0,1\right]$ is the restitution coefficient for the determination of percentage of lost energy in the inelastic collision.
While $r=1$ means no energy loss (elastic collision), $r=0$ refers to total loss of all internal energy of particle phase $\epsilon_s\rho_se_s =\frac{3}{2}p_s$, where $p_s=\frac{\epsilon_s\rho_s}{2\lambda_s}$ is the granular pressure.

In order to evaluate the acceleration, the external force of particle phase has to be determined firstly. Here, the drag force $\textbf{D}$ and the buoyancy force $\textbf{F}_b$ are considered, which stand for the force applied on the particle phase by gas phase. Different drag force models can be used, and here the following model is taken,
\begin{gather}\label{drag force model}
\textbf{D} = \frac{m_s}{\tau_{st}}\left(\textbf{U}_g-\textbf{u}\right),
\end{gather}
where $\textbf{U}_g$ is the macroscopic velocity of gas phase, and $\tau_{st}$ is the particle internal response time, which can be written as,
\begin{equation}\label{taust equation}
\tau_{st}=\frac{4}{3}\frac{\rho_s d_s}{C_d\rho_g|\textbf{U}_g-\textbf{u}|}.
\end{equation}
$C_d$ is the drag coefficient, and the Kliatchko drag model is employed to obtain $C_d$ in this paper \cite{Gasparticle-drag-Kliatchko-fuks1955mechanics},
\begin{equation}
C_d = \left\{\begin{aligned}
&\frac{24}{Re_s}+\frac{4}{Re_s^{1/3}}, &  & Re_s \le 1000, \\
&0.424, &  & Re_s > 1000,
\end{aligned} \right.
\end{equation}
where $d_s$ is the diameter of solid particle, and $\mu_g$ is the dynamic viscosity of gas phase. $Re_s = |\textbf{U}_g-\textbf{u}| d_s/\nu_g$ is the particle Reynolds number, and $\nu_g=\mu_g/\rho_g$ is the kinematic viscosity of gas phase. Besides, another interactive force considered is the buoyancy force, which can be modeled as,
\begin{gather}\label{buoyancy force model}
\textbf{F}_b = -\frac{m_s}{\rho_{s}} \nabla_x p_g,
\end{gather}
where $p_g$ is the pressure of gas phase. Therefore, the acceleration of particle caused by the inter-phase force can be written as,
\begin{gather}\label{particle phase acceleration term}
\textbf{a}=\frac{\textbf{D} + \textbf{F}_b}{m_s}.
\end{gather}

When the collision between solid particles are elastic with $r=1$, in the continuum flow regime the hydrodynamic equations become the in Euler equations which can be obtained based on the Chapman-Enskog asymptotic analysis,
\begin{align}\label{particle phase Euler equ}
&\frac{\partial \left(\epsilon_s\rho_s\right)}{\partial t}
+ \nabla_x \cdot \left(\epsilon_s\rho_s \textbf{U}_s\right) = 0,\nonumber \\
&\frac{\partial \left(\epsilon_s\rho_s \textbf{U}_s\right)}{\partial t}
+ \nabla_x \cdot \left(\epsilon_s\rho_s \textbf{U}_s \textbf{U}_s + p_s \mathbb{I} \right)
= \frac{\epsilon_{s}\rho_{s}\left(\textbf{U}_g - \textbf{U}_s\right)}{\tau_{st}}
- \epsilon_{s} \nabla_x p_g
+ \epsilon_{s}\rho_{s} \textbf{G} , \\
&\frac{\partial \left(\epsilon_s\rho_s E_s\right)}{\partial t}
+ \nabla_x \cdot \left(\left(\epsilon_s\rho_s E_s  + p_s\right) \textbf{U}_s \right)
= \frac{\epsilon_{s}\rho_{s}\textbf{U}_s \cdot \left(\textbf{U}_g - \textbf{U}_s\right)}{\tau_{st}}
- \epsilon_{s} \textbf{U}_s \cdot \nabla_x p_g
+ \epsilon_{s}\rho_{s} \textbf{U}_s \cdot \textbf{G}
- \frac{3p_s}{\tau_{st}}.\nonumber
\end{align}
When the collision is inelastic with $r=0$, the governing equations would be the pressureless Euler equations.

Besides, the heat conduction between the particle and gas phase will be considered, which is associated with the temperature change.
For particle phase, the material temperature of particle phase is denoted as $T^{m}_{s}$, which is different from the granular temperature $T_s$.
The governing equation for $T^{m}_{s}$ in 3D can be written as
\begin{gather}\label{particle phase Tm equ}
\frac{\partial \left(\epsilon_s\rho_sC_sT^m_s\right)}{\partial t}
+ \nabla_x \cdot \left(\epsilon_s\rho_sC_sT^m_s \textbf{U}_s\right)
= \epsilon_s\rho_sC_s \frac{T_g-T^m_s}{\tau_T}
+ \frac{Q_{loss}}{\tau_s},
\end{gather}
where $C_s$ is the specific heat capacity of particle phase. $\tau_T=\frac{\rho_sC_sr_s}{3h}$ is the relaxation time for heat conduction, where $r_s=d_s/2$ is the radius of one solid particle and $h$ is the convection coefficient between solid material and gas phase. $p_s$ is the granular pressure of particle phase. The first term on right hand side stands for the heat conduction between particle phase and gas phase, while the second term on the right hand side is the lost energy due to the inelastic collision between solid particles. Here we assume that all the lost energy in inelastic collision is transferred into internal energy of the solid particles.

In summary, the evolution of particle phase is governed by Eq.\eqref{particle phase kinetic equ} and Eq.\eqref{particle phase Tm equ}.

\subsection{Governing equation for gas phase}
The gas phase is regarded as continuum flow and the macroscopic governing equations are the Navier-Stokes (NS) equations. In this work, the gas phase is solved by GKS, which a NS solver based on kinetic equation. The kinetic equation for gas phase can be written as,
\begin{gather}\label{gas phase kinetic equ}
\frac{\partial f_{g}}{\partial t}
+ \nabla_x \cdot \left(\textbf{u}f_{g}\right)
- \nabla_u \cdot \left(\textbf{a}f_{s}\right)
= \frac{g_{g}-f_{g}}{\tau_{g}},
\end{gather}
where $\textbf{u}$ is the velocity, $\textbf{a}$ is the acceleration caused by the interactive force between particle phase and gas phase, such as  $\textbf{a}=\frac{\textbf{D} + \textbf{F}_b}{m_s}$ determined by the external force, $\tau_g$ is the relaxation time for gas phase, $f_{g}$ is the distribution function of gas phase, and $g_{g}$ is the corresponding equilibrium state (Maxwellian distribution). In the current work, the gravitational acceleration for gas phase $\textbf{G}$ is neglected.

The local equilibrium state $g_{g}$ can be written as,
\begin{gather*}
g_{g}=\widetilde{\rho_g}\left(\frac{\lambda_g}{\pi}\right)^{\frac{K+3}{2}}e^{-\lambda_g\left[(\textbf{u}-\textbf{U}_g)^2+\bm{\xi}^2\right]},
\end{gather*}
where $\widetilde{\rho_g}=\epsilon_g\rho_g$ is the apparent density of gas phase. $\epsilon_g$ is the volume fraction of particle phase, which satisfies the relation $\epsilon_s+\epsilon_g=1$, and $\rho_g$ is the density of gas phase. $\lambda_g$ is determined by gas temperature through $\lambda_g = \frac{m_g}{2k_BT_g}$, where $m_g$ is the mass of one molecule for gas phase, $\textbf{U}_g$ is the macroscopic velocity of gas phase. $K$ is the internal degree of freedom.

The collision term satisfies the compatibility condition
\begin{equation}\label{gas phase compatibility condition}
\int \frac{g_g-f_g}{\tau_g} \bm{\psi} \text{d}\Xi=0,
\end{equation}
where $\bm{\psi}=\left(1,\textbf{u},\displaystyle \frac{1}{2}(\textbf{u}^2+\bm{\xi}^2)\right)^T$, the internal variables $\bm{\xi}^2=\xi_1^2+...+\xi_K^2$, $\text{d}\Xi=\text{d}\textbf{u}\text{d}\bm{\xi}$, $K$
is the internal degree of freedom, and $\gamma$
is the specific heat ratio.
Under continuum flow regime, the Navier-Stokes equations for gas flow can be recovered from the above kinetic equation based on the Chapman-Enskog asymptotic analysis,
\begin{align}\label{gas phase macroscopic equ}
&\frac{\partial \left(\widetilde{\rho_g}\right)}{\partial t}
+ \nabla_x \cdot \left(\widetilde{\rho_g} \textbf{U}_g\right)= 0,\nonumber \\
&\frac{\partial \left(\widetilde{\rho_g} \textbf{U}_g\right)}{\partial t}
+ \nabla_x \cdot \left(\widetilde{\rho_g} \textbf{U}_g \textbf{U}_g + \widetilde{p_g}\mathbb{I} - \widetilde{\mu_g} \bm{\sigma}\right)
= -\frac{\epsilon_{s}\rho_{s}\left(\textbf{U}_g - \textbf{U}_s\right)}{\tau_{st}}
+ \epsilon_{s} \nabla_x p_g , \\
&\frac{\partial \left(\widetilde{\rho_g} E_g\right)}{\partial t}
+ \nabla_x \cdot \left(\left(\widetilde{\rho_g} E_g  + \widetilde{p_g}\right) \textbf{U}_g
- \widetilde{\mu_g} \bm{\sigma}\cdot\textbf{U}_g + \widetilde{\kappa} \nabla_x T_g \right)
= -\frac{\epsilon_{s}\rho_{s}\textbf{U}_s \cdot \left(\textbf{U}_g - \textbf{U}_s\right)}{\tau_{st}}
+ \epsilon_{s} \textbf{U}_s \cdot \nabla_x p_g \nonumber \\
& ~~~~~~~~~~~~~~~~~~~~~~~~~~~~~~~~~~~~~~~~~~~~~~~~~~~~~~~~~~~~~~~~~~~~~~~~~~~~
+ \frac{3p_s}{\tau_{st}} -\epsilon_s\rho_sC_s \frac{T_g-T^m_s}{\tau_T}, \nonumber
\end{align}
where $p_g=\rho_gRT_g$ is the pressure of gas phase and $\widetilde{p_g}=\widetilde{\rho_g}RT_g$, the strain rate tensor $\bm{\sigma}$ is
\begin{gather*}
\bm{\sigma} = \nabla_x\textbf{U}_g + \left(\nabla_x\textbf{U}_g\right)^T
- \frac{2}{3} \nabla_x \cdot \textbf{U}_g \mathbb{I},
\end{gather*}
and
\begin{gather*}
\widetilde{\mu_g} = \tau_{g} \widetilde{p_g}, ~~~~ \widetilde{\kappa} = \frac{5}{2} R \tau_{g} \widetilde{p_g}.
\end{gather*}

Besides, the heat conduction for gas phase can be written as,
\begin{gather}\label{gas phase T heat conduction}
\frac{\text{d} \left(\widetilde{\rho_g}C_gT_g\right)}{\text{d} t}
= -\epsilon_s\rho_sC_s \frac{T_g-T^m_s}{\tau_T},
\end{gather}
where $C_g$ is the specific heat capacity of gas phase.

In summary, the evolution of gas phase is governed by Eq.\eqref{gas phase kinetic equ} and Eq.\eqref{gas phase T heat conduction}.

\section{Numerical scheme for gas-particle system}
\subsection{Unified gas-kinetic wave-particle method for particle phase}
In this subsection, the evolution of particle phase by UGKWP method is introduced. Generally, the particle phase kinetic equation Eq.\eqref{particle phase kinetic equ} is split as,
\begin{align}
\label{particle phase kinetic equ without acce}
\mathcal{L}_{s1} &:~~ \frac{\partial f_{s}}{\partial t}
+ \nabla_x \cdot \left(\textbf{u}f_{s}\right)
= \frac{g_{s}-f_{s}}{\tau_{s}}, \\
\label{particle phase kenetic equ only acce}
\mathcal{L}_{s2} &:~~ \frac{\partial f_{s}}{\partial t}
+ \nabla_u \cdot \left(\textbf{a}f_{s} + \textbf{G}f_{s}\right)
= 0.
\end{align}

Firstly, we focus on $\mathcal{L}_{s1}$ part. Consider the particle phase kinetic equation without external force,
\begin{gather*}
\frac{\partial f_{s}}{\partial t}
+ \nabla_x \cdot \left(\textbf{u}f_{s}\right)
= \frac{g_{s}-f_{s}}{\tau_{s}}.
\end{gather*}
For brevity, the subscript $s$ standing for the solid particle phase will be neglected in this subsection. The integration solution of the kinetic equation can be written as,
\begin{equation}\label{particle phase integration solution}
f(\textbf{x},t,\textbf{u}, \bm{\xi})=\frac{1}{\tau}\int_0^t g(\textbf{x}',t',\textbf{u}, \bm{\xi})e^{-(t-t')/\tau}\text{d}t'\\
+e^{-t/\tau}f_0(\textbf{x}-\textbf{u}t, \textbf{u}, \bm{\xi}),
\end{equation}
where $\textbf{x}'=\textbf{x}+\textbf{u}(t'-t)$ is the trajectory of particles, $f_0$ is the initial gas distribution function at time $t=0$, and $g$ is the corresponding equilibrium state.

In UGKWP, both the macroscopic variables and the gas distribution function need to be updated. Generally, in the finite volume framework, the cell-averaged macroscopic variables $\textbf{W}_i$ of cell $i$ can be updated by the following equation,
\begin{gather}
\textbf{W}_i^{n+1} = \textbf{W}_i^n - \frac{1}{\Omega_i} \sum_{S_{ij}\in \partial \Omega_i}\textbf{F}_{ij}S_{ij},
\end{gather}
where $\textbf{W}_i=\left(\rho_i, \rho_i \textbf{U}_i, \rho_i E_i\right)$ is the cell-averaged macroscopic variables,
\begin{gather*}
\textbf{W}_i = \frac{1}{\Omega_{i}}\int_{\Omega_{i}} \textbf{W}\left(\textbf{x}\right) \text{d}\Omega,
\end{gather*}
 $\Omega_i$ is the volume of cell $i$, $\partial\Omega_i$ denotes the set of interface of cell $i$, $S_{ij}$ is one interface of cell $i$, $\textbf{F}_{ij}$ denotes the macroscopic fluxes across the interface $S_{ij}$, which can be written as
\begin{align}\label{particle phase Flux equation}
\textbf{F}_{ij}=\int_{0}^{\Delta t} \int \textbf{u}\cdot\textbf{n}_{ij} f_{ij}(\textbf{x},t,\textbf{u}, \bm{\xi}) \bm{\psi} \text{d}\Xi\text{d}t
\end{align}
where $\textbf{n}_{ij}$ denotes the normal vector of interface $S_{ij}$, $f_{ij}\left(t\right)$ is the time-dependent distribution function on the interface $S_{ij}$, $\bm{\psi}=(1,\textbf{u},\displaystyle \frac{1}{2}(\textbf{u}^2+\bm{\xi}^2))^T$, and $\text{d}\Xi=\text{d}\textbf{u}\text{d}\bm{\xi}$.

Substitute the time-dependent distribution function Eq.\eqref{particle phase integration solution} into Eq.\eqref{particle phase Flux equation}, and the flux can be obtained,
\begin{align*}
\textbf{F}_{ij}
&=\int_{0}^{\Delta t} \int \textbf{u}\cdot\textbf{n}_{ij} f_{ij}(\textbf{x},t,\textbf{u}, \bm{\xi}) \bm{\psi} \text{d}\Xi\text{d}t\\
&=\int_{0}^{\Delta t} \int\textbf{u}\cdot\textbf{n}_{ij} \left[ \frac{1}{\tau}\int_0^t g(\textbf{x}',t',\textbf{u}, \bm{\xi})e^{-(t-t')/\tau}\text{d}t' \right] \bm{\psi} \text{d}\Xi\text{d}t\\
&+\int_{0}^{\Delta t} \int\textbf{u}\cdot\textbf{n}_{ij} \left[ e^{-t/\tau}f_0(\textbf{x}-\textbf{u}t),\textbf{u}, \bm{\xi} \right] \bm{\psi} \text{d}\Xi\text{d}t\\
&\overset{def}{=}\textbf{F}^{eq}_{ij} + \textbf{F}^{fr}_{ij}.
\end{align*}

The procedure of obtaining the local equilibrium state $g_0$ at the cell interface as well as the construction of $g\left(t\right)$ is the same as that in GKS. Here, the construction of $g\left(t\right)$ with acceleration term is simply introduced.

For second order accuracy, the equilibrium state $g$ around the cell interface is written as,
\begin{gather*}
g\left(\textbf{x}',t',\textbf{u},\bm{\xi}\right)=g_0\left(\textbf{x},\textbf{u},\bm{\xi}\right)
\left(1 + \overline{\textbf{a}} \cdot \textbf{u}\left(t'-t\right) + \bar{A}t'\right),
\end{gather*}
where $\overline{\textbf{a}}=\left[\overline{a_1}, \overline{a_2}, \overline{a_3}\right]^T$, $\overline{a_i}=\frac{\partial g}{\partial x_i}/g$, $i=1,2,3$,  $\overline{A}=\frac{\partial g}{\partial t}/g$, and $g_0$ is the local equilibrium on the interface.
More specifically, the coefficients of spatial derivatives $\overline{a_i}$ can be obtained from the associated derivatives of the macroscopic variables,
\begin{equation*}
\left\langle \overline{a_i}\right\rangle=\partial \textbf{W}_0/\partial x_i,
\end{equation*}
where $i=1,2,3$, and $\left\langle...\right\rangle$ means the moments of the Maxwellian distribution functions,
\begin{align*}
\left\langle...\right\rangle=\int \bm{\psi}\left(...\right)g\text{d}\Xi.
\end{align*}
The coefficients of temporal derivative $\overline{A}$ in 3D can be determined by the compatibility condition,
\begin{equation*}
\left\langle \overline{\textbf{a}} \cdot \textbf{u}+\overline{A} \right\rangle =
\left[\begin{array}{c}
0\\
\textbf{0}\\
-\frac{Q_{loss}}{\tau_s}
\end{array}\right].
\end{equation*}
where $Q_{loss}=\frac{\left(1-r^2\right)3p_s}{2}$, and the last term is caused by the lost energy due to the particle-particle inelastic collision.

With the determination of all coefficients in $g\left(\textbf{x}',t',\textbf{u},\bm{\xi}\right)$ for the equilibrium state, its integration becomes
\begin{gather}
f^{eq}(\textbf{x},t,\textbf{u},\bm{\xi}) \overset{def}{=} \frac{1}{\tau}\int_0^t g(\textbf{x}',t',\textbf{u},\bm{\xi})e^{-(t-t')/\tau}\text{d}t' \nonumber\\
= c_1 g_0\left(\textbf{x},\textbf{u},\bm{\xi}\right)
+ c_2 \overline{\textbf{a}} \cdot \textbf{u} g_0\left(\textbf{x},\textbf{u},\bm{\xi}\right)
+ c_3 A g_0\left(\textbf{x},\textbf{u},\bm{\xi}\right),
\end{gather}
with coefficients,
\begin{align*}
c_1 &= 1-e^{-t/\tau}, \\
c_2 &= \left(t+\tau\right)e^{-t/\tau}-\tau, \\
c_3 &= t-\tau+\tau e^{-t/\tau},
\end{align*}
and thereby the integrated flux over a time step for equilibrium state can be obtained,
\begin{gather*}
\textbf{F}^{eq}_{ij}
=\int_{0}^{\Delta t} \int \textbf{u}\cdot\textbf{n}_{ij} f_{ij}^{eq}(\textbf{x},t,\textbf{u},\bm{\xi})\bm{\psi}\text{d}\Xi\text{d}t.
\end{gather*}

Besides, the flux contributed by the free transport of $f_0$ is calculated by tracking the particles sampled from $f_0$. Therefore, the updating of the cell-averaged macroscopic variables can be written as,
\begin{gather}\label{particle phase equ_updateW_ugkp}
\textbf{W}_i^{n+1} = \textbf{W}_i^n - \frac{1}{\Omega_i} \sum_{S_{ij}\in \partial \Omega_i}\textbf{F}^{eq}_{ij}S_{ij}
+ \frac{\textbf{w}_{i}^{fr}}{\Omega_{i}}
+ \Delta t \textbf{S}_{i},
\end{gather}
where $\textbf{w}^{fr}_i$ is the net free streaming flow of cell $i$, standing for the flux contribution of the free streaming of particles, and the term $\textbf{S}_{i} = \left[0,\textbf{0},-\frac{Q_{loss}}{\tau_s}\right]^T$ is the source term in 3D due to the inelastic collision for solid particle phase.

Now it is about how to obtain the net free streaming flow $\textbf{w}^{fr}_i$. The evolution of particle should also satisfy the integral solution of the kinetic equation, which can be written as,
\begin{equation}
f(\textbf{x},t,\textbf{u},\bm{\xi})
=\left(1-e^{-t/\tau}\right)g^{+}(\textbf{x},t,\textbf{u},\bm{\xi})
+e^{-t/\tau}f_0(\textbf{x}-\textbf{u}t,\textbf{u},\bm{\xi}),
\end{equation}
where $g^{+}$ is named as the hydrodynamic distribution function with analytical formulation. The initial distribution function $f_0$ have a probability of $e^{-t/\tau}$ to free transport and $1-e^{-t/\tau}$ to colliding with other particles. The post-collision particles satisfies the distribution $g^+\left(\textbf{x},\textbf{u},t\right)$. The free transport time before the first collision with other particles is called $t_c$. The cumulative distribution function of $t_c$ is,
\begin{gather}\label{particle phase wp cumulative distribution}
F\left(t_c < t\right) = 1 - e^{-t/ \tau},
\end{gather}
and therefore $t_c$ can be sampled as $t_c=-\tau \text{ln}\left(\eta\right)$, where $\eta$ is a random number generated from a uniform distribution $U\left(0,1\right)$. Then the free streaming time $t_f$ for particle $k$ is determined by,
\begin{gather}
t_f = \min \left[-\tau\text{ln}\left(\eta\right), \Delta t\right],
\end{gather}
where $\Delta t$ is the time step. Therefore, in one time step, all the particles can be divided into two groups: the collisionless particles and the collisional particles, which are determined by the relation between of the time step $\Delta t$ and free streaming time $t_f$. Specifically, if $t_f=\Delta t$ for one particle, it is collisionless particle, and the trajectory of this particle is fully tracked in the whole time step. On the contrary, if $t_f<\Delta t$ for the particle, it is the collisional particle and its trajectory will be tracked until $t_f$. Subsequently, this particle is eliminated in the simulation, and the mass, momentum and energy of this particle are merged into the macroscopic quantities of the relevant cell. The particle trajectory in the free streaming process within the time $t_f$ is tracked by
\begin{gather}
\textbf{x} = \textbf{x}^n + \textbf{u}^n t_f .
\end{gather}

Then term $\textbf{w}_{i}^{fr}$ can be calculated by counting the particles passing through the interfaces of cell $i$,
\begin{gather}
\textbf{w}_{i}^{fr} = \sum_{k\in P\left(\partial \Omega_{i}^{+}\right)} \bm{\phi}_k - \sum_{k\in P\left(\partial \Omega_{i}^{-}\right)} \bm{\phi}_k,
\end{gather}
where, $P\left(\partial \Omega_{i}^{+}\right)$ is the particle set moving into the cell $i$ during one time step, $P\left(\partial \Omega_{i}^{-}\right)$ is the particle set moving out of the cell $i$ during one time step, $k$ is the particle index in one specific set, and $\bm{\phi}_k=\left[m_{k}, m_{k}\textbf{u}_k, \frac{1}{2}m_{k}(\textbf{u}^2_k+e_{k})\right]^T$ is the mass, momentum and energy carried by particle $k$. Therefore, $\textbf{w}_{i}^{fr}/\Omega_{i}$ is the net conservative quantities caused by the free streaming of all tracked particles. Now, all the terms in Eq.\eqref{particle phase equ_updateW_ugkp} have been determined and the macroscopic variables $\textbf{W}_i$ can be updated.

The trajectories of all the particles have been tracked during the time interval $\left(0, t_f\right)$. For the collisionless particles $t_f=\Delta t$, they still survive at the end of one time step; while the collisional particles $t_f<\Delta t$ are deleted after their first collision with other particles and the overall effect of the collisions is to make the collisional particles follow the local equilibrium distribution. Therefore, the macroscopic variables of the collisional particles in cell $i$ at the end of each time step can be directly obtained based on the conservation law,
\begin{gather}
\textbf{W}^h_i = \textbf{W}^{n+1}_i - \textbf{W}^p_i,
\end{gather}
where $\textbf{W}^p_i$ are the total conservative variables from remaining collisionless particles at the end of one time step. Besides, the macroscopic variables $\textbf{W}^h_i$ is coming from eliminated collisional particles, which can be recovered by re-sampling the collisional particles according to $\textbf{W}^h_i$ based on a Maxwellian distribution. Now the update of both macroscopic variables as well the microscopic particles have been finished. The above method is so-called unified gas-kinetic particle (UGKP) method.

The UGKWP method further develops UGKP as follows.
In UGKP method, all particles can be divided into two types: the collisionless particles, surviving at the end of one time step, and collisional particles, deleted after the first collision and re-sampled at the end of one time step. To further improve the efficiency, only the collisionless particles, which are capable of surviving in the next whole time step, need to be sampled from $\textbf{W}^h_i$.
The collisional particles from $\textbf{W}^h_i$ in the next time step can be represented as hydrodynamic wave with analytical solution.
More specifically, in the next time step the transport flux from these un-sampled particles can be evaluated analytically.
According to the cumulative distribution Eq.\eqref{particle phase wp cumulative distribution}, the proportion of the collisionless particles is $e^{-\Delta t/\tau}$, and therefore in UGKWP the macroscopic variables of the particles required to be sampled from the hydrodynamic variables $\textbf{W}^{h}_i$ in cell $i$ are,
\begin{gather}
\textbf{W}^{hp}_i = e^{-\Delta t/\tau} \textbf{W}^{h}_i.
\end{gather}
Note that, the free transport time of all the sampled particles is $t_f=\Delta t$ in UGKWP. Now, the net flow determined by the free streaming of particles are denoted as $\textbf{w}_{i}^{fr,p}$, which can be calculated with the same way,
\begin{gather}
\textbf{w}_{i}^{fr,p} = \sum_{k\in P\left(\partial \Omega_{i}^{+}\right)} \bm{\phi}_k - \sum_{k\in P\left(\partial \Omega_{i}^{-}\right)} \bm{\phi}_k.
\end{gather}
Besides, the fluxes contributed by the free transport of the collisional particles are evaluated as,
\begin{align*}
\textbf{F}^{fr,wave}_{ij}
&=\textbf{F}^{fr,UGKS}_{ij}(\textbf{W}^h_i) - \textbf{F}^{fr,DVM}_{ij}(\textbf{W}^{hp}_i) \\
&=\int_{0}^{\Delta t} \int \textbf{u} \cdot \textbf{n}_{ij} \left[ e^{-t/\tau}f_0(\textbf{x}-\textbf{u}t,\textbf{u},\bm{\xi})\right] \bm{\psi} \text{d}\Xi\text{d}t\\
&-e^{-\Delta t/\tau}\int_{0}^{\Delta t} \int \textbf{u} \cdot \textbf{n}_{ij} \left[g_0^h\left(\textbf{x},\textbf{u},\bm{\xi}\right) - t\textbf{u} \cdot g_\textbf{x}^h\left(\textbf{x},\textbf{u},\bm{\xi}\right) \right] \bm{\psi}\text{d}\Xi\text{d}t\\
&=\int \textbf{u} \cdot \textbf{n}_{ij} \left[ \left(q_4  - \Delta t e^{-\Delta t/\tau}\right) g_0^h \left(\textbf{x},\textbf{u},\bm{\xi}\right)
+ \left(q_5 + \frac{\Delta t^2}{2}e^{-\Delta t/\tau}\right) \textbf{u} \cdot g_\textbf{x}^h\left(\textbf{x},\textbf{u},\bm{\xi}\right) \right]\bm{\psi}\text{d}\Xi,
\end{align*}
with the coefficients,
\begin{align*}
q_4&=\tau\left(1-e^{-\Delta t/\tau}\right), \\
q_5&=\tau\Delta te^{-\Delta t/\tau} - \tau^2\left(1-e^{-\Delta t/\tau}\right).
\end{align*}

The macroscopic flow variables in UGKWP can be updated,
\begin{gather}\label{particle phase wp final update W}
\textbf{W}_i^{n+1} = \textbf{W}_i^n
- \frac{1}{\Omega_i} \sum_{S_{ij}\in \partial \Omega_i}\textbf{F}^{eq}_{ij}S_{ij}
- \frac{1}{\Omega_i} \sum_{S_{ij}\in \partial \Omega_i}\textbf{F}^{fr,wave}_{ij}S_{ij}
+ \frac{\textbf{w}_{i}^{fr,p}}{\Omega_{i}}
+ \Delta t \textbf{S}_{i}.
\end{gather}

The second part is evaluating the effect of the acceleration term according to the associated governing equation Eq.\eqref{particle phase kenetic equ only acce},
\begin{gather*}
\frac{\partial f_{s}}{\partial t}
+ \nabla_u \cdot \left(\textbf{a}f_{s} + \textbf{G}f_{s}\right)
= 0,
\end{gather*}
where velocity-dependent acceleration term caused by the inter-phase forces has the following form,
\begin{gather*}
\textbf{a} = \frac{\textbf{D}+\textbf{F}_b}{m_s} = \frac{\textbf{U}_g - \textbf{u}}{\tau_{st}} - \frac{1}{\rho_{s}} \nabla_x p_g.
\end{gather*}

Take moment with $\bm{\psi}=\left(1,\textbf{u},\displaystyle \frac{\textbf{u}^2+\bm{\xi}^2}{2}\right)^T$ for Eq.\eqref{particle phase kenetic equ only acce},
\begin{gather*}
\int \bm{\psi}
\left( \frac{\partial f_{s}}{\partial t}
+ \textbf{a} \cdot \nabla_u f_{s}
+ f_{s}\nabla_u \cdot \textbf{a}
+ \textbf{G} \cdot \nabla_u f_{s}
+ f_{s}\nabla_u \cdot \textbf{G}
\right) \text{d}\Xi = 0,
\end{gather*}
and in Euler regime $f_s = g_s + \mathcal{O}\left(\tau_{s}\right)$, we can obtain,
\begin{gather*}
\frac{\partial \textbf{W}_s}{\partial t} + \left(\textbf{Q} + \textbf{Q}_G\right)= 0,
\end{gather*}
where
\begin{gather*}
\textbf{W}_s=\left[\begin{array}{c}
\epsilon_s\rho_s\\
\epsilon_s\rho_s \textbf{U}_s\\
\epsilon_s\rho_s E_s
\end{array}
\right], ~~
\textbf{Q}=\left[\begin{array}{c}
0 \\
\frac{\epsilon_s\rho_s\left(\textbf{U}_s-\textbf{U}_g\right)}{\tau_{st}}+\epsilon_s \nabla_x p_g\\
\frac{\epsilon_s\rho_{s}\textbf{U}_s \cdot \left(\textbf{U}_s-\textbf{U}_g\right)}{\tau_{st}} +3\frac{p_s}{\tau_{st}} + \epsilon_s\textbf{U}_s \cdot \nabla_x p_g
\end{array}\right],~~
\textbf{Q}_G=\left[\begin{array}{c}
0 \\
- \epsilon_{s}\rho_{s} \textbf{G}\\
- \epsilon_{s}\rho_{s} \textbf{U}_s \cdot \textbf{G}
\end{array}\right],
\end{gather*}
and $p_s=\frac{\epsilon_s\rho_s}{2\lambda_s}$ is the granular pressure of particle phase.

When the first order forward Euler method is employed for time marching, the cell-averaged macroscopic variables for three-dimensional problems can be updated by
\begin{gather}\label{update macroscopic variable of acceleration wave}
\textbf{W}^{n+1}_s = \textbf{W}_s - \left(\textbf{Q} + \textbf{Q}_G\right) \Delta t,
\end{gather}
and the modifications on velocity and location of the remaining free streaming particles are
\begin{align}
\textbf{u}^{n+1} &= \textbf{u} + \left(\textbf{a} + \textbf{G}\right)t_f,\\
\textbf{x}^{n+1} &= \textbf{x} + \frac{\left(\textbf{a} + \textbf{G}\right)}{2} t_f^2 .\label{displacement by acceleartion term}
\end{align}

Finally, the evolution of the material temperature $T^m_s$ of the solid phase is calculated. Two factors need to be considered. The first one is the effect of the heat conduction between particle phase and gas phase. The exchanged energy rate $q_{cond}$ between particle phase and gas phase due to the heat conduction can be evaluated as,
\begin{gather*}
q_{cond} = \epsilon_s\rho_sC_s \frac{T_g-T^m_s}{\tau_T}
\end{gather*}
The second factor is the lost energy due to inelastic collision for particle phase, which is,
\begin{gather*}
q_{loss} = \frac{Q_{loss}}{\tau_s}=\frac{\left(1-r^2\right)3p_s}{2\tau_s},
\end{gather*}
where $r\in\left[0,1\right]$ is the restitution coefficient, determining the percentage of lost energy in inelastic collision. Here we assume that all the lost energy of inelastic collision is transferred to the thermal energy of solid particles. With the determination of $q_{cond}$ and $q_{loss}$, Eq.\eqref{particle phase Tm equ} for the evolution of material temperature of solid particles $T^m_s$ is
\begin{gather*}
\frac{\partial \left(\epsilon_s\rho_sC_sT^m_s\right)}{\partial t}
+ \nabla_x \cdot \left(\epsilon_s\rho_sC_sT^m_s \textbf{U}_s\right)
= \epsilon_s\rho_sC_s \frac{T_g-T^m_s}{\tau_T}
+ \frac{\left(1-r^2\right)3p_s}{2\tau_s}.
\end{gather*}
In this paper, the Lax-Friedrichs method is employed to calculate the flux and then the material temperature $T^m_s$ is updated in the finite volume framework. Now the update for particle phase in one time step has been finished.

\subsection{Gas kinetic scheme for gas phase}

 The gas phase is governed by Navier-Stokes equations and the corresponding scheme is the GKS, which is the limiting scheme of UGKWP in the
  continuum flow regime. In general, the gas phase kinetic equation Eq.\eqref{gas phase kinetic equ} is split into two parts,
\begin{align}
\label{gas phase kinetic equ without acce}
\mathcal{L}_{g1} &:~~ \frac{\partial f_{g}}{\partial t}
+ \nabla_x \cdot \left(\textbf{u}f_{g}\right)
= \frac{g_{g}-f_{g}}{\tau_{g}}, \\
\label{gas phase kenetic equ only acce}
\mathcal{L}_{g2} &:~~ \frac{\partial f_{g}}{\partial t}
- \nabla_u \cdot \left(\textbf{a}f_{s}\right)
= 0.
\end{align}

Firstly, the kinetic equation without acceleration term $\mathcal{L}_{g1}$ is solved,
\begin{gather*}
\frac{\partial f_{g}}{\partial t}
+ \nabla_x \cdot \left(\textbf{u}f_{g}\right)
= \frac{g_{g}-f_{g}}{\tau_{g}}.
\end{gather*}
Here the collision term satisfies the compatibility condition
\begin{equation*}
\int \frac{g-f}{\tau} \bm{\psi} \text{d}\Xi=0,
\end{equation*}
where $\bm{\psi}=(1,\textbf{u},\displaystyle \frac{1}{2}(\textbf{u}^2+\bm{\xi}^2))^T$, the internal variables $\bm{\xi}^2=\xi_1^2+...+\xi_K^2$, and $\text{d}\Xi=\text{d}\textbf{u}\text{d}\bm{\xi}$. $K$ is the internal degree of freedom of each molecule which is related to the
specific heat ratio $\gamma$.

Based on Eq.\eqref{gas phase kinetic equ without acce}, the solution of $f$ at a cell interface can be written as,
\begin{equation}\label{gas phase equ_integral1}
f(\textbf{x},t,\textbf{u},\bm{\xi})=\frac{1}{\tau}\int_0^t g(\textbf{x}',t',\textbf{u},\bm{\xi})e^{-(t-t')/\tau}\text{d}t'\\
+e^{-t/\tau}f_0(\textbf{x}-\textbf{u}t,\textbf{u},\bm{\xi}),
\end{equation}
where $\textbf{x}'=\textbf{x}+\textbf{u}(t'-t)$ is the particle trajectory, $f_0$ is the initial gas distribution function at time $t=0$, and $g$ is the corresponding equilibrium state.
The initial gas distribution function $f_0$ can be constructed as
\begin{equation}\label{gas phase equ_f0}
f_0=f_0^l(\textbf{x},\textbf{u})H(x)+f_0^r(\textbf{x},\textbf{u})(1-H(x)),
\end{equation}
where $H(x)$ is the Heaviside function, $f_0^l$ and $f_0^r$ are the
initial gas distribution functions on the left and right side of one cell interface, which can be determined by the corresponding macroscopic variables.
The initial gas distribution function $f_0^k$, $k=l,r$, is constructed as
\begin{equation*}
f_0^k=g^k\left(1+\textbf{a}^k \cdot \textbf{x}-\tau(\textbf{a}^k \cdot \textbf{u}+A^k)\right),
\end{equation*}
where $g^l$ and $g^r$ are the Maxwellian distribution functions on the left and right hand sides of a cell interface, which are determined by the corresponding conservative variables $\textbf{W}^l$ and $\textbf{W}^r$. The coefficients $\textbf{a}^l=\left[a^l_1, a^l_2, a^l_3\right]^T$, $\textbf{a}^r=\left[a^r_1, a^r_2, a^r_3\right]^T$, are related to the spatial derivatives in normal and tangential directions, which are obtained from the derivatives of initial macroscopic flow variables,
\begin{equation*}
\left\langle a^l_i\right\rangle=\partial \textbf{W}^l/\partial x_i,
\left\langle a^r_i\right\rangle=\partial \textbf{W}^r/\partial x_i,
\end{equation*}
where $i=1,2,3$, and $\left\langle...\right\rangle$ means the moments of the Maxwellian distribution functions,
\begin{align*}
\left\langle...\right\rangle=\int \bm{\psi}\left(...\right)g\text{d}\Xi.
\end{align*}
Based on the Chapman-Enskog expansion, the non-equilibrium part of the distribution function satisfies
\begin{equation*}
\left\langle \textbf{a}^l \cdot\textbf{u}+A^l\right\rangle = 0,~
\left\langle \textbf{a}^r \cdot\textbf{u}+A^r\right\rangle = 0,
\end{equation*}
from which the coefficients $A^l$ and $A^r$ are fully determined.
The equilibrium state $g$ around the cell interface is modeled as,
\begin{equation}\label{gas phase equ_g}
g=g_0\left(1+\overline{\textbf{a}}\cdot\textbf{x}+\bar{A}t\right),
\end{equation}
where $\overline{\textbf{a}}=\left[\overline{a}_1, \overline{a}_2, \overline{a}_3\right]^T$, $g_0$ is the local equilibrium of the cell interface. More specifically, $g$ can be determined by the compatibility condition,
\begin{align*}
\int\bm{\psi} g_{0}\text{d}\Xi=\textbf{W}_0
&=\int_{u>0}\bm{\psi} g^{l}\text{d}\Xi+\int_{u<0}\bm{\psi} g^{r}\text{d}\Xi, \nonumber \\
\int\bm{\psi} \overline{a_i} g_{0}\text{d}\Xi=\partial \textbf{W}_0/\partial x_i
&=\int_{u>0}\bm{\psi} a^l_i g^{l}\text{d}\Xi+\int_{u<0}\bm{\psi} a^r_i g^{r}\text{d}\Xi,
\end{align*}
$i=1,2,3$, and
\begin{equation*}
\left\langle \overline{\textbf{a}} \cdot \textbf{u}+\bar{A}\right\rangle = 0.
\end{equation*}
After determining all parameters in the initial gas distribution function $f_0$ and the equilibrium state $g$, substituting Eq.\eqref{gas phase equ_f0} and Eq.\eqref{gas phase equ_g} into Eq.\eqref{gas phase equ_integral1} the time-dependent distribution function $f(\textbf{x}, t, \textbf{u},\bm{\xi})$ at a cell interface can be expressed as,
\begin{align}\label{gas phase equ_finalf}
f(\textbf{x}, t, \textbf{u},\bm{\xi})
&=c_1 g_0+ c_2 \overline{\textbf{a}}\cdot\textbf{u}g_0 +c_3 {\bar{A}} g_0\nonumber\\
&+\left[c_4 g^r +c_5 \textbf{a}^r\cdot\textbf{u} g^r + c_6 A^r g^r\right] (1-H(u)) \\
&+\left[c_4 g^l +c_5 \textbf{a}^l\cdot\textbf{u} g^l + c_6 A^l g^l\right] H(u) \nonumber,
\end{align}
with coefficients,
\begin{align*}
c_1 &= 1-e^{-t/\tau}, \\
c_2 &= \left(t+\tau\right)e^{-t/\tau}-\tau, \\
c_3 &= t-\tau+\tau e^{-t/\tau}, \\
c_4 &= e^{-t/\tau}, \\
c_5 &= -\left(t+\tau\right)e^{-t/\tau}, \\
c_6 &= -\tau e^{-t/\tau}.
\end{align*}
Then, the integrated flux over a time step can be obtained,
\begin{align}
\textbf{F} =\int_{0}^{\Delta t} \int\textbf{u}\cdot\textbf{n} f(\textbf{x},t,\textbf{u},\bm{\xi})\bm{\psi}\text{d}\Xi\text{d}t,
\end{align}
where $\textbf{n}$ is the unit vector in the outer normal direction of the associated cell interface.
Then, the cell-averaged conservative variables of cell $i$ can be updated as follows,
\begin{gather}
\textbf{W}_i^{n+1} = \textbf{W}_i^n
- \frac{1}{\Omega_i} \sum_{S_{ij}\in \partial \Omega_i}\textbf{F}_{ij}S_{ij},
\end{gather}
where $\Omega_i$ is the volume of cell $i$, $\partial\Omega_i$ denotes the set of interfaces of cell $i$, $S_{ij}$ is the area of the cell interface, $\textbf{F}_{ij}$ denotes the projected macroscopic fluxes in the normal direction, and $\textbf{W}_{g}=\left[\widetilde{\rho_g},\widetilde{\rho_g} \textbf{U}_g, \widetilde{\rho_g} E_g\right]^T$ are the cell-averaged conservative variables for gas phase.
With the interactive force between gas and particle phase, the increased momentum and energy in gas phase can be calculated as
\begin{gather}
\textbf{W}^{n+1}_g = \textbf{W}_g + \textbf{Q}\Delta t,
\end{gather}
where
\begin{gather*}
\textbf{W}_g=\left[\begin{array}{c}
\widetilde{\rho_g}\\
\widetilde{\rho_g} \textbf{U}_g\\
\widetilde{\rho_g} E_g
\end{array}
\right], ~~
\textbf{Q}=\left[\begin{array}{c}
0 \\
\frac{\epsilon_s\rho_s\left(\textbf{U}_s-\textbf{U}_g\right)}{\tau_{st}}+\epsilon_s \nabla_x p_g\\
\frac{\epsilon_s\rho_{s}\textbf{U}_s\cdot\left(\textbf{U}_s-\textbf{U}_g\right)}{\tau_{st}} + 3\frac{p_s}{\tau_{st}} + \epsilon_s\textbf{U}_s \cdot \nabla_x p_g
\end{array}\right].
\end{gather*}

Finally, the heat conduction between gas flow and solid particles will also affect the temperature of gas flow, which is governed by  Eq.\eqref{gas phase T heat conduction},
\begin{gather*}
\frac{\text{d} \left(\widetilde{\rho_g}C_gT_g\right)}{\text{d} t} = -\epsilon_s\rho_sC_s \frac{T_g-T^m_s}{\tau_T}.
\end{gather*}
The temperature of gas phase $T_g$ can be updated by the analytical solution from the above equation.

With the updates of macroscopic flow variables in the particle and gas phases, the volume fractions $\epsilon_s$ and $\epsilon_g$ are
updated as follows:
\begin{itemize}
	\item [1)] Update the apparent density of solid particle phase and gas phase,  $\epsilon_{s}\rho_{s}$ and $\widetilde{\rho_g}$.
	\item [2)]
	The volume fraction of solid particle phase is given by
	\begin{equation*}
	\epsilon_{s} = \frac{\epsilon_{s}\rho_{s}}{\rho_{s}},
	\end{equation*}
	where $\rho_{s}$ is a material dependent constant of the solid particle.
	\item [3)]
	The volume fraction of gas phase is calculated according to the relation between $\epsilon_{s}$ and $\epsilon_g$,
	\begin{equation*}
	\epsilon_g = 1 - \epsilon_s.
	\end{equation*}
\end{itemize}

\subsection{Limiting case analysis}

The limiting cases of the gas-particle two-phase system will be analyzed. In general, the gas-particle flow regime is determined by two dimensionless numbers, i.e., the Knudsen number $Kn_s$ and Stokes number $St_s$.
The Knudsen number is defined by the ratio of collision time of solid particles to the characteristic time of macroscopic flow,
\begin{gather}\label{particle phase Kn_s}
Kn_s = \frac{\tau_s}{t_{ref}}.
\end{gather}
Specifically, $\tau_s$ is the time interval between collisions of solid particles, or the called relaxation time of particle phase.
According to the previous studies \cite{Gasparticle-MOM-Fox-passalacqua2010fully, Gasparticle-momentmethod-Fox2013computational}, in this paper $\tau_s$ is taken as,
\begin{gather}\label{particle phase tau_s}
\tau_s = \frac{\sqrt{\pi}d_s}{12\epsilon_sg_0}\sqrt{2\lambda_s},
\end{gather}
where $d_s$ is the diameter of solid particle, $\epsilon_s$ is the volume fraction of solid phase, and $\lambda_s=\frac{1}{2RT_s}$ is
 related to the granular temperature of solid phase $T_s$. $g_0$ is the radial distribution function with the following form,
\begin{gather}
g_0 = \frac{2-c}{2\left(1-c\right)^3},
\end{gather}
where $c=\epsilon_s/\epsilon_{s,max}$ is the ratio of the volume fraction $\epsilon_{s}$ to the maximum allowed volume fraction of solid phase $\epsilon_{s,max}$. In dilute regime, $\epsilon_s \ll \epsilon_{s,max}$, which leads to $c \to 0$ and $g_0 \to 1$. On the contrary, when $\epsilon_s \to\epsilon_{s,max}$, we have $g_0 \to \infty$ and thus $\tau_s \to 0$, which usually occurs in dense or highly dense flow. In this paper, we mainly focus on dilute flow, and therefore $g_0$ can be neglected. Besides, $t_{ref}$ is the characteristic time scale, which is defined as the ratio flow characteristic length to the flow characteristic velocity, $t_{ref}=L_{ref}/U_{ref}$. $Kn_s$ determines the flow regime of particle phase. When $Kn_s$ is very large, i.e., $\tau_s \to \infty$, the collision time is much large than the characteristic time scale, and no  particle-particle collision occurs. Numerically, the fluxes associated with wave, such as $\textbf{F}^{eq}_{ij}$ and $\textbf{F}^{fr,wave}_{ij}$ in Eq.\eqref{particle phase wp final update W}, become zero. Therefore, the collisionless particle transport solution with the interaction with the gas phase alone is obtained by UGKWP. On the contrary, when $Kn_s$ is very small, i.e., $\tau_s \to 0$, the particle phase will get to the continuum flow regime
and only the wave part will be left in the update of solutions in Eq.\eqref{particle phase wp final update W}.

Another dimensionless number $St_s$ is defined by,
\begin{gather}\label{particle phase tau_st}
St_s = \frac{\tau_{st}}{t_{ref}},
\end{gather}
where $\tau_{st}$ is the particle internal response time, which can be obtained from the drag force, and $t_{ref}$ is the same with the above definition. $St_s$ determines the interactive coupling between the particle phase and gas phase.
When $St_s$ is very large, i.e., $\tau_{st} \to \infty$, the drag fore term $\textbf{D}$ is nearly zero, where the particle phase is decoupled from the gas phase. In this case, the particle phase cannot be driven by gas flow and PTC phenomena will occur.
On the other hand, a small $St_s \to 0 $ leads to $\textbf{u} \to \textbf{U}_g$, where the solid particle will have the same velocity as gas flow and the single mixture phase will follow the fluid dynamic equations.

\section{Numerical test cases}
\subsection{Wind sand shock tube}
\begin{figure}[htbp]
	\centering
	\subfigure{
		\includegraphics[height=4.5cm]{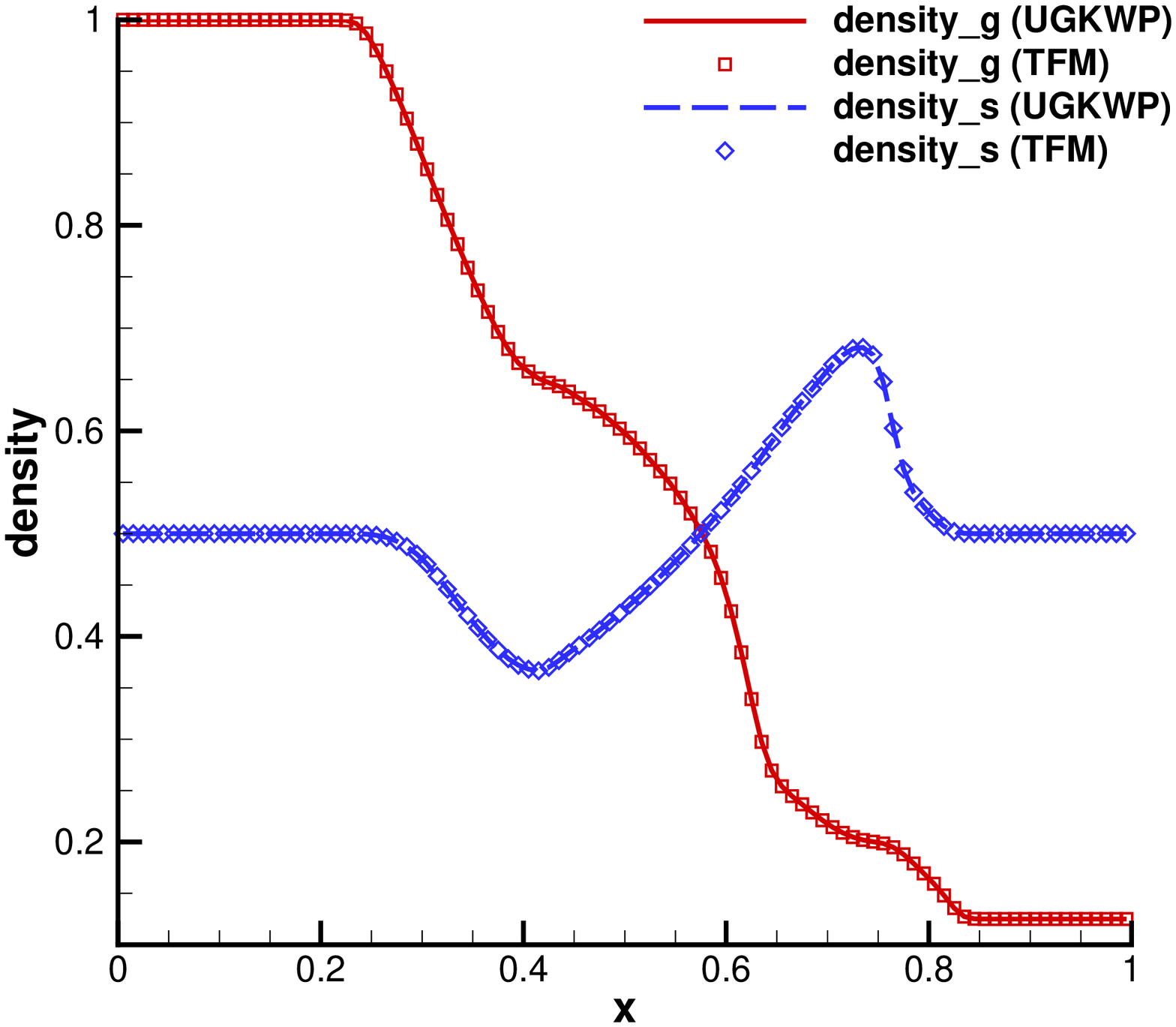}	
	}
	\quad
	\subfigure{
		\includegraphics[height=4.5cm]{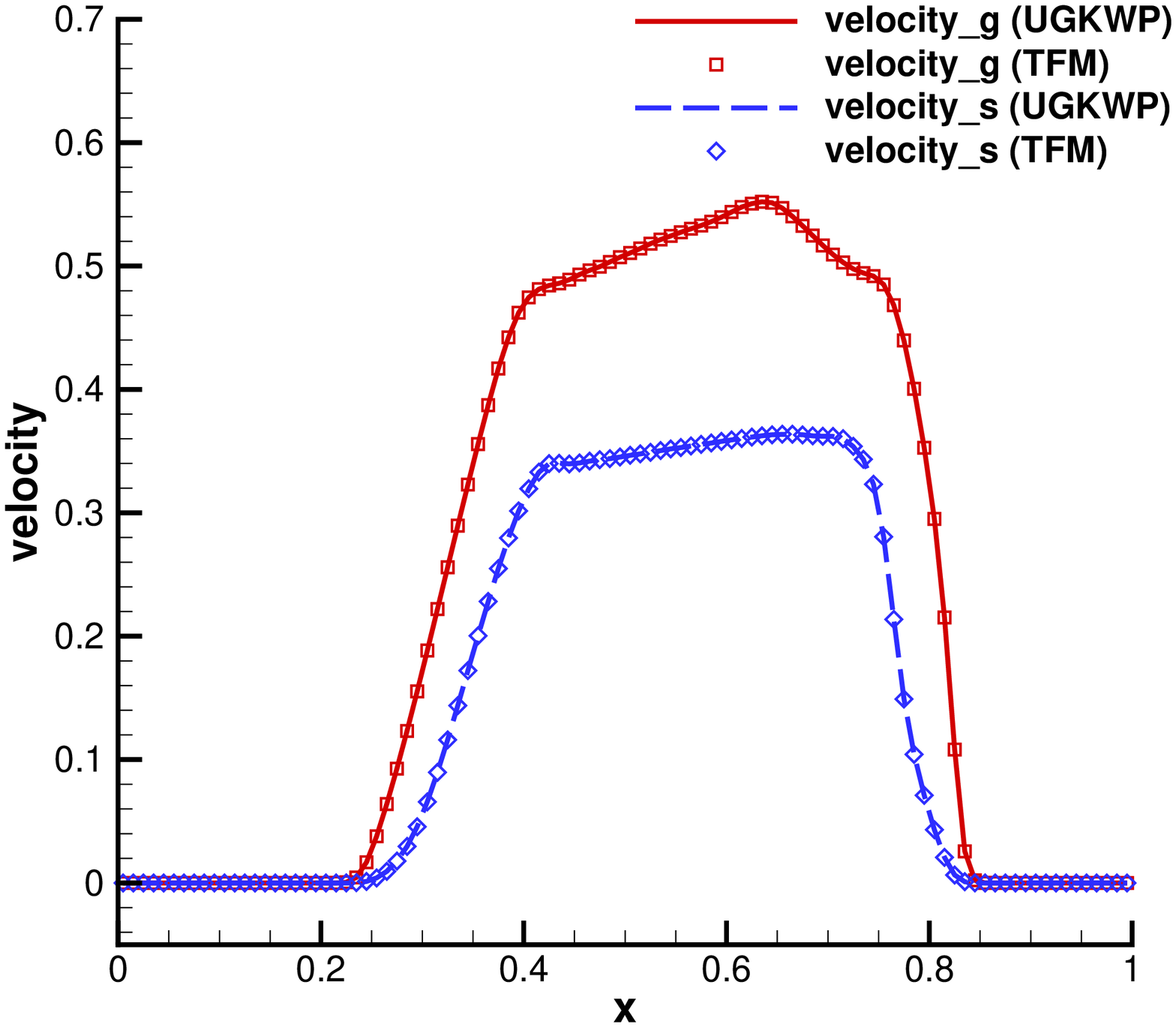}	
	}

	\subfigure{
		\includegraphics[height=4.5cm]{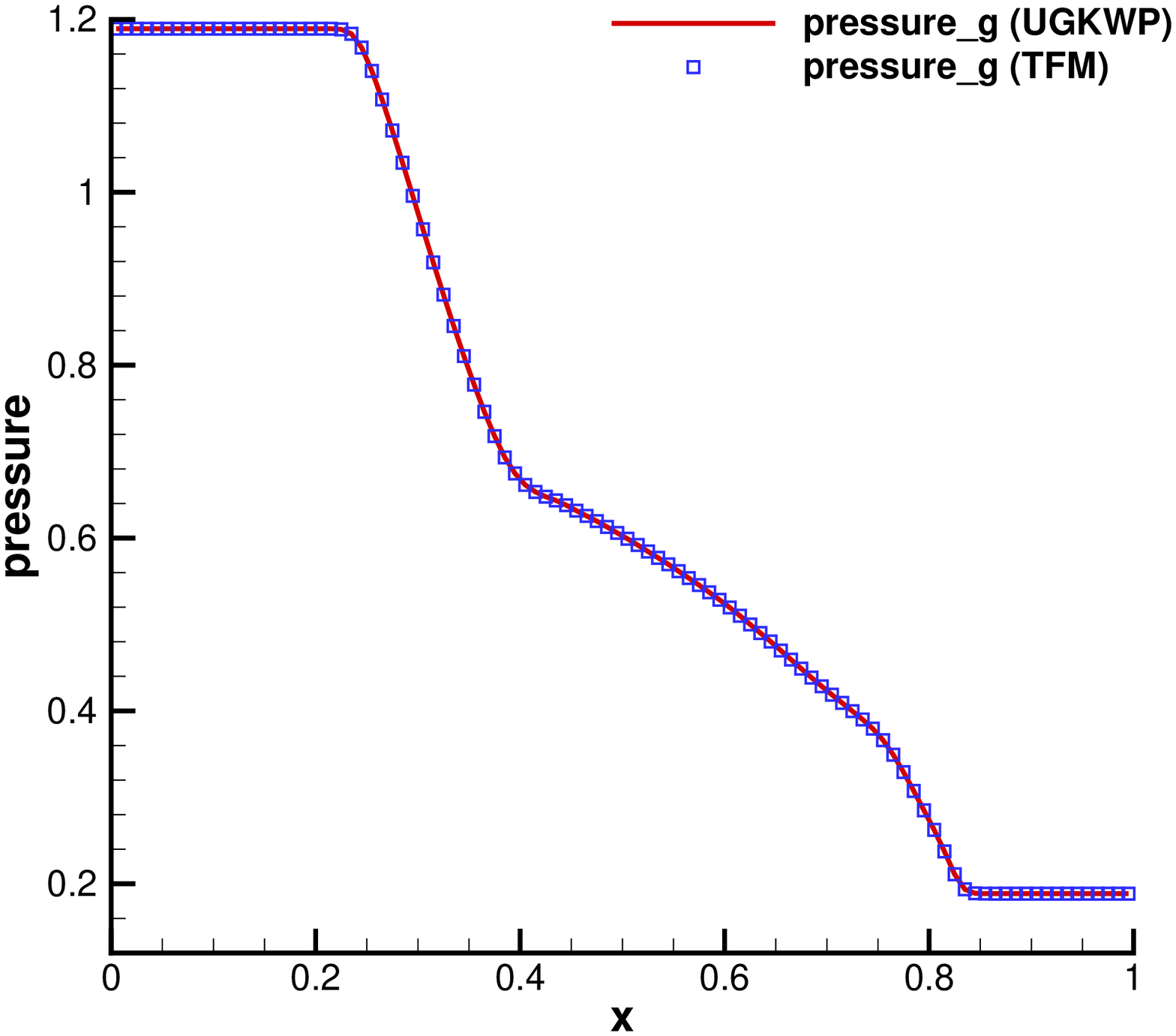}	
	}
	\quad
	\subfigure{
		\includegraphics[height=4.5cm]{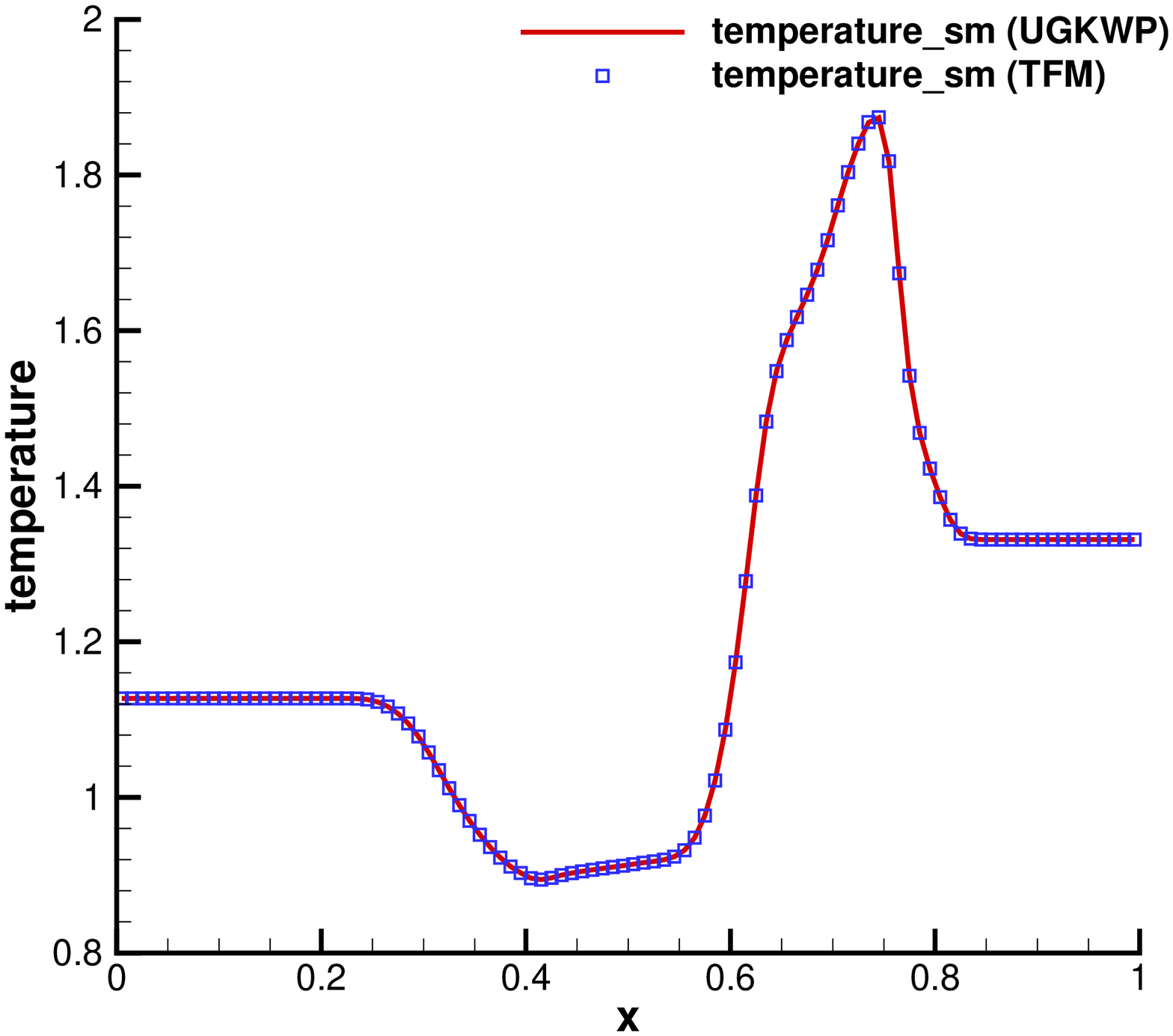}	
	}
	\caption{Wind sand shock tube problem. Solutions from the UGKWP and TFM in highly collisional regime with $t=0.2$ with $Kn_s=10^{-5}$, $\tau_{st}=0.1$, $\tau_T=0.1$. Apparent density of gas and particle phase, velocity of gas and particle phase, pressure of gas phase, and material temperature of solid particles are included.}
	\label{wind sand Kn1m5}
\end{figure}

\begin{figure}[htbp]
	\centering
	\subfigure{
		\includegraphics[height=4.5cm]{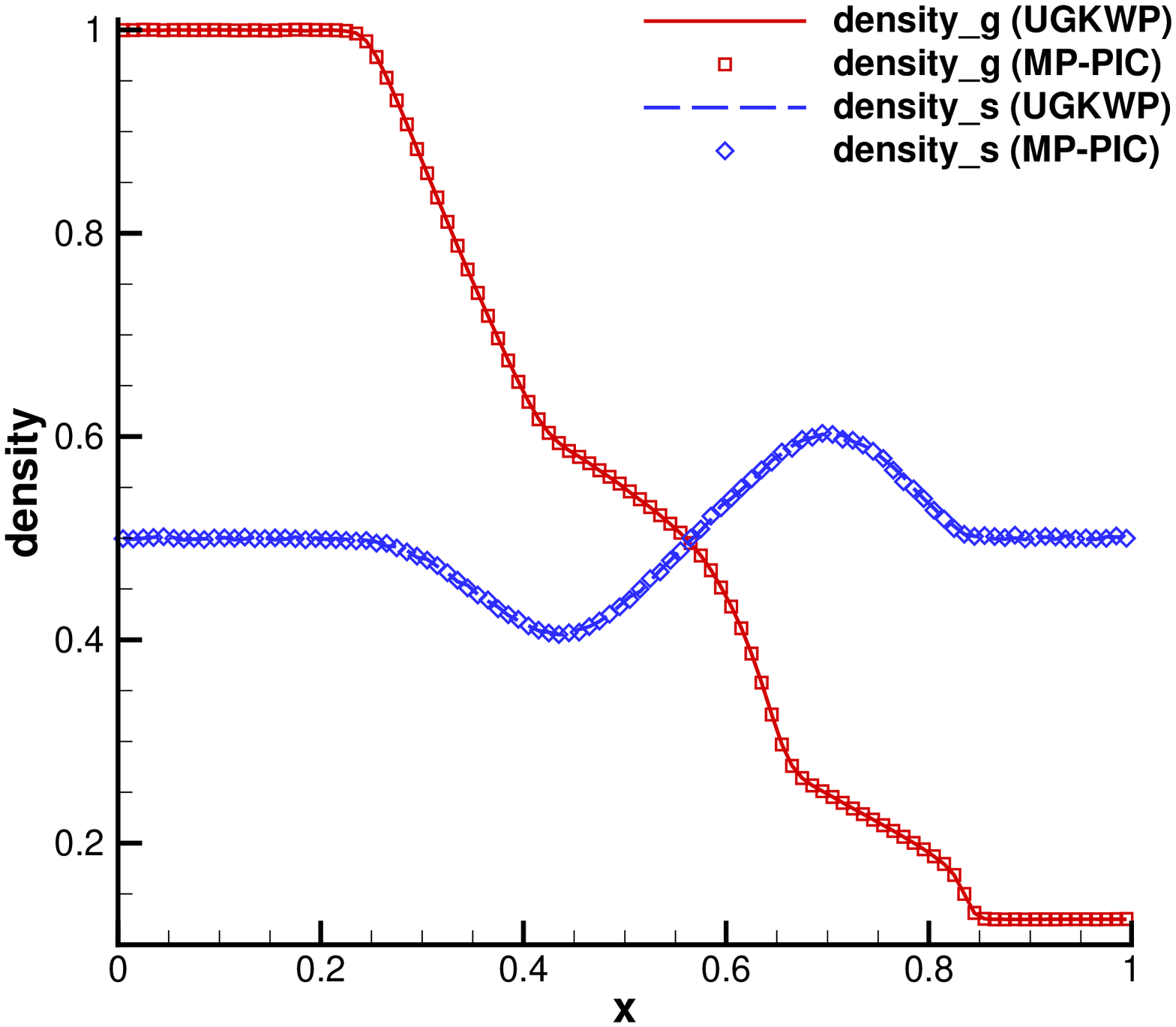}	
	}
	\quad
	\subfigure{
		\includegraphics[height=4.5cm]{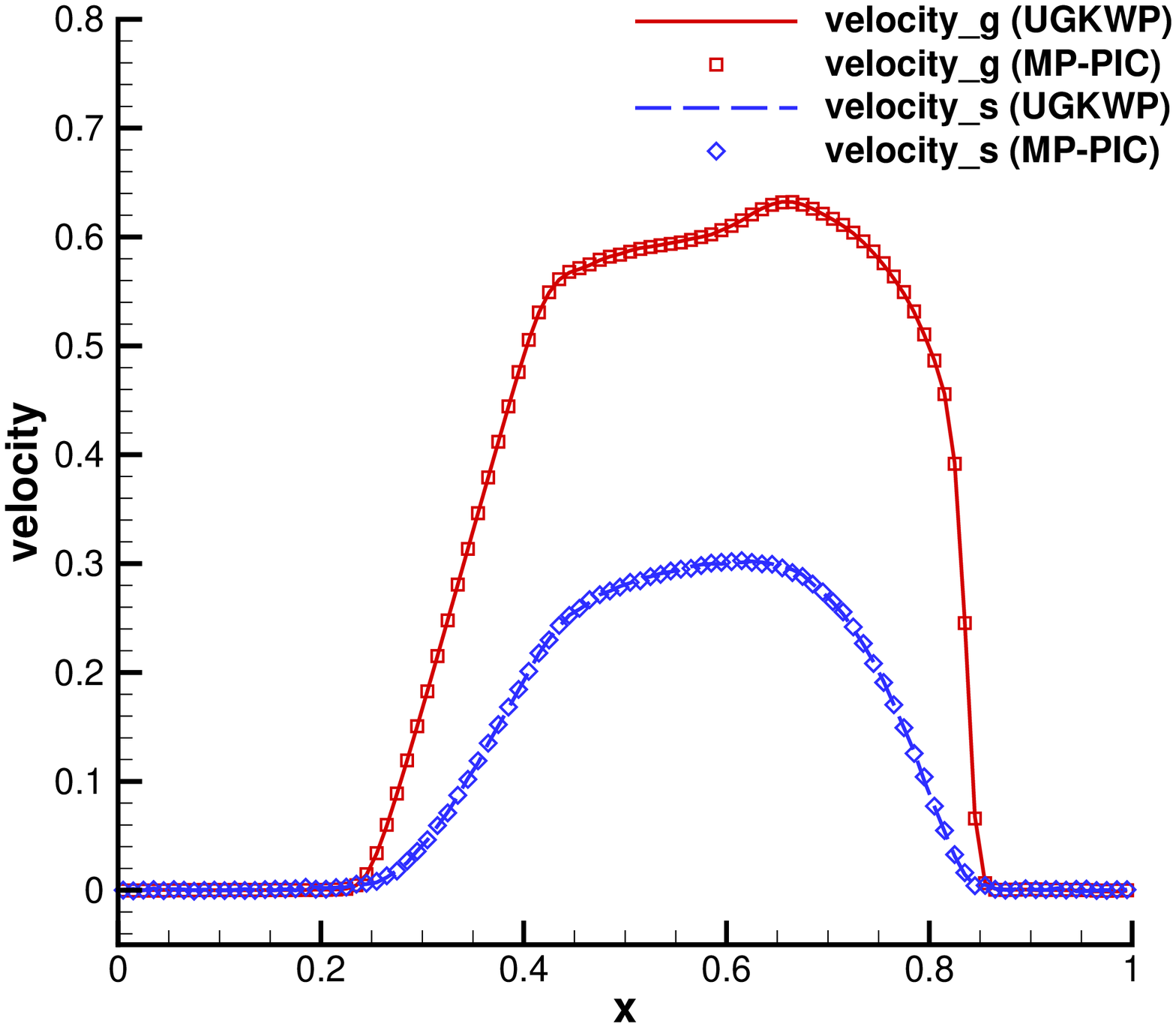}	
	}
	
	\subfigure{
		\includegraphics[height=4.5cm]{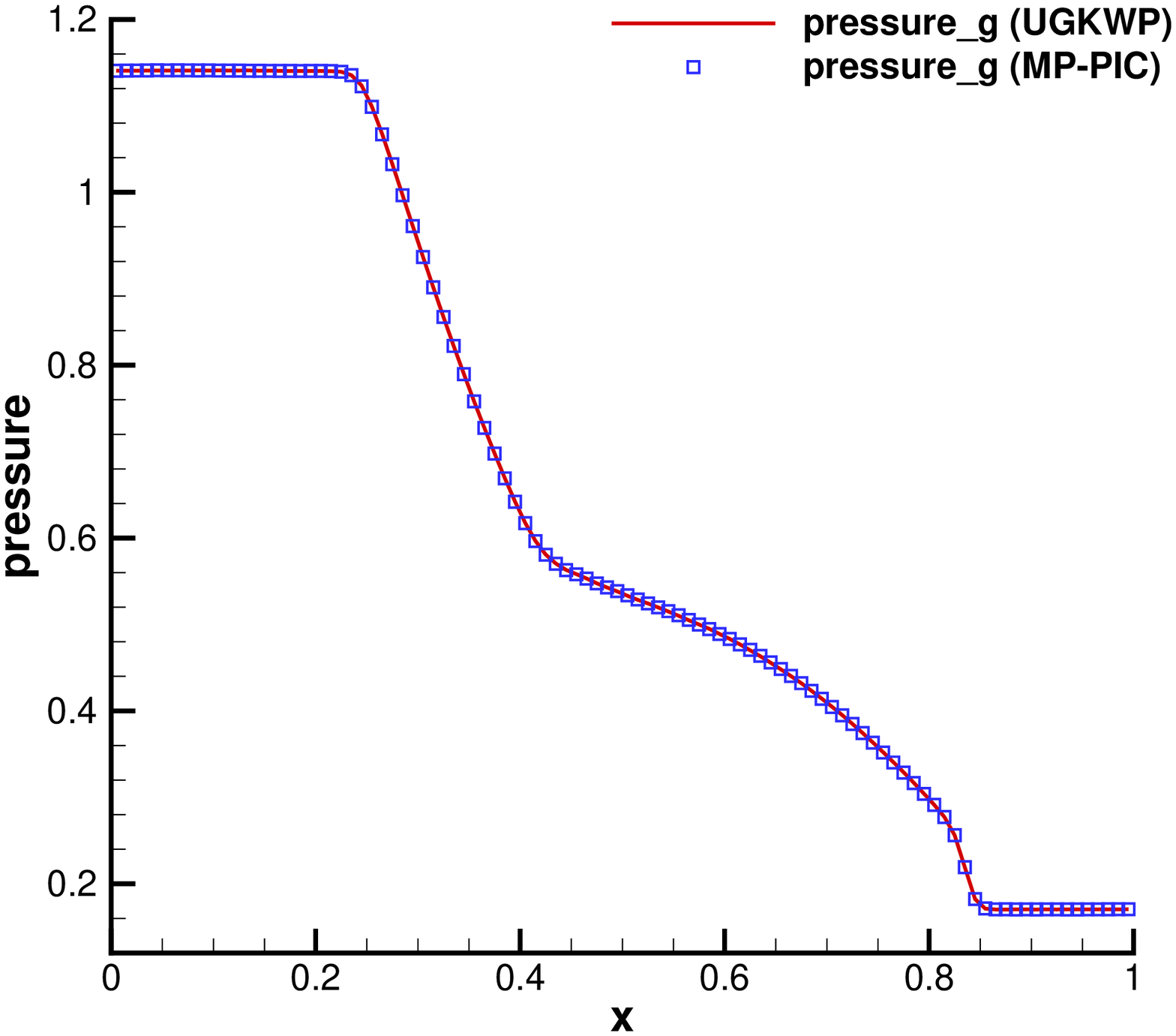}	
	}
	\quad
	\subfigure{
		\includegraphics[height=4.5cm]{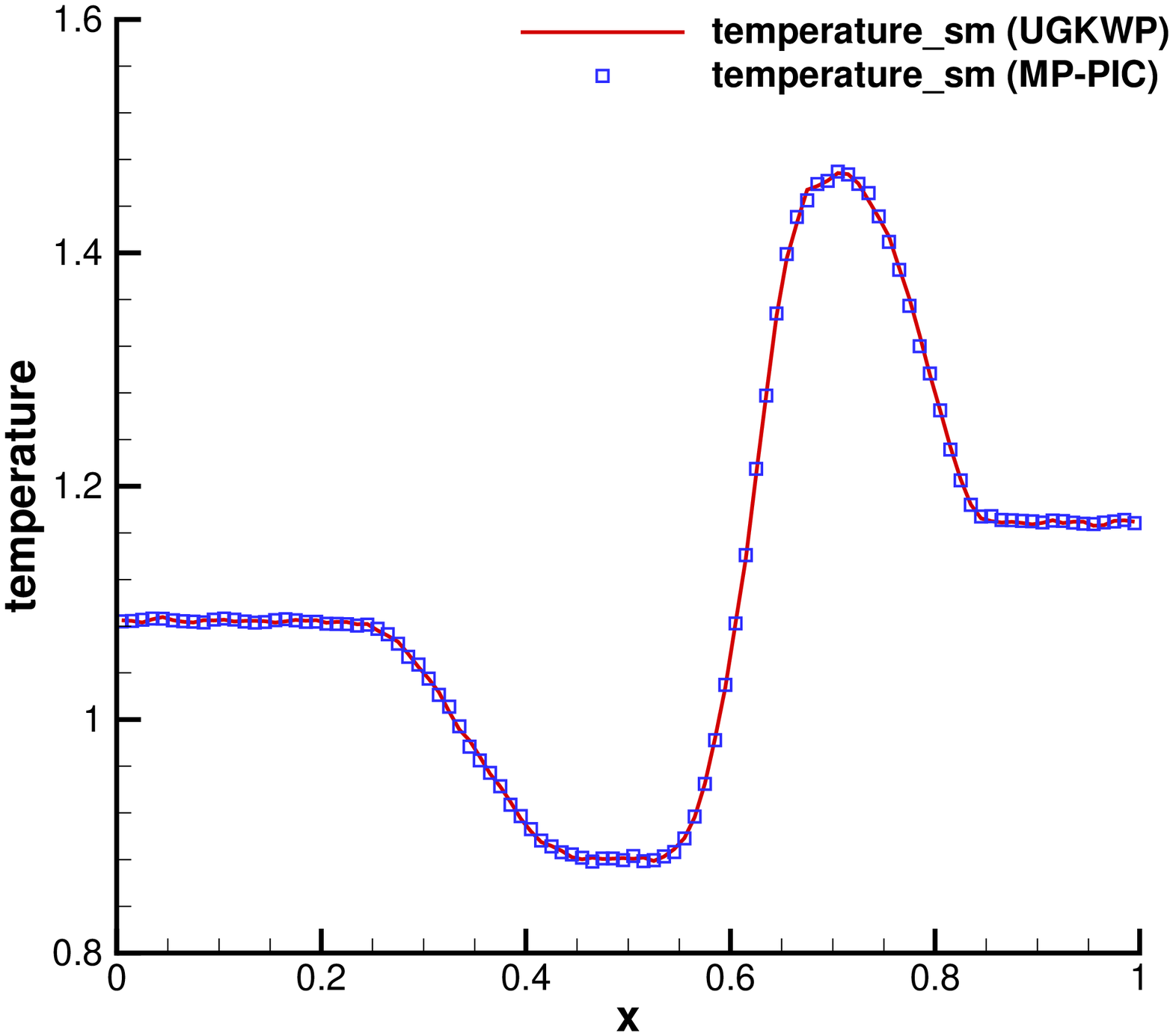}	
	}
	\caption{Wind sand shock tube problem. Solutions by the UGKWP and MP-PIC method in the collisionless limit at $t=0.2$, $Kn_s  =\infty$, $\tau_{st}=0.2$, $\tau_T=0.1$. Apparent density of gas and particle phase, velocity of gas and particle phase, pressure of gas phase, and material temperature of solid particles are included.}
	\label{wind sand Kn1}
\end{figure}

The 1D wind sand shock tube problem is tested to show the multiscale solution captured by the UGKWP. The initial condition is
\begin{equation*}
(\epsilon_s\rho_s,U_s,p_s,T^m_s)=\left\{\begin{aligned}
&(0.5, 0, 0.75, 1.0), & & 0\le x \le 0.5,\\
&(0.5, 0, 0.75, 0.8),  & & 0.5 < x \le 1,
\end{aligned} \right.
\end{equation*}
for solid particle phase, and  		
\begin{equation*}
(\epsilon_g\rho_g,U_g,p_g)=\left\{\begin{aligned}
&(1, 0, 1), & & 0\le x \le 0.5,\\
&(0.125, 0, 0.1),  & & 0.5 < x \le 1,
\end{aligned} \right.
\end{equation*}		
for gas phase. The computational domain is $\left[0,1\right]$ and the mesh number is 100. The non-dimensional heat capacity of particle phase and gas phase is $C_s=0.1$ and $C_g=0.2$, respectively. The relaxation time of the heat conduction $\tau_{T}$ between gas flow and solid particles is assumed as $0.1$. In this case, the collision between solid particles are assumed as elastic collision, i.e., $r=1$, and the gravity is neglected for solid particle phase. The target of this case it to validate the ability of the UGKWP to recover the solutions of TFM and MP-PIC in different limiting flow regimes for solid particle phase. Firstly, the high collision regime with $Kn_s=10^{-5}$ is considered,
and the solid particles are in the continuum flow regime with local equilibrium distribution determined by the macroscopic variables.
Therefore, the methods based on macroscopic flow variables, such as TFM, could give correct solution for the evolution of particle phase. The particle internal response time is assumed as a constant $\tau_{st}=0.1$. The results from the UGKWP and TFM are presented in Figure \ref{wind sand Kn1m5}, where good agreements have been obtained in the highly collisional regime.
Then, the limit of collisionless regime is tested by assuming an infinity solid phase Knudsen number $Kn_s$. Under such a flow regime, no inter-particle collision exists and the solid  particle takes free transport. In this case, a larger $\tau_{st}=0.2$ is taken.
The results from the particle MP-PIC and UGKWP  are shown in Figure \ref{wind sand Kn1}, where almost identical solutions are obtained in the  particle phase collisionless regime.

\subsection{Impinging particle jets problem}
\begin{figure}[htbp]
	\centering
	\subfigure{
		\includegraphics[height=6.5cm]{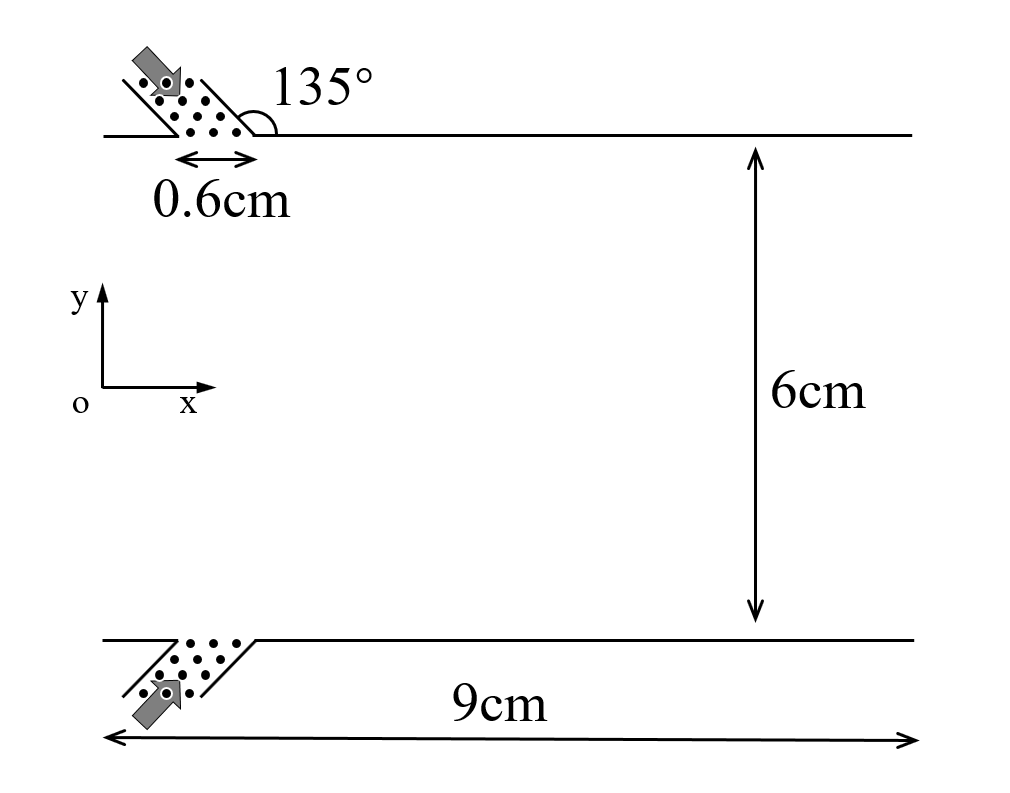}
	}		
	\caption{The sketch of particle jets impinging problem.}
	\label{Particle jet impinging sketch}		
\end{figure}

A significant non-equilibrium behavior of gas-particle flow is particle trajectory crossing (PTC), which occurs in the collisionless regime \cite{Gasparticle-momentmethod-Fox2013computational}. It is easy to understand that if no collision exists, all the solid particles will keep their initial velocity and thus cross with each other when they meet. In contrast, in a finite $Kn$ number, the collision between particles will lead to the multi-particle velocity, and scatter the particle cloud.
In this section, two particle jets impinging problem is tested for the performance of the UGKWP in the capturing of the above representative phenomena of gas-particle flow. The associated sketch is given in Figure \ref{Particle jet impinging sketch}. The whole computational domain is $L\times H=9cm\times6cm$ and a uniformed grid with $180\times120$ is used. Initially, two particle beams impinge the computational domain at $x=0.6cm$ from the up and down boundary symmetrically. The angle between particle beam with the boundary is $135^{\circ}$ and the length of particle beam is $0.424cm$. Besides, the absolute velocity of particle beam is $|\textbf{U}|=1m/s$, and the granular temperature $T_s$ is zero. The reference time in this case is taken as $t_{ref}=L/U$. The reflection boundary condition is employed at the up and down boundaries except at the inlet of the impinging particle beam. The right boundary is the outflow boundary condition. In this case, the interaction between particle phase with gas phase is neglected and thus only the particle phase is simulated.

\begin{figure}[htbp]
	\centering
	\subfigure{
		\includegraphics[height=5cm]{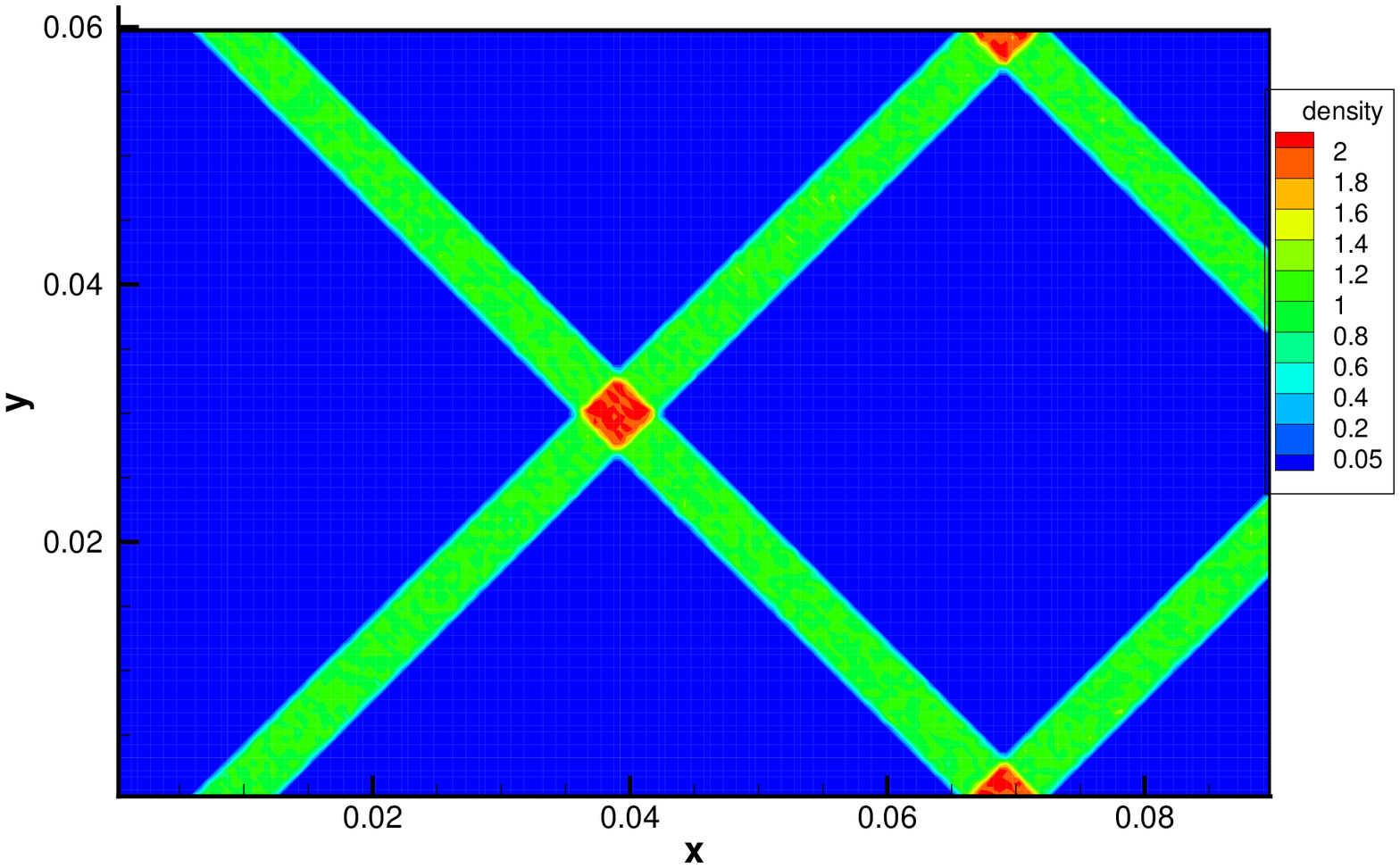}	
	}
	\quad
	\subfigure{
		\includegraphics[height=5cm]{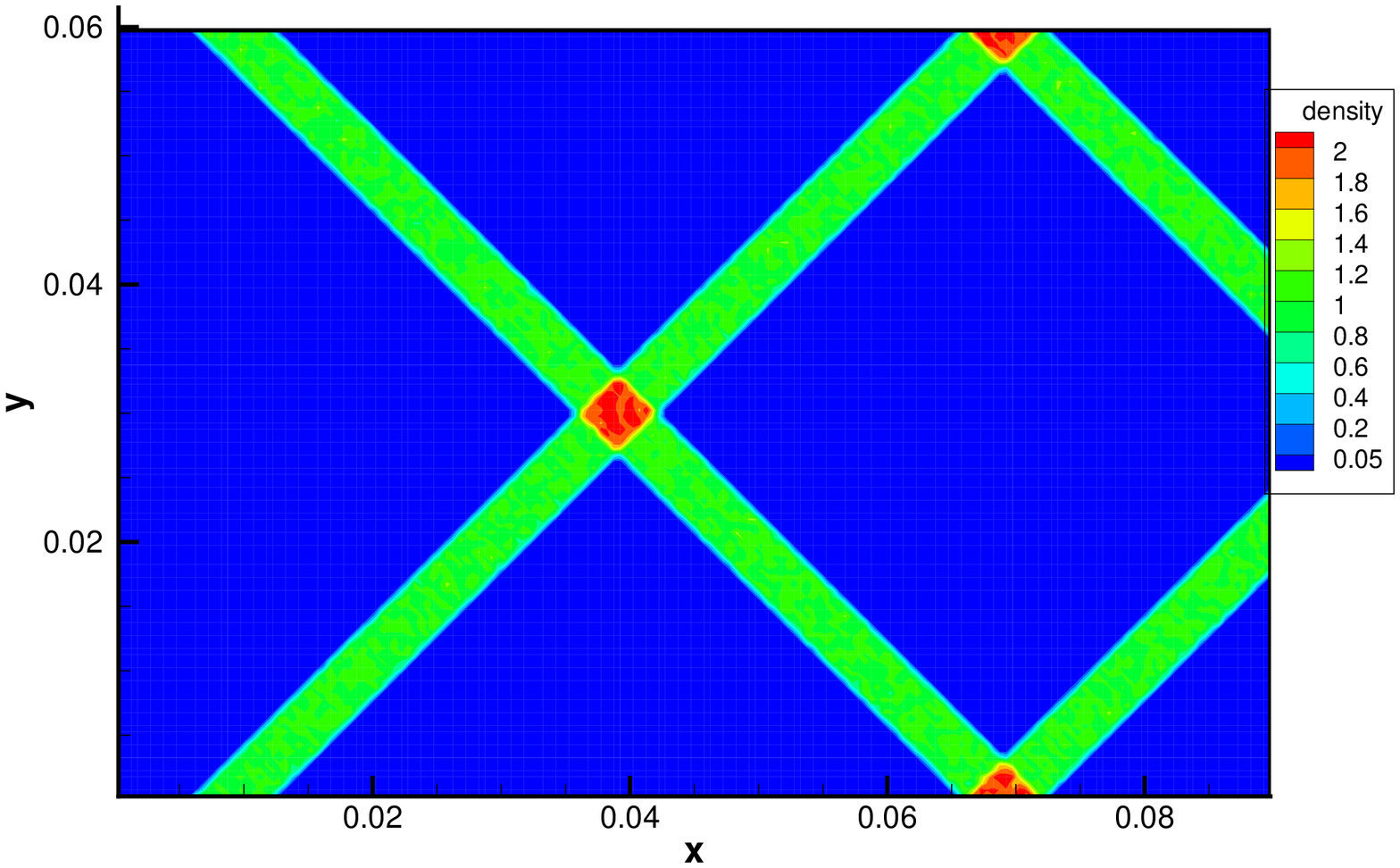}	
	}		
	\caption{Apparent density for impinging particle jets problem by the UGKWP in collisionless regime with $Kn =\infty$ at $t=0.12$. The left figure is elastic collision and the right is inelastic collision.}
	\label{Particle jet impinging Kn infinity}		
\end{figure}

\begin{figure}[htbp]
	\centering
	\subfigure{
		\includegraphics[height=5cm]{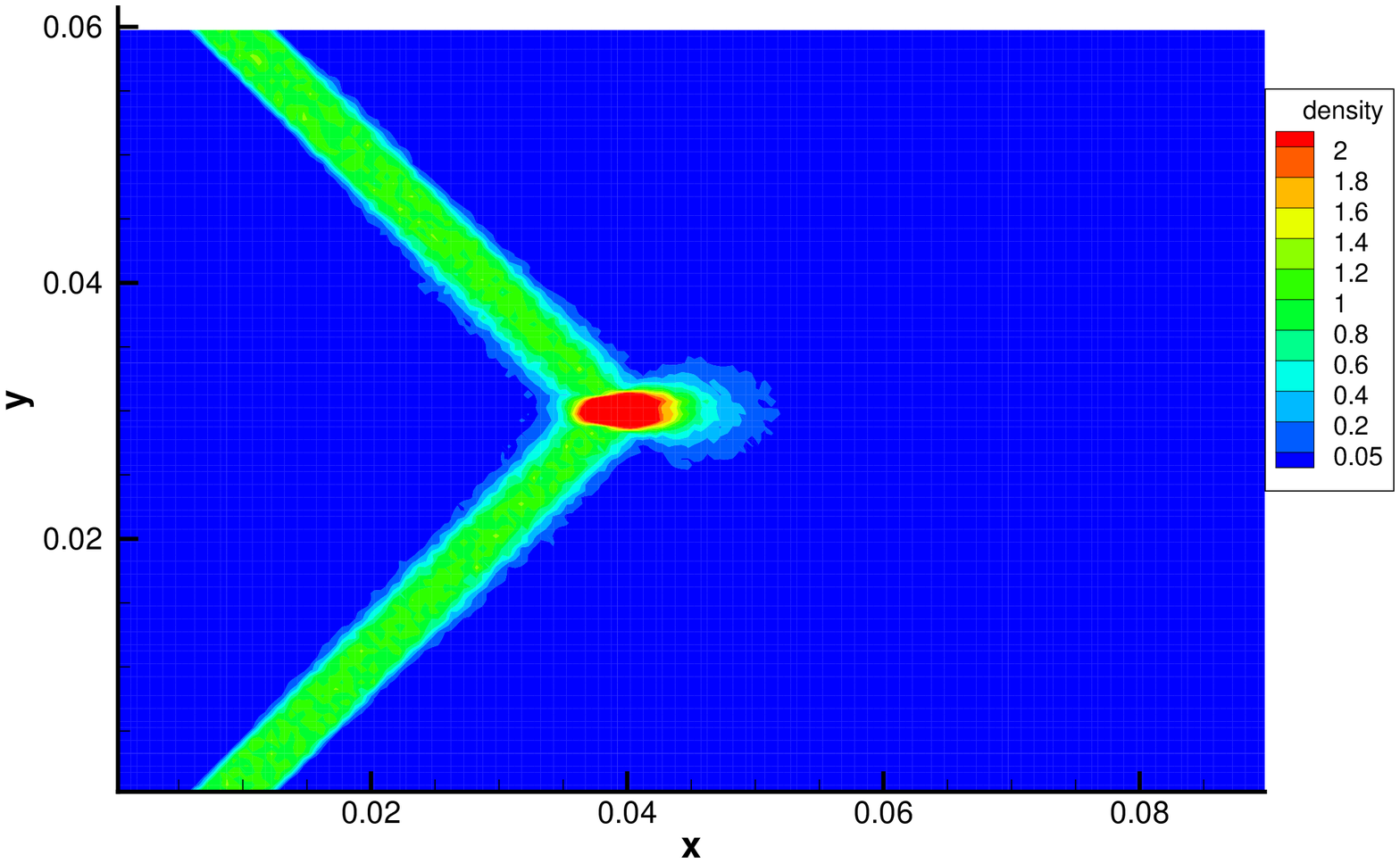}	
	}
	\quad
	\subfigure{
		\includegraphics[height=5cm]{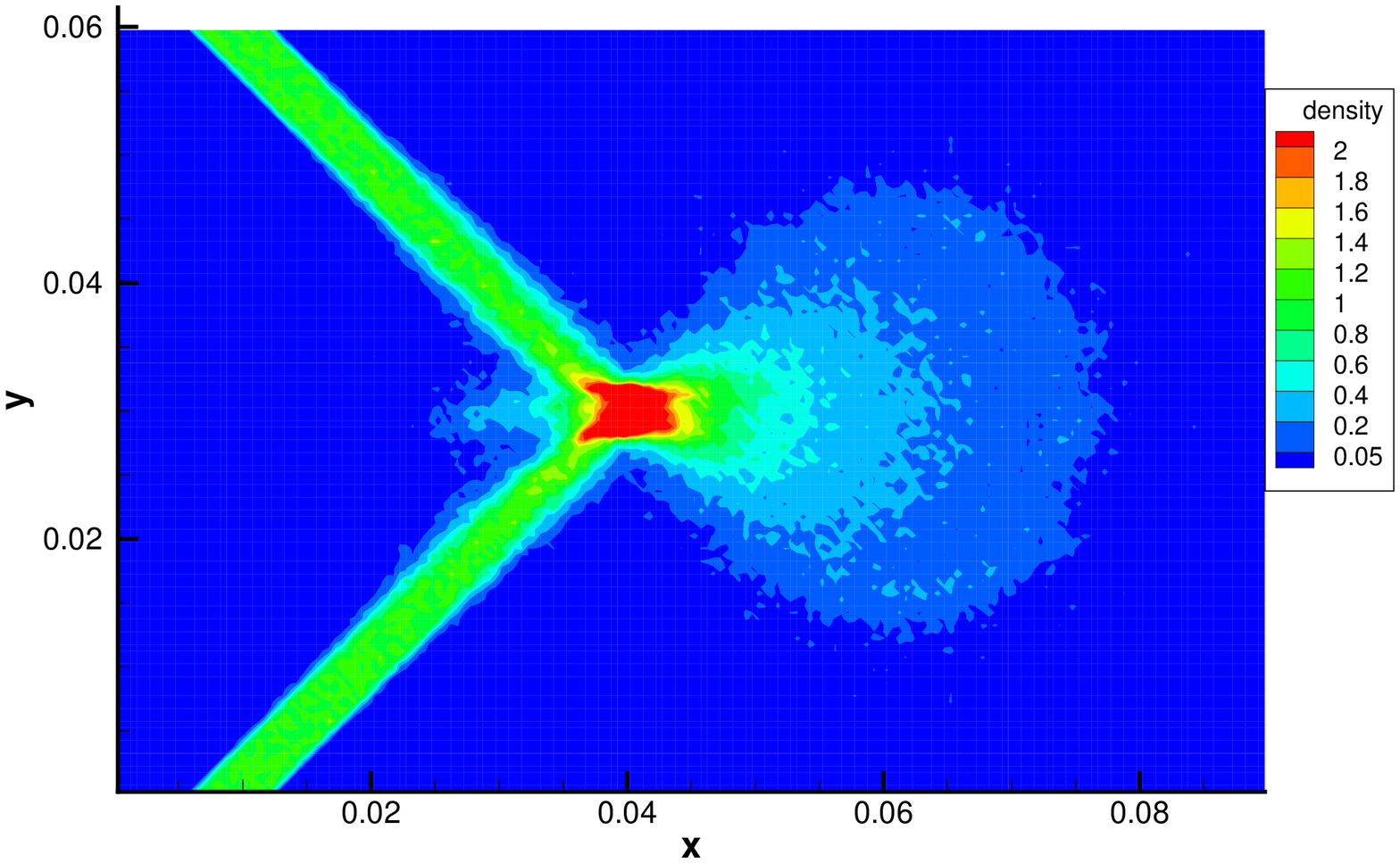}	
	}		
	\caption{Apparent density for the impinging particle jets problem by UGKWP with $Kn=1$ at $t=0.05$ and $t=0.08$ and elastic collision.}		
	\label{Particle jet impinging Kn 1 elastic}	
\end{figure}

\begin{figure}[htbp]
	\centering
	\subfigure{
		\includegraphics[height=5cm]{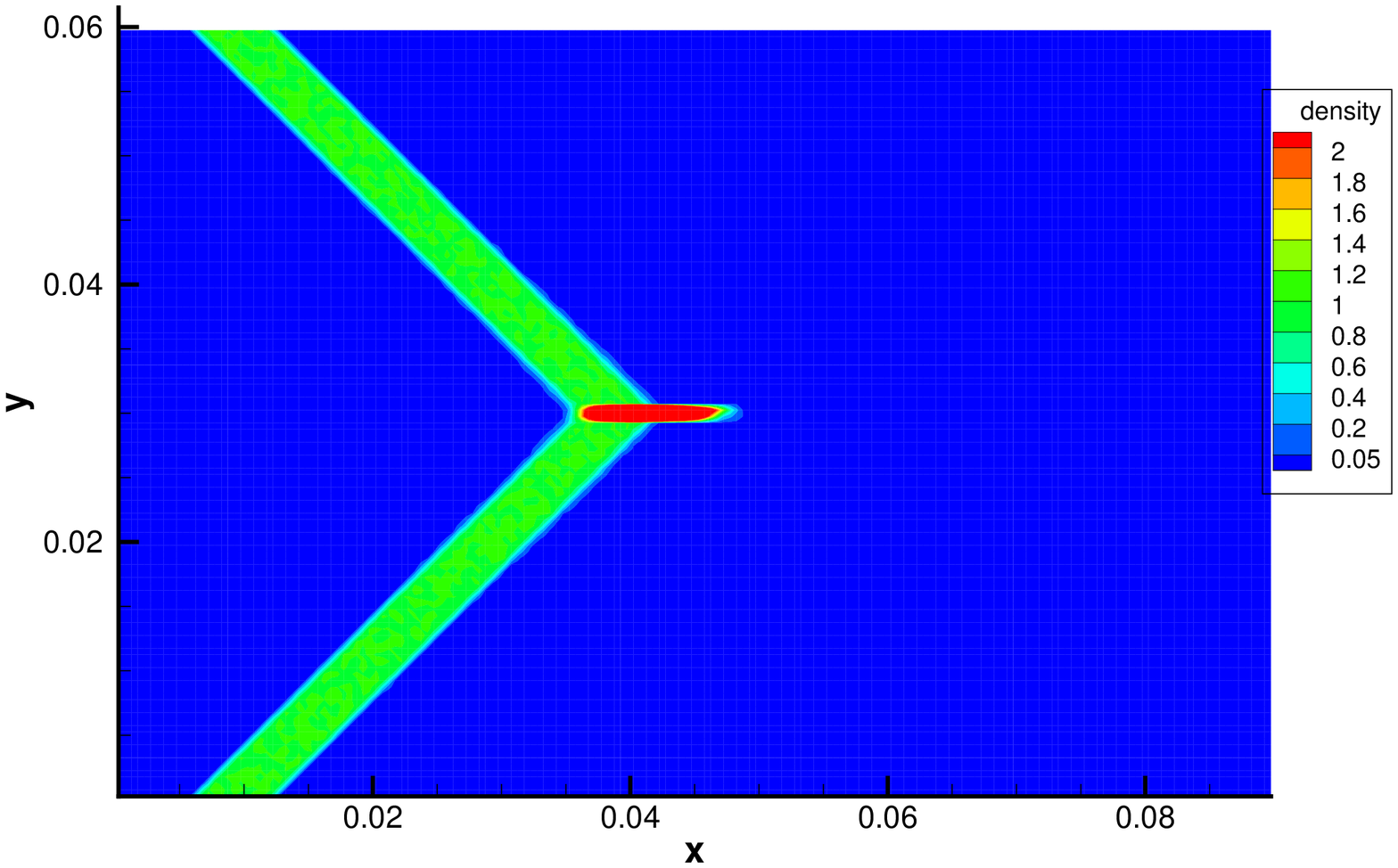}	
	}
	\quad
	\subfigure{
		\includegraphics[height=5cm]{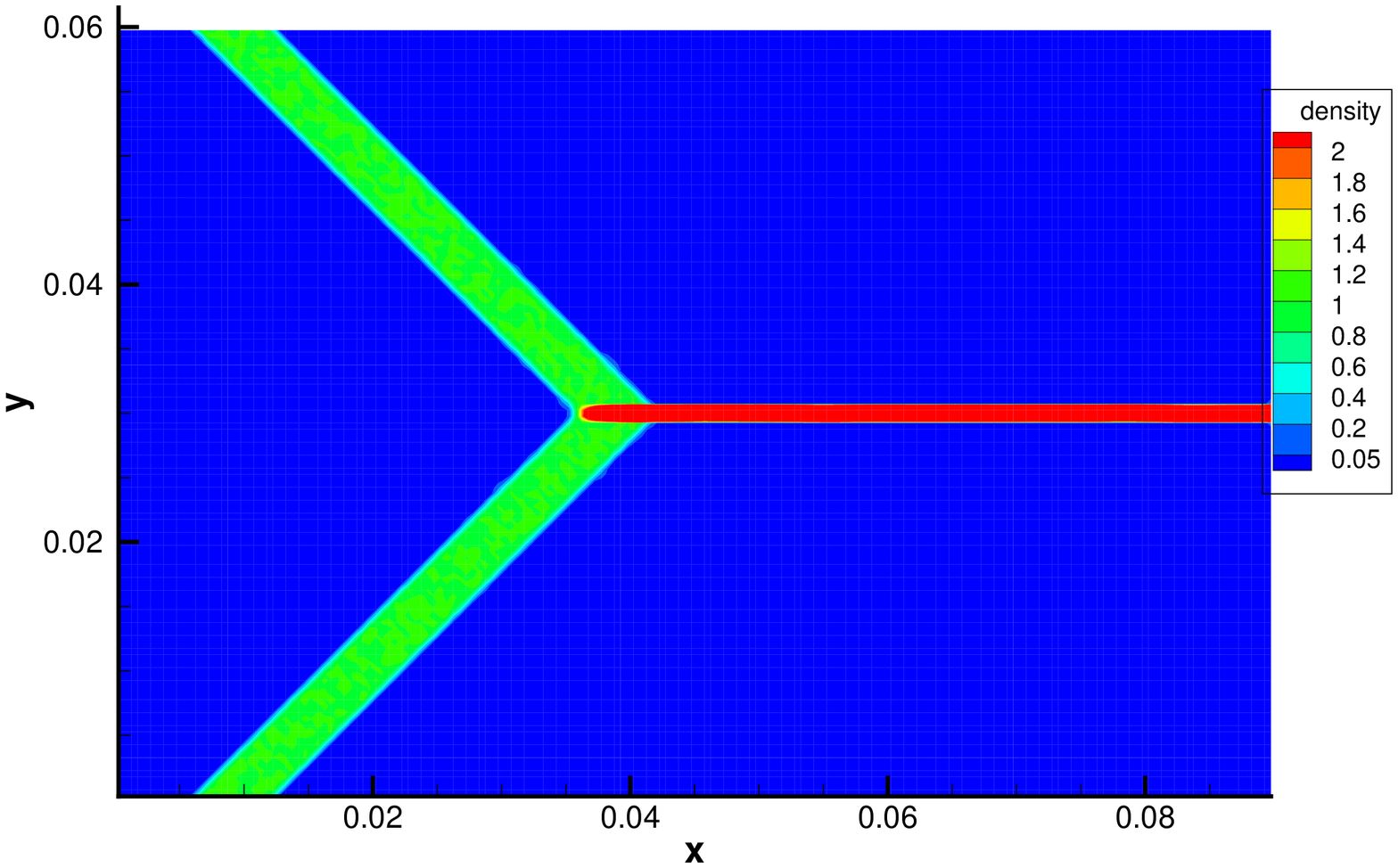}	
	}		
	\caption{Apparent density for the impinging particle jets problem by the UGKWP with $Kn=1$ at $t=0.05$ and $t=0.12$ and inelastic collision.}	
	\label{Particle jet impinging Kn 1 inelastic}		
\end{figure}

The collisionless regime is considered first at $Kn_s =\infty$. In this regime, the solid particle will keep the initial velocity until they meet the wall. Figure \ref{Particle jet impinging Kn infinity} shows that the particle behaviors, PTC and wall reflection are captured exactly. Since no collision exists, both the elastic collision and inelastic collision models show the same result.
Then, at $Kn_s =1$ with  the influence of both transport and collision, the results from the elastic collision model are shown in Figure \ref{Particle jet impinging Kn 1 elastic}. Due to the effect of elastic particle-particle collision, the solid particles will get scatted at the encountering of two impinging jets. In contrast, for the inelastic collision, the particles will get together after collision and move horizontally to the right, as shown in Figure \ref{Particle jet impinging Kn 1 inelastic}. During this process, the granular temperature is lost in the inelastic collision, and the post-collision particles have the same velocity.

\begin{figure}[htbp]
	\centering
	\subfigure{
		\includegraphics[height=5cm]{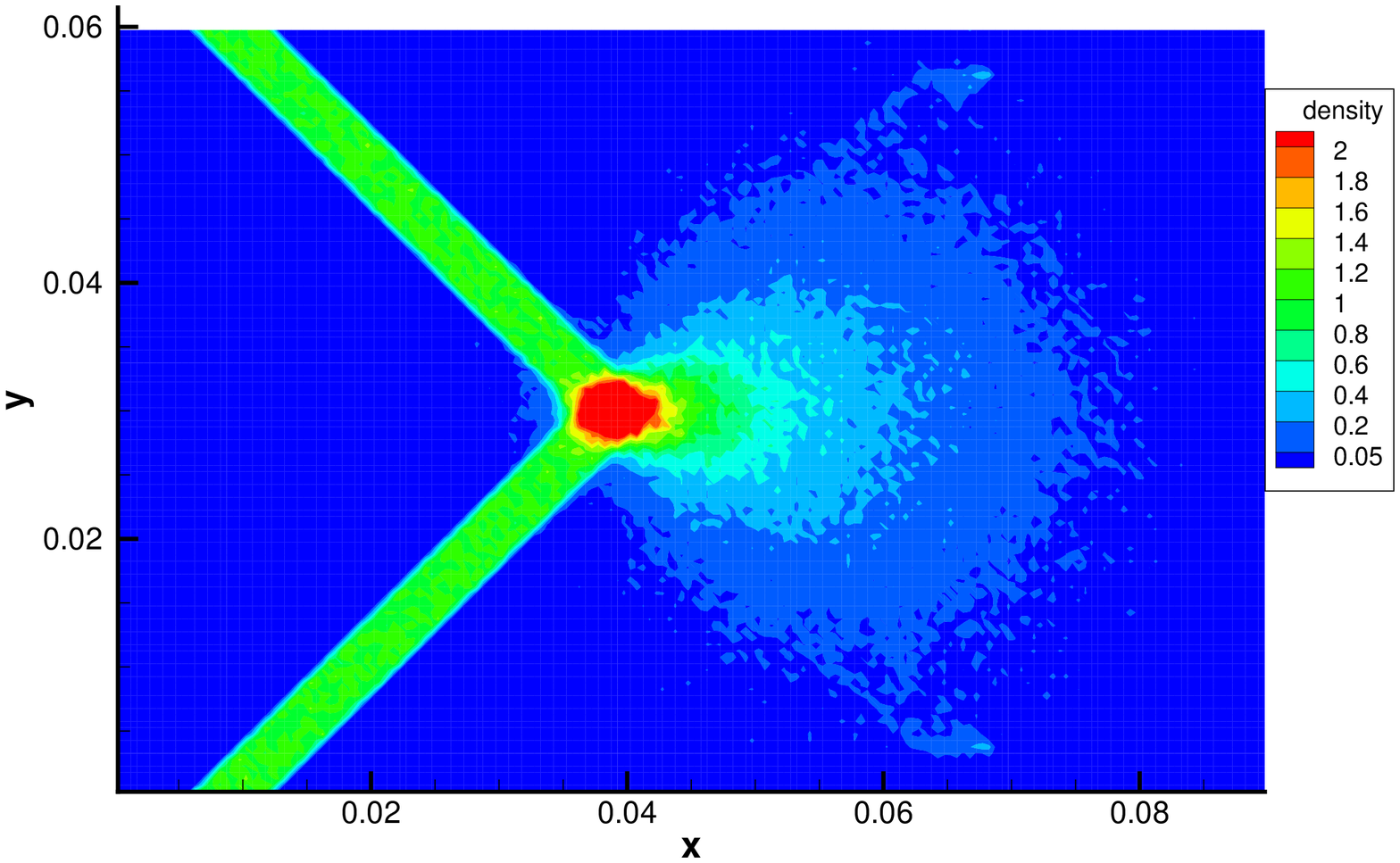}	
	}
	\quad
	\subfigure{
		\includegraphics[height=5cm]{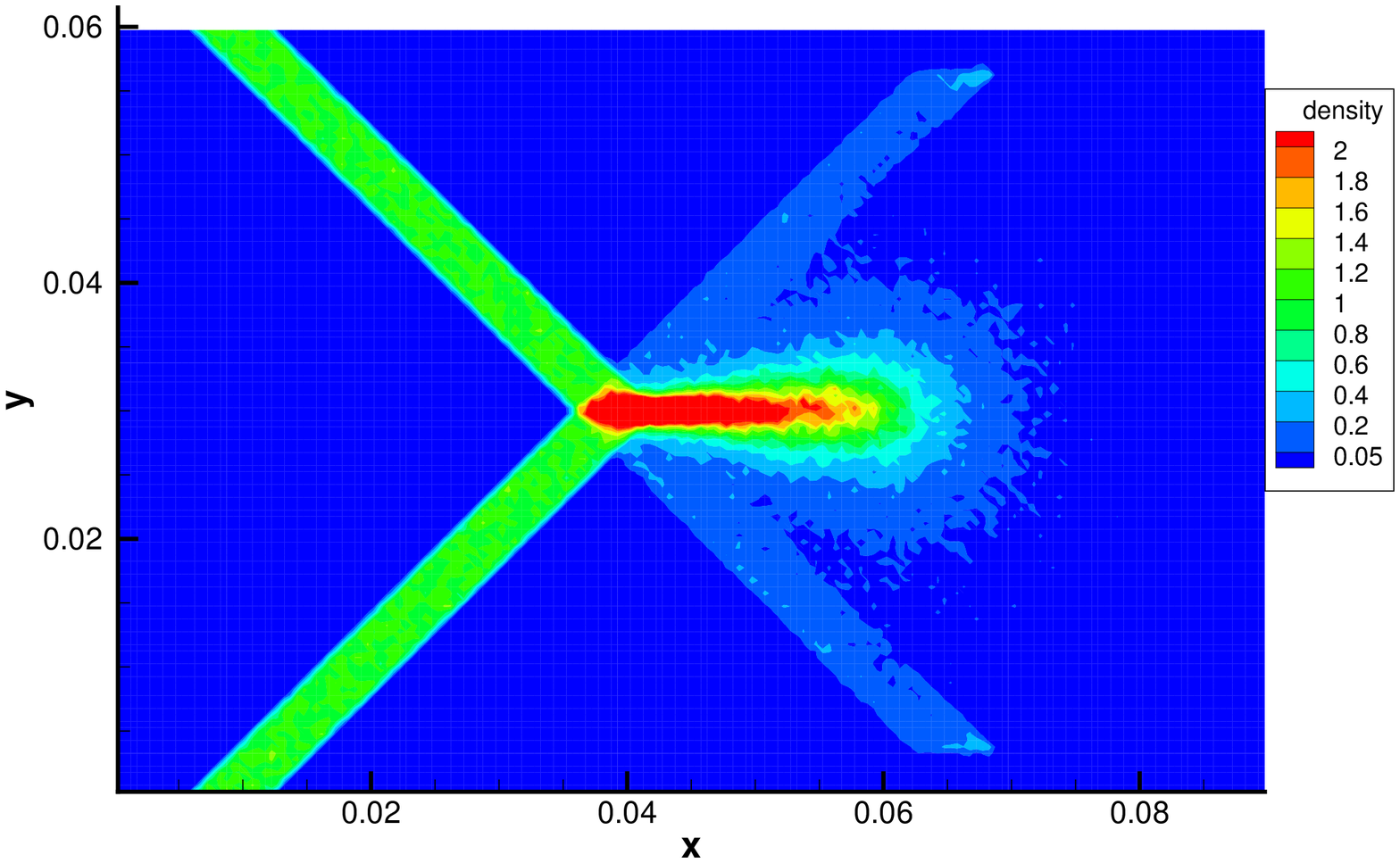}	
	}		
	\caption{Apparent density for the impinging particle jets problem by the UGKWP with $Kn=10^2$ at $t=0.08$. The left figure is elastic collision and the right is inelastic collision.}
	\label{Particle jet impinging Kn 100}			
\end{figure}

 At the Knudsen number $Kn_s=10^2$, the solutions are shown in Figure \ref{Particle jet impinging Kn 100}.
 For the elastic particle collision, besides the scattering, a small portion of particles still keeps their initial velocity,
 which is not found in the solution at $Kn_s=1$. In comparison with $Kn_s=1$ case, the free transport of particles at $Kn_s=10^2$ will play
 a more significant role on the particle evolution. The same phenomenon can be obtained more obviously in the inelastic collision case.
 Figure \ref{Particle jet impinging Kn 100} shows clearly the effect from the particle free transport and collision at such a Knudsen number.

\subsection{Taylor-Green vortex}
\begin{figure}[htbp]
	\centering
	\subfigure{
		\includegraphics[height=5.5cm]{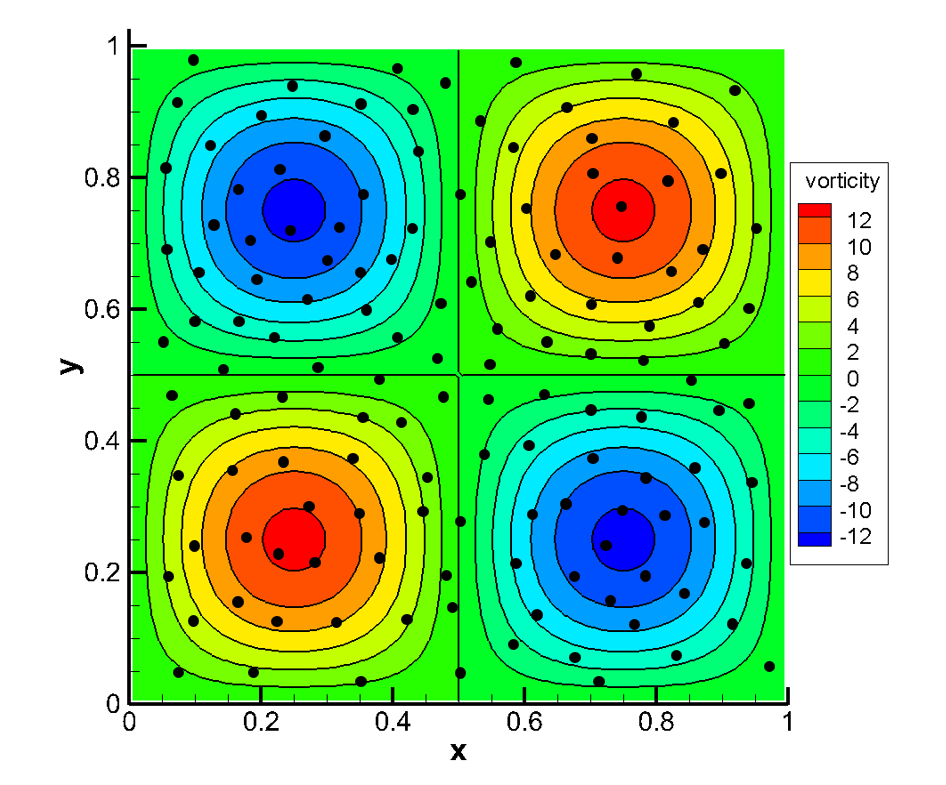}	
	}
	\quad
	\subfigure{
		\includegraphics[height=5.5cm]{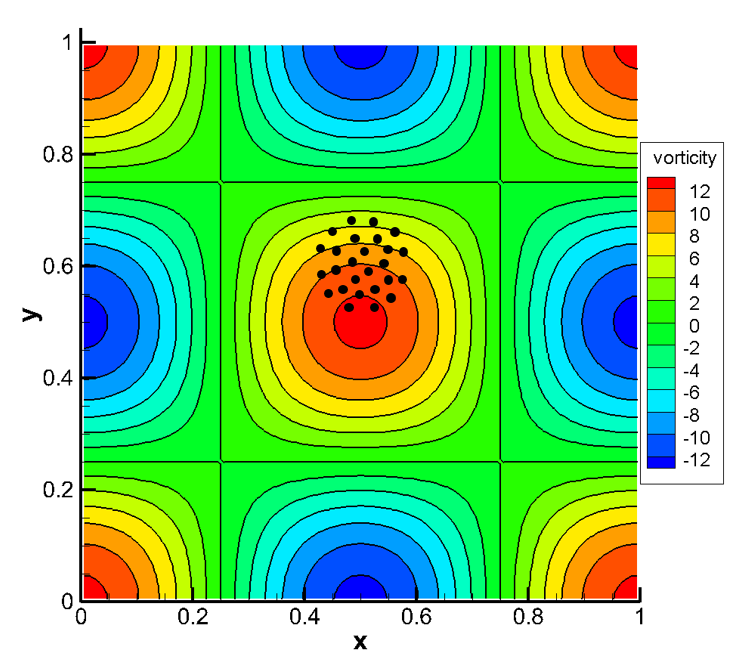}	
	}
	\caption{The sketch of the initial distribution of apparent density of solid phase (black points) and the initial vorticity of gas phase (contours). The left figure is Type 1 and the right figure is Type 2.}	
	\label{Taylor-Green flow particle distribution IC}	
\end{figure}

The preferential concentration of dispersed particles in gas-particle two-phase flow is investigated. In general, the particles prefer to concentrate on the region where the vorticity of gas phase is lower \cite{Gasparticle-Taylor-Green-vortex-Stephane2007evaluation, Gasparticle-Taylor-Green-vortex-Moment-method-Fox2008quadrature}. The two-dimensional Taylor-Green vortex flow solution is chosen for the gaseous flow field. Two initial conditions are considered in this paper, which are shown in Figure \ref{Taylor-Green flow particle distribution IC}. For Type 1, the solid particles are uniformly distributed initially in the whole computational domain,
\begin{gather*}
\rho_g=1.0,~~
U_g=\text{cos}\left(2\pi x\right)\text{sin}\left(2\pi y\right),~~
V_g=-\text{sin}\left(2\pi x\right)\text{cos}\left(2\pi y\right),\\
p_g=10.0+\left[\text{cos}\left(4\pi x\right)+\text{cos}\left(4\pi y\right)\right]/4,
\end{gather*}
for gas phase, and
\begin{gather*}
\epsilon_s\rho_s=1.0,~~
U_s=U_g,~~
V_s=V_g,~~
T_s=1.0^{-8},
\end{gather*}
for solid particle phase. While for Type 2, the solid particles are uniformly located within a circle centered at $\left(0.5, 1-1.25/\pi\right)$ with radius $r=0.25/\pi$, and the gaseous flow is translated with a distance of $0.5\pi$ in both directions compared with Type 1. For both initial conditions, the computational domain is $[0,1]\times[0,1]$ covered by $100\times100$ uniform grids. The periodic boundary condition is employed for all boundaries. The Reynolds number of gas phase is taken as $Re=400$, which leads to $\mu=2.5\times10^{-3}$. The reference time is taken as $t_{ref}=1$ for this test case. According to the previous research, the critical Stokes number is $St_{c}=1/\left(8\pi\right)$ \cite{Gasparticle-Taylor-Green-vortex-Stephane2007evaluation}. If the Stokes number below $St_{c}$, the particle will concentrate around the low vorticity area; while if the Stokes number is larger than $St_{c}$, the particle trajectory crossing will occur.

\begin{figure}[htbp]
	\centering
	\subfigure{
		\includegraphics[height=6.5cm]{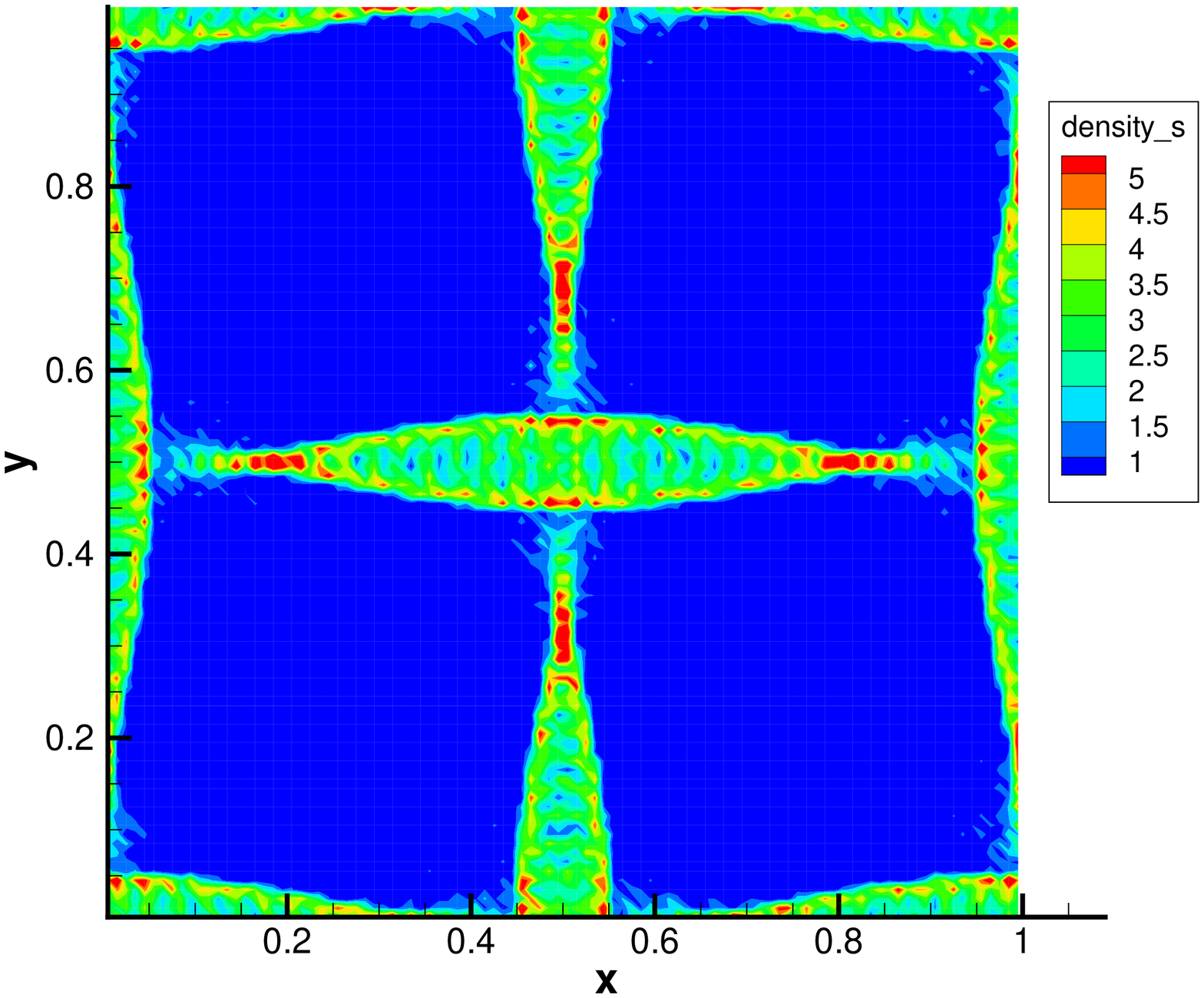}	
	}
	\quad
	\subfigure{
		\includegraphics[height=6.5cm]{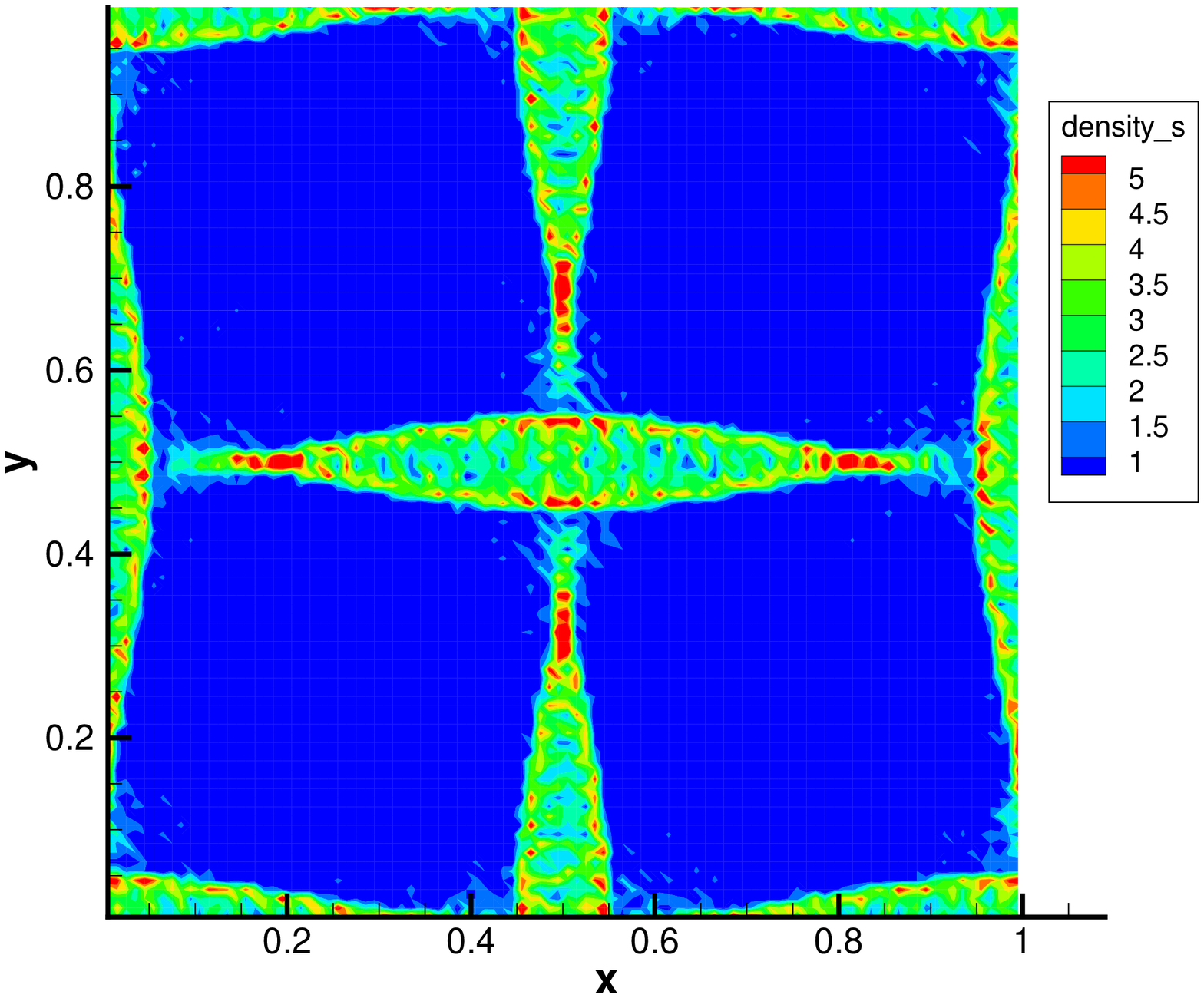}	
	}
	\caption{Apparent density of particle phase for Taylor-Green vortex problem in collisionless regime at $Kn_s=10^{4}$ and $St=0.3$ at $t=0.3$. Left: UGKWP solution;  Right: MP-PIC solution.}
	\label{Taylor-Green flow IC 1 Kn 10p4 taust 0.3 at t 0.3}		
\end{figure}

\begin{figure}[htbp]
	\centering
	\subfigure{
		\includegraphics[height=6.5cm]{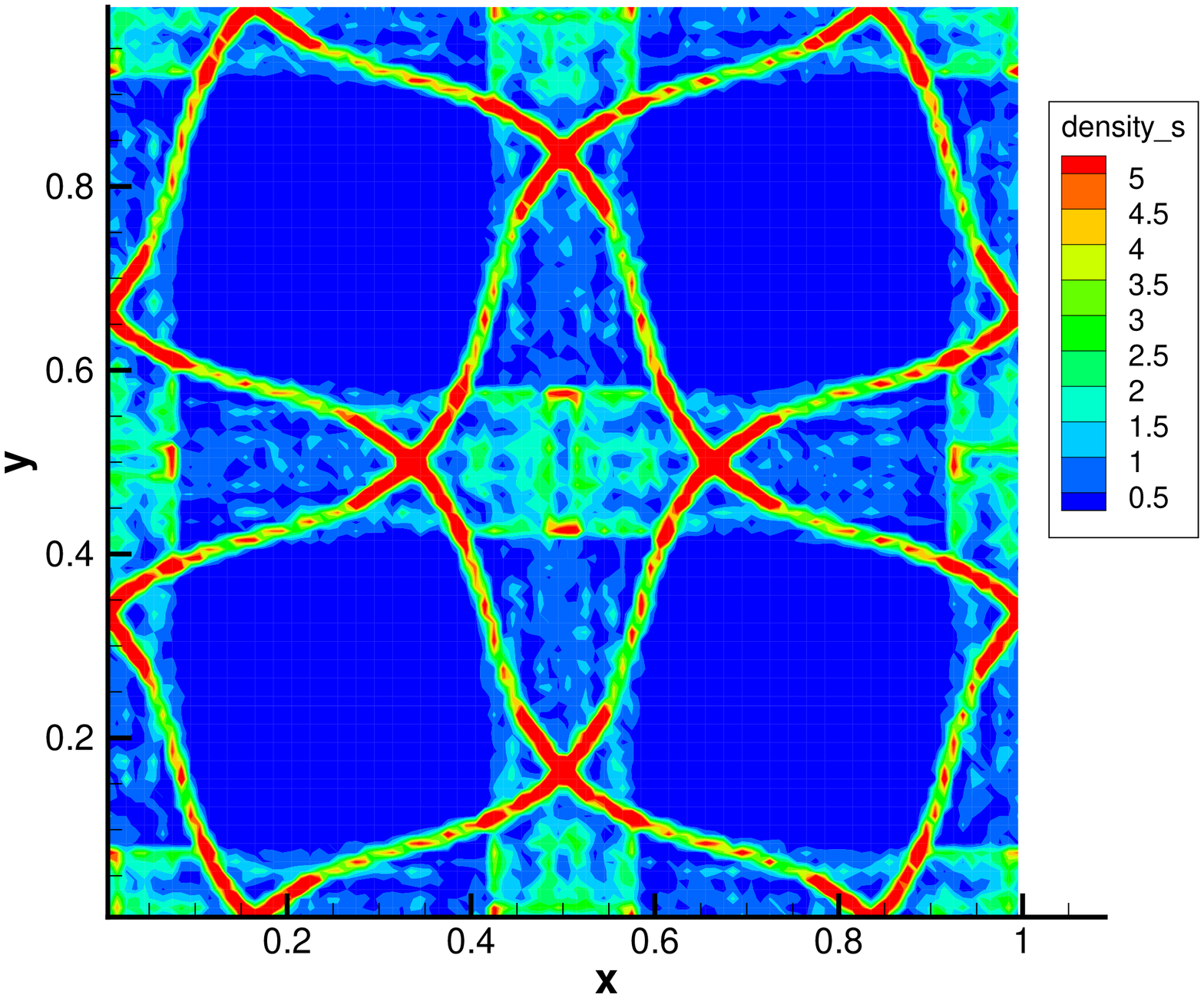}	
	}
	\quad
	\subfigure{
		\includegraphics[height=6.5cm]{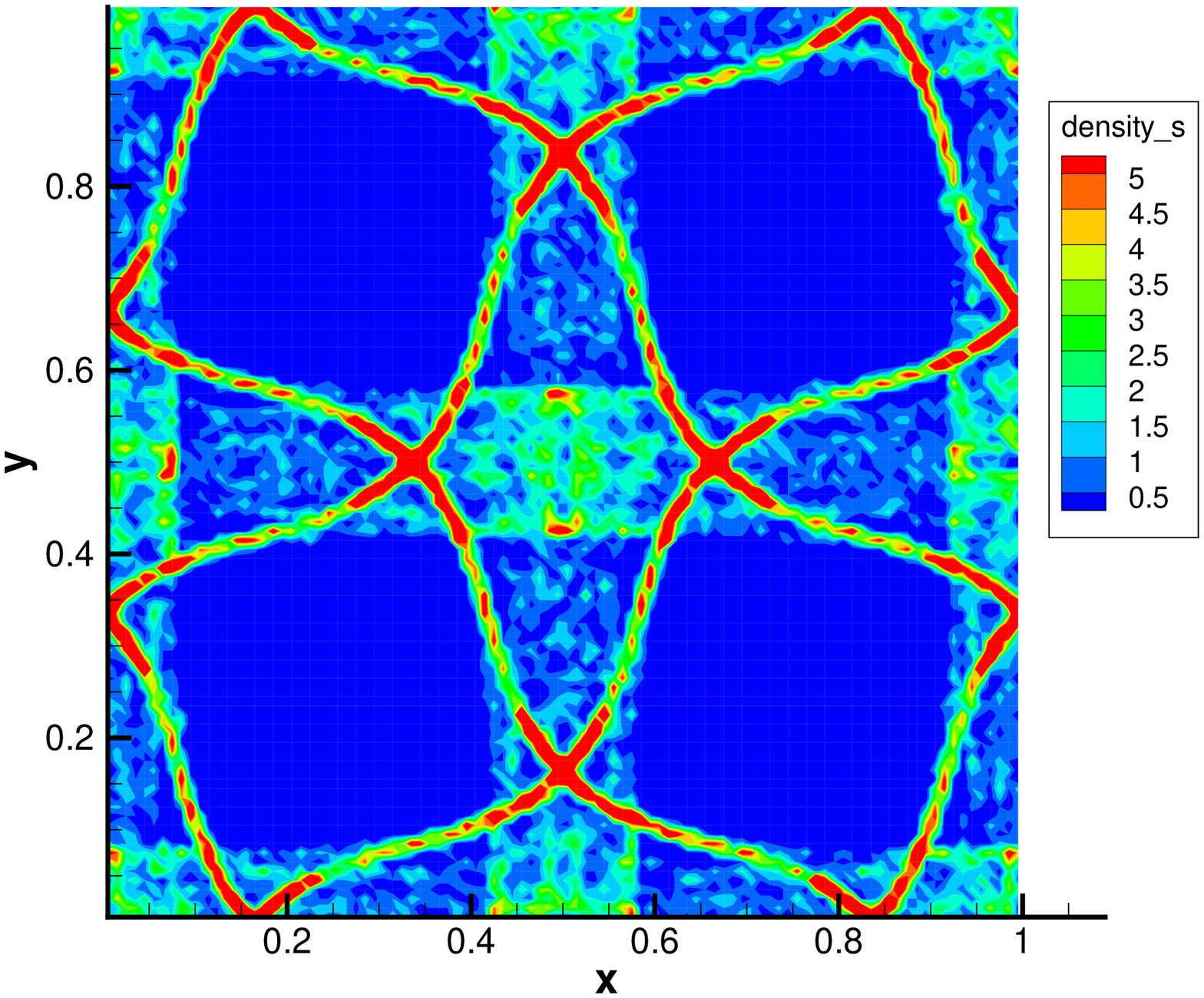}	
	}
	\caption{Apparent density of particle phase for Taylor-Green vortex problem in collisionless regime with $Kn_s=10^{4}$ and $St=0.3$ at $t=2.5$. Left: UGKWP solution; Right: MP-PIC solution.}
	\label{Taylor-Green flow IC 1 Kn 10p4 taust 0.3 at t 2.5}			
\end{figure}

\begin{figure}[htbp]
	\centering
	\subfigure{
		\includegraphics[height=6.5cm]{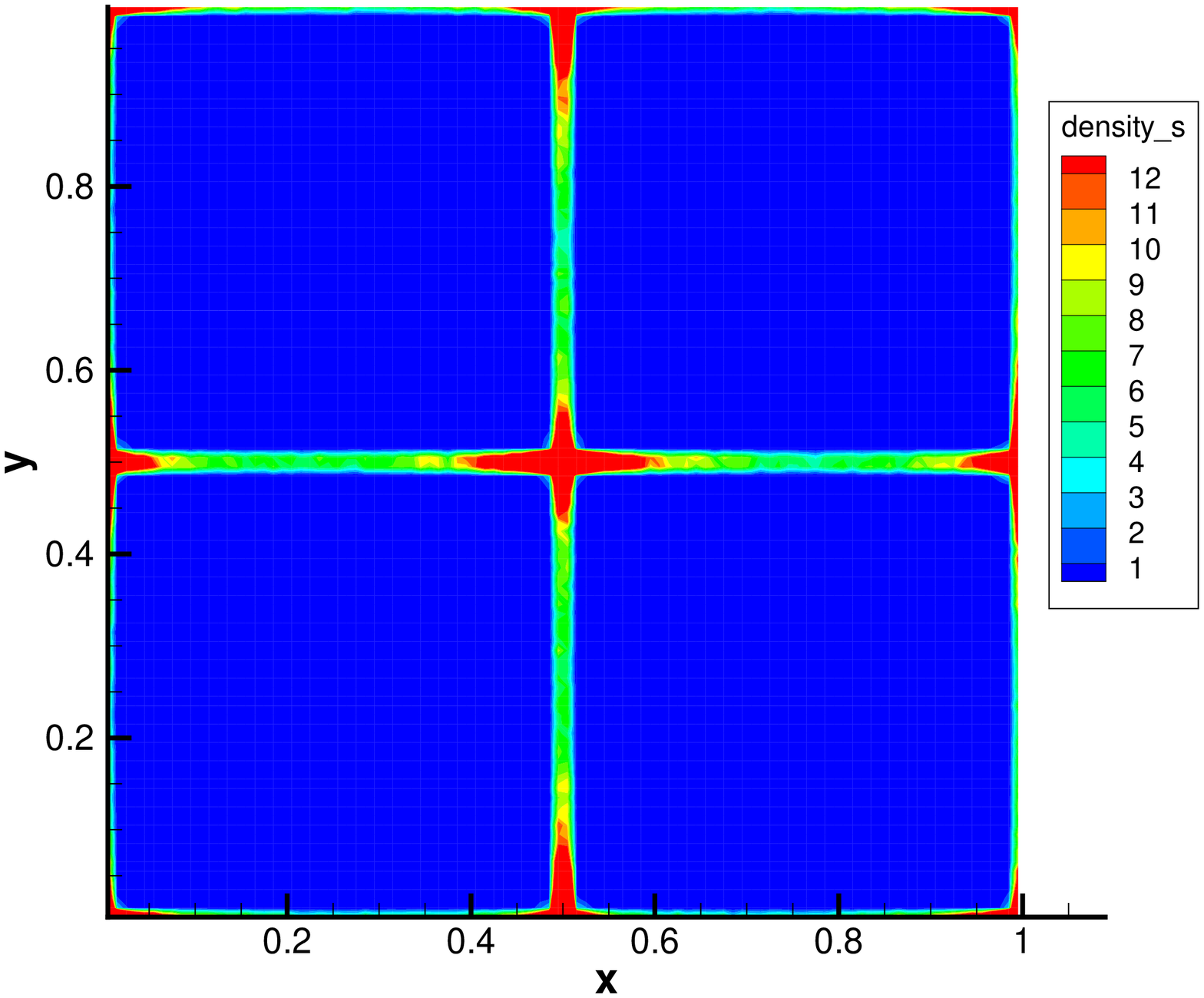}	
	}
	\quad
	\subfigure{
		\includegraphics[height=6.5cm]{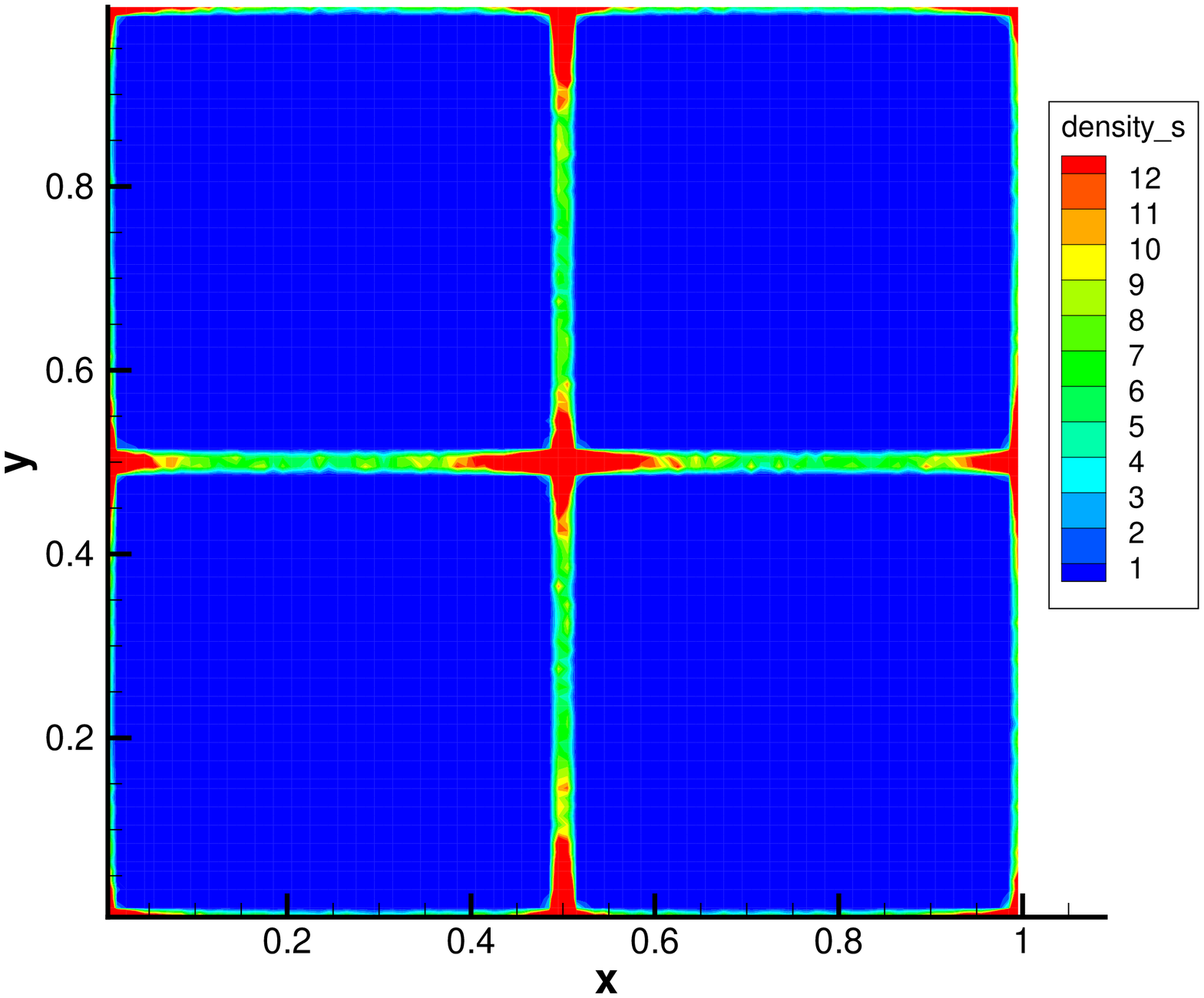}	
	}
	\caption{Apparent density of particle phase for Taylor-Green vortex problem in collisionless regime with $Kn_s=10^{4}$ and $St=0.039$ at $t=2.5$. Left: UGKWP solution; Right: MP-PIC solution.}
	\label{Taylor-Green flow IC 1 Kn 10p4 taust 0.039 at t 2.5}			
\end{figure}

\begin{figure}[htbp]
	\centering
	\subfigure{
		\includegraphics[height=6.5cm]{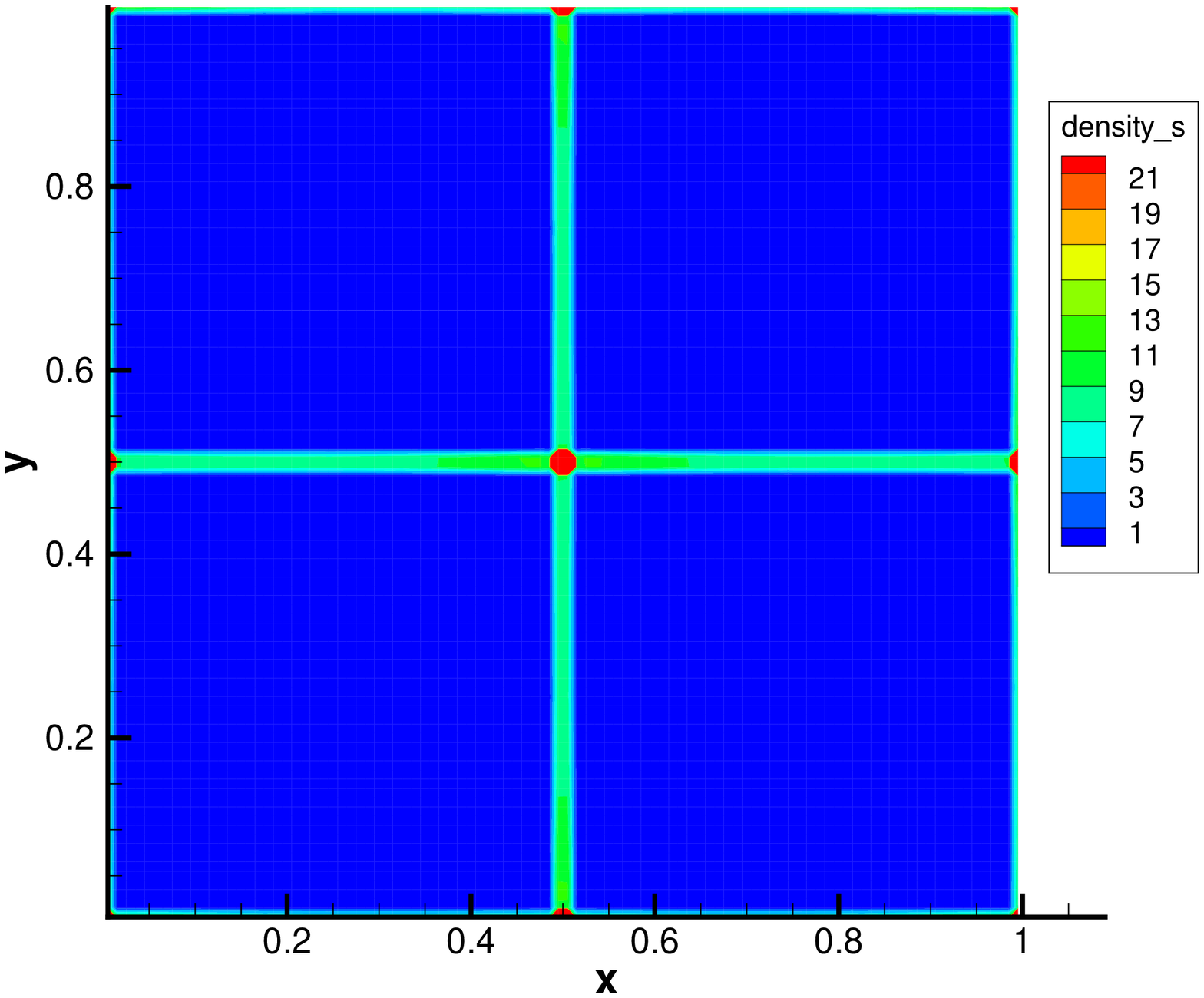}	
	}
	\quad
	\subfigure{
		\includegraphics[height=6.5cm]{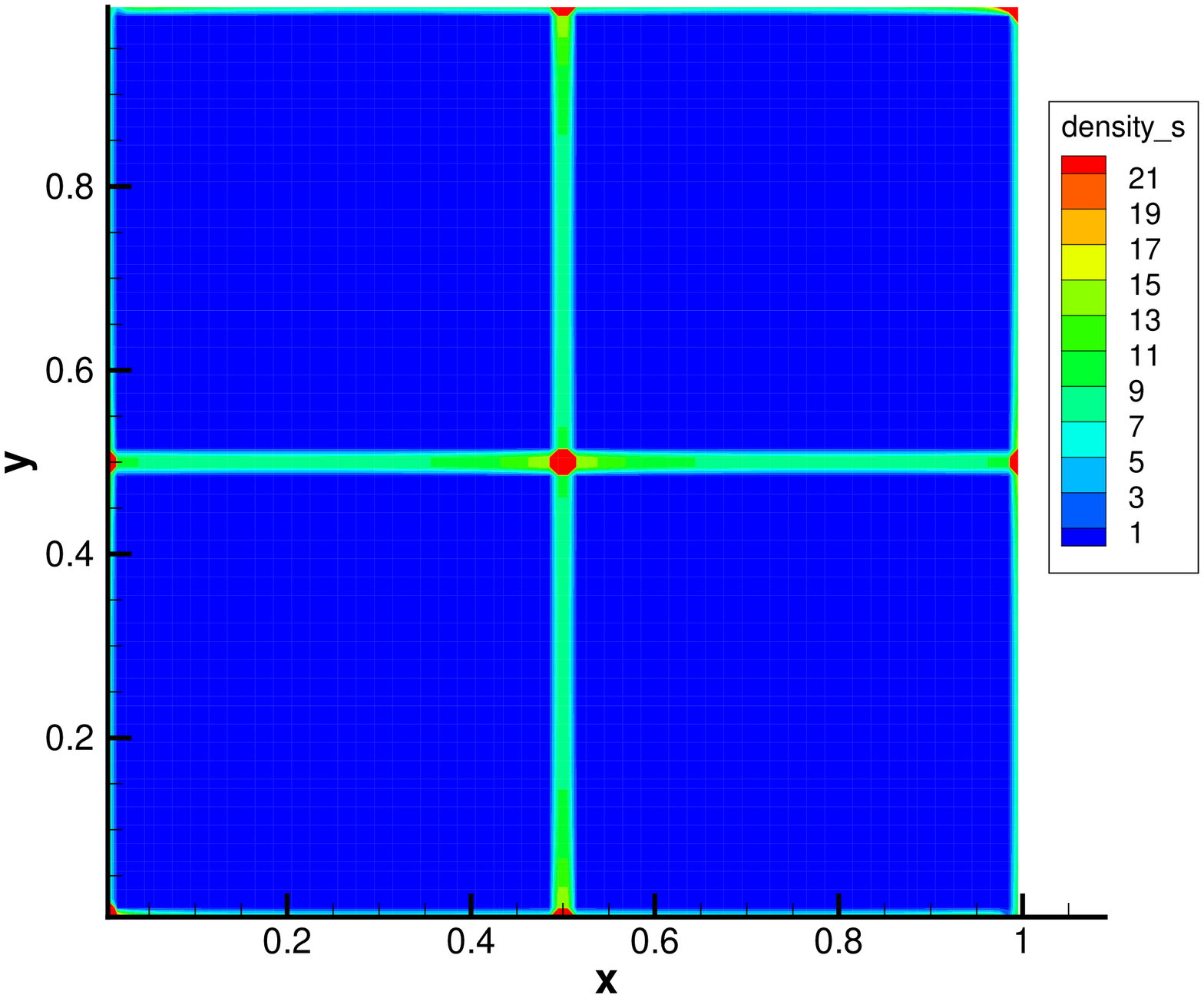}	
	}
	\caption{Apparent density of particle phase for Taylor-Green vortex problem in high collision regime with $Kn_s=10^{-6}$ and $St=0.3$ at $t=0.6$. Left: UGKWP solution; Right: TFM solution.}	
	\label{Taylor-Green flow IC 1 Kn 10m6 taust 0.3 at t 0.6}		
\end{figure}

For the Type 1 initial condition, firstly we consider a collisionless regime with $Kn_s=10^4$. At $St=0.3$, the apparent density of solid phase at $t=0.3$ and $t=2.5$ are shown in Figure \ref{Taylor-Green flow IC 1 Kn 10p4 taust 0.3 at t 0.3} and Figure \ref{Taylor-Green flow IC 1 Kn 10p4 taust 0.3 at t 2.5}, respectively. When $0.3>St_c$, the velocity of particle-phase will not keep mono-kinetic
and the particle trajectory crossing will occur accordingly. In contrast, when $St=0.039<St_c$, the solid particles will concentrate at the low-vorticity region, as presented in Figure \ref{Taylor-Green flow IC 1 Kn 10p4 taust 0.039 at t 2.5}.
 In this regime, both the UGKWP and MP-PIC give the same solutions.
 Then, at $Kn_s=10^{-6}$ and $St=0.3$,  the effect of collision is dominant in the determination of the evolution results.
 When the inelastic collision is employed for the particle-particle collision, the granular temperature will get lost after collision and thus post-collision particles will have the same velocity.
 All particles will accumulate and move towards to the low-vorticity region even though $St=0.3$ is larger than  $St_{c}$, as shown in Figure \ref{Taylor-Green flow IC 1 Kn 10m6 taust 0.3 at t 0.6}. In this highly collisional regime, the results by TFM are also presented and show the same solutions as the UGKWP.

\begin{figure}[htbp]
	\centering
	\subfigure{
		\includegraphics[height=6.5cm]{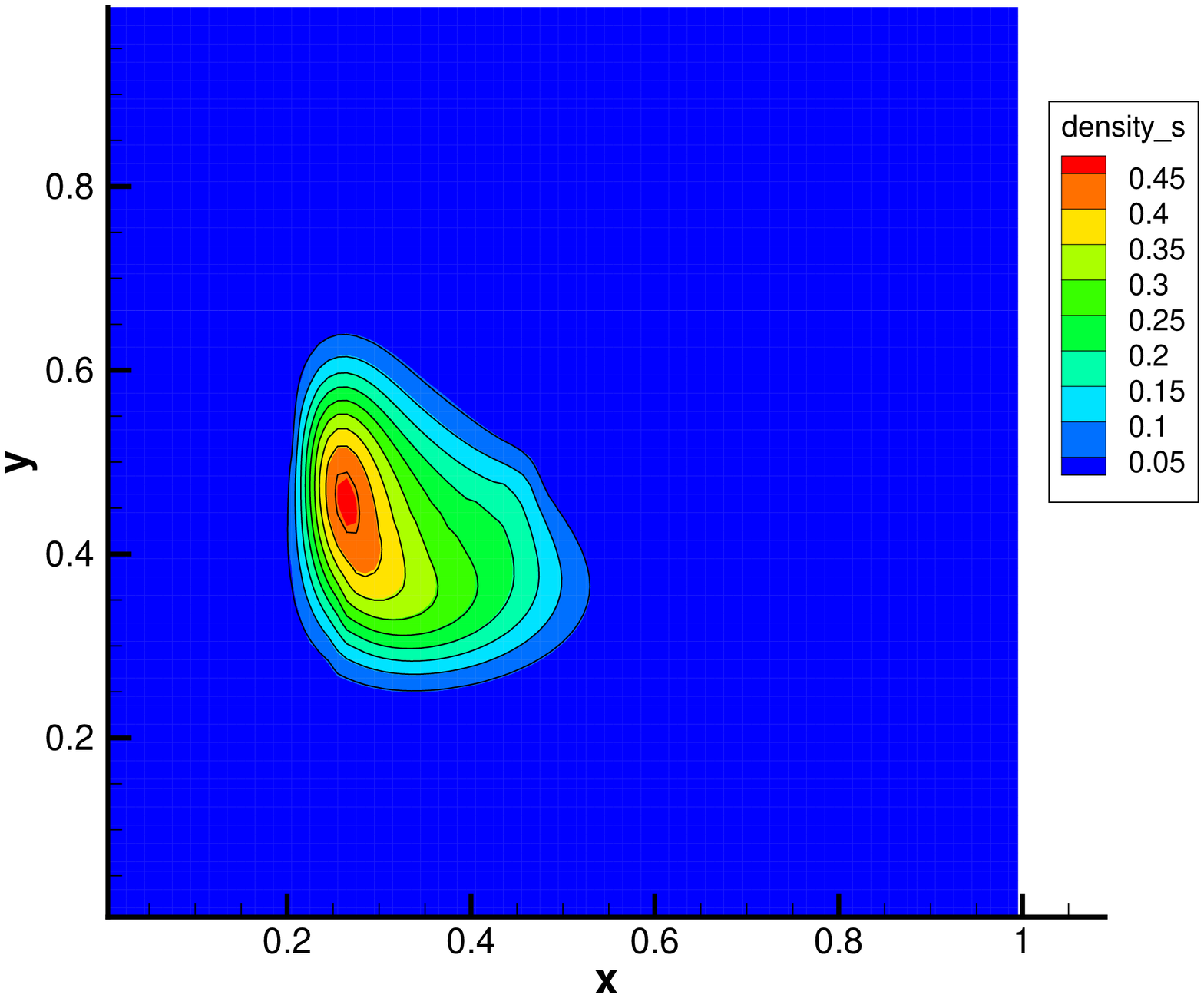}	
	}
	\quad
	\subfigure{
		\includegraphics[height=6.5cm]{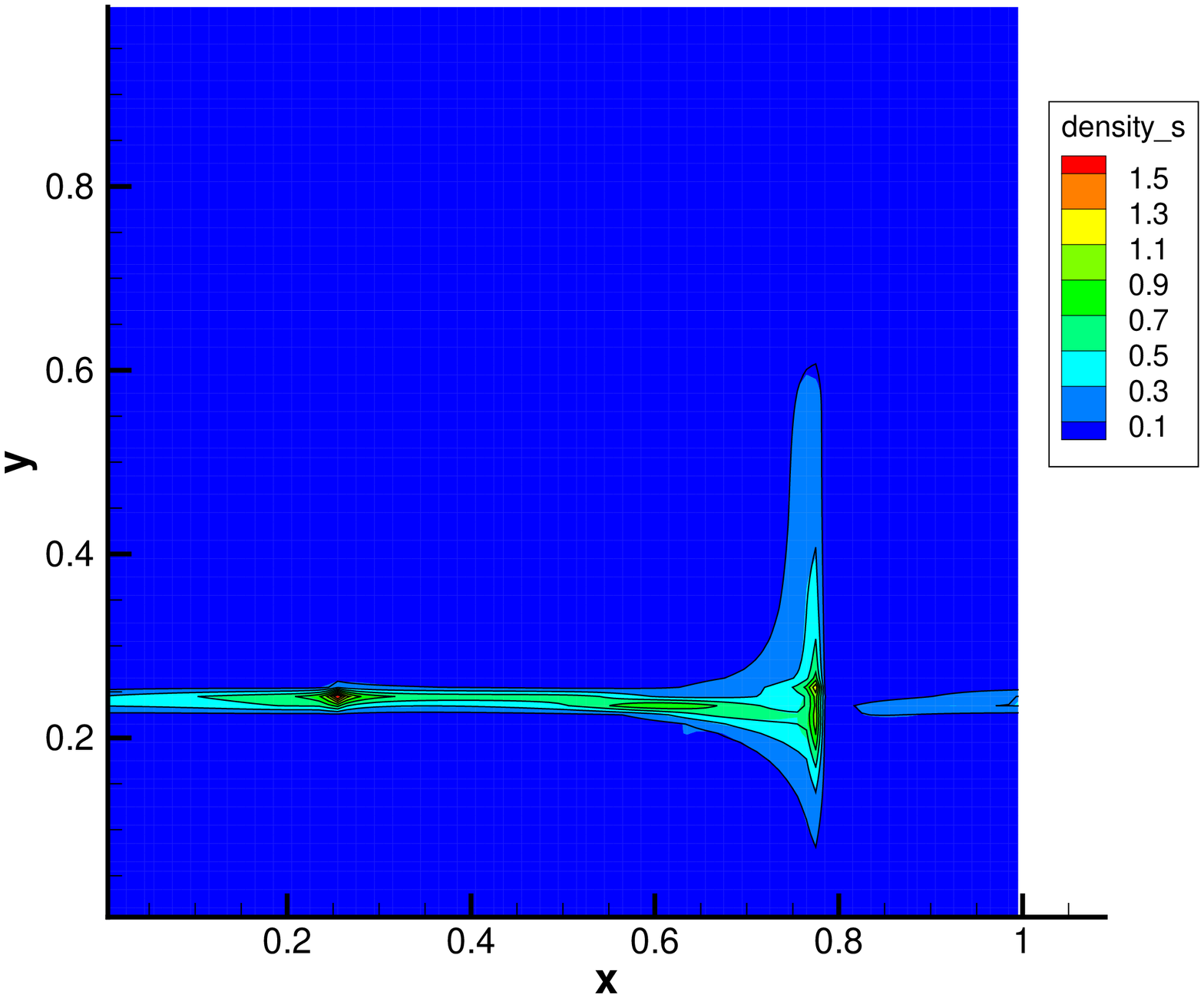}	
	}
	\caption{Apparent density of particle phase for Taylor-Green vortex problem in high collision regime with $Kn_s=10^{-6}$ and $St=0.1$ at $t=0.5$ (left) and $t=1.2$ (right) for the UGKWP (flood) and TFM (line) solutions.}		
	\label{Taylor-Green flow IC 2 Kn 10m6 taust 0.1}
\end{figure}

\begin{figure}[htbp]
	\centering
	\subfigure{
		\includegraphics[height=6.5cm]{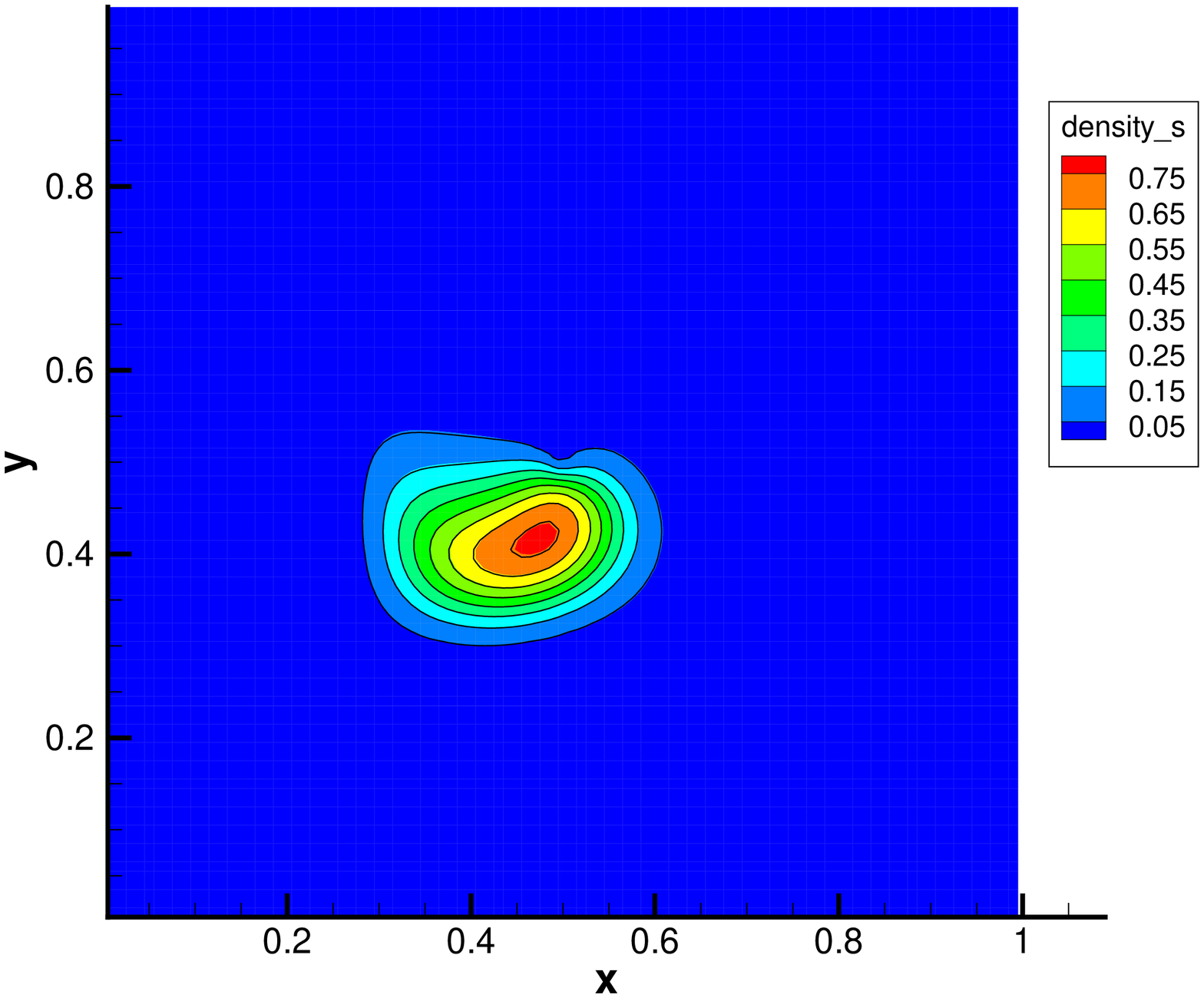}	
	}
	\quad
	\subfigure{
		\includegraphics[height=6.5cm]{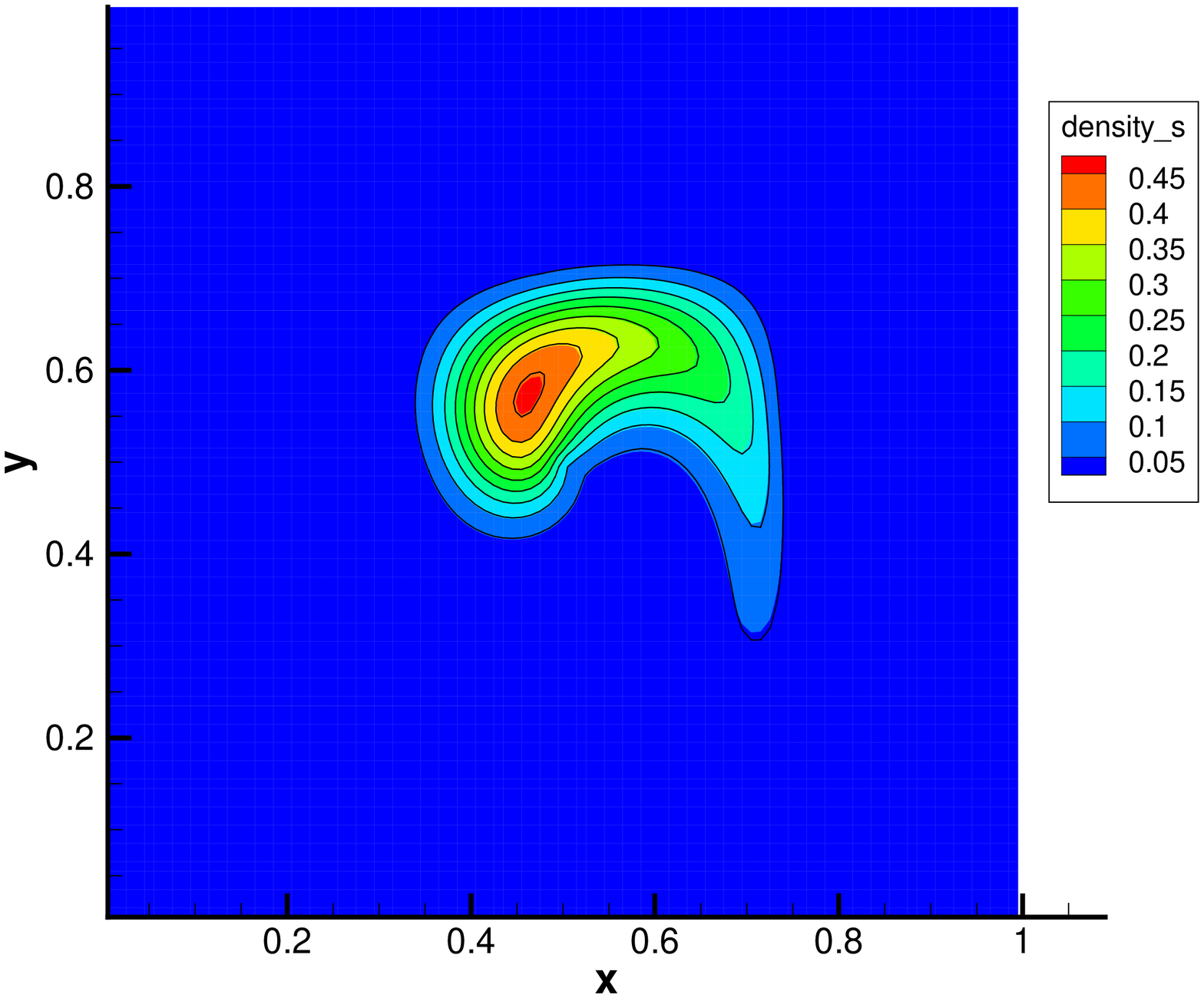}	
	}
	\caption{Apparent density of particle phase for Taylor-Green vortex problem in high collision regime with $Kn_s=10^{-6}$ and $St=0.001$ at $t=0.5$ (left) and $t=1.2$ (right) for the UGKWP (flood) and TFM (line) solutions.}
	\label{Taylor-Green flow IC 2 Kn 10m6 taust 0.001}		
\end{figure}

For the Type 2 initial condition, in the highly collision regime the Knudsen number $Kn_{s}=10^{-6}$ and two Stokes numbers $St=0.1$ and $St=0.001$ are considered and the results are given in Figure \ref{Taylor-Green flow IC 2 Kn 10m6 taust 0.1} and Figure \ref{Taylor-Green flow IC 2 Kn 10m6 taust 0.001}. The same conclusion as Type 1 case can be made. When $St<St_c$, the solid particles will stay in the original vortex and PTC will not occur. When $St>St_c$, the solid particles will have the ability to leave from the original vortex and go to the neighboring one. In the highly collisional regime, the results by TFM and UGKWP are consistent.

\subsection{Upward particle-laden jet into a cross-flow}

\begin{figure}[htbp]
	\centering
	\subfigure{
		\includegraphics[height=5.5cm]{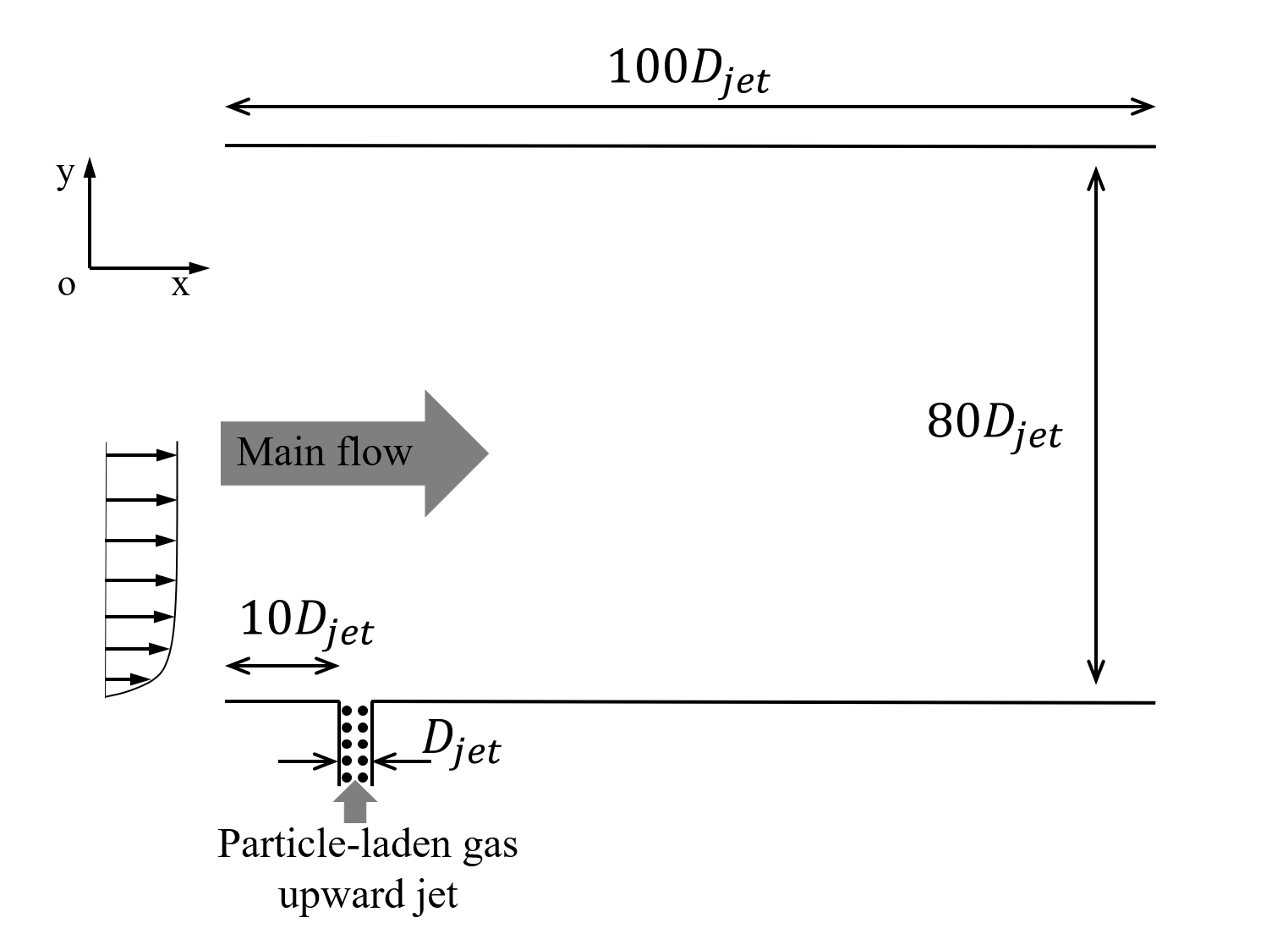}
	}		
	\quad
	\subfigure{
		\includegraphics[height=5.5cm]{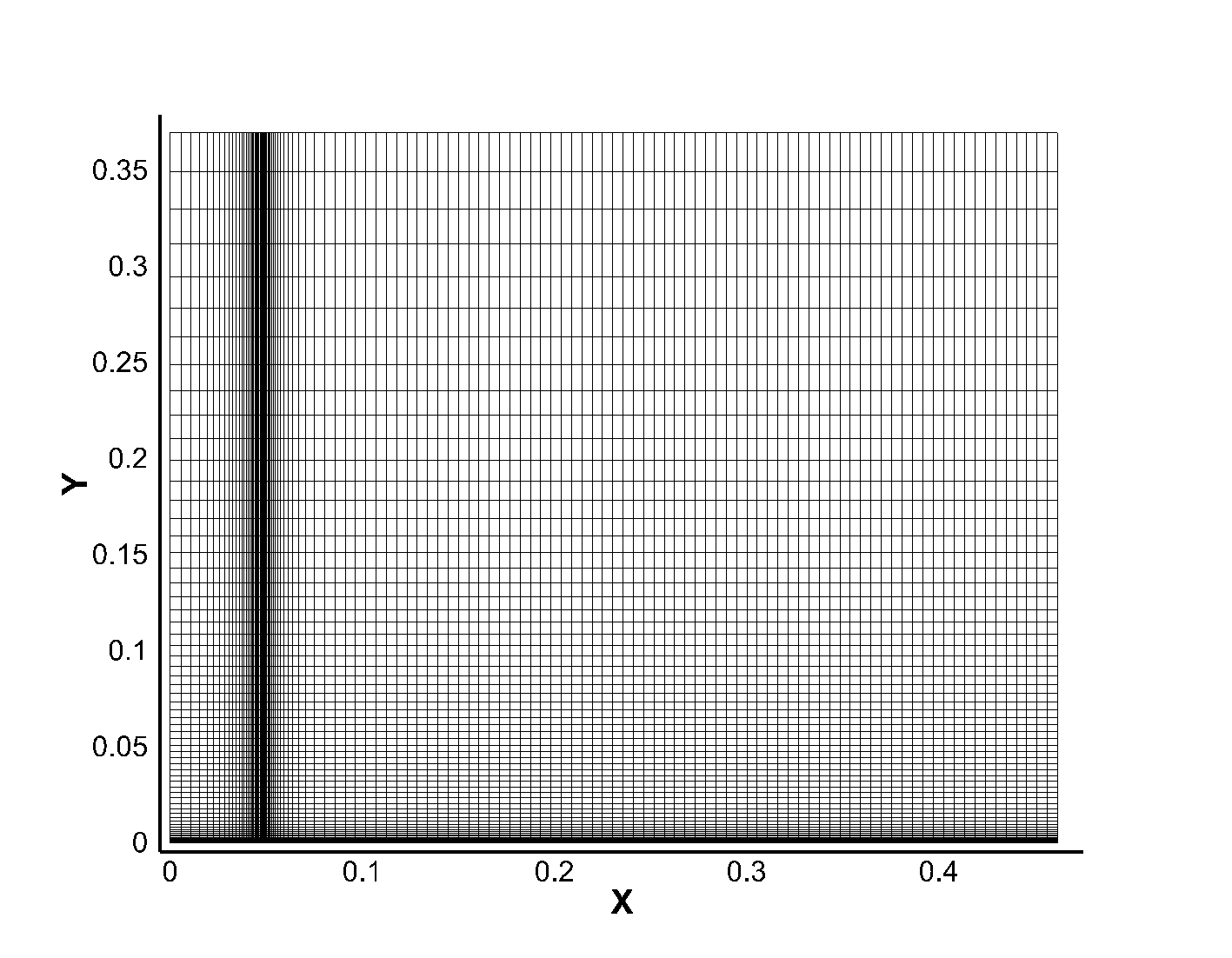}
	}		
	\caption{The sketch and mesh of the test for the upward particle-laden jet in a cross flow field.}
	\label{Particle jet crossflow sketch}		
\end{figure}

The 2D upward particle-laden jet into a cross-flow is considered for further investigation of gas-particle system.
Many studies about the interaction between the upward jet and the cross-flow have been conducted under different conditions \cite{Gasparticle-jet-crossflow-han1992numerical, Gasparticle-jet-crossflow-Stefanl2015state, Gasparticle-jet-crossflow-park2021particle}.
This problem is challenge when the upward jet carries solid particles.
Figure \ref{Particle jet crossflow sketch} shows a sketch of the simulated problem.
A particle-laden gas jet is injecting from the bottom of the computational domain.
The diameter of the upward jet is $D_{jet}$ is $4.62 mm$ and it is located at $10D_{jet}$.
The whole computational domain $L\times H$ is $100D_{jet}\times80D_{jet}$.
The density and diameter of the solid particles carried by the upward gas jet are $\rho_s=2638 kg/m^3$ and $d_s=15 \mu m$, respectively.
Initially, the volume fraction $\epsilon_{s}$ of particle phase in the upward jet is $0.0087$, and the corresponding apparent density $\epsilon_{s}\rho_{s}$ is $22.95 kg/m^3$. The jet velocity of the upward gas $V_{jet}$ is $26.38 m/s$ and the solid particle has the same velocity as the gas initially. The density of the gas in the jet is $1.21 kg/m^3$. The gravity for the solid particles is $9.8m/s^2$.
The initial granular temperature of solid particle phase $T_s$ is assumed to be zero. The cross-flow, or called main flow in the domain, moves from left to right with density $1.1 kg/m^3$ and temperature $298 K$. The velocity of main flow $U_{\infty}$ is $16.80 m/s$, which leads to a velocity ratio $V_{jet}/U_{\infty} =1.57$.  A non-uniform mesh with $120\times65$ grid points is used, which is shown in Figure \ref{Particle jet crossflow sketch}. The mesh is refined in the region near the wall and the upward jet, and the size of the first layer mesh is set as $0.20 mm$.
Non-slip wall boundary condition is applied for bottom boundary except the upward jet region, and free boundary condition is used for both right and up boundaries. The reference time $t_{ref}$ is defined as the ratio of the height of the computational domain to the upward jet velocity $H/V_{jet}$. The Stokes number is about $0.12$ and the Knudsen number $Kn_s = 1.1\times10^{-2}$ for the initial particle phase. Considering that the gas phase flow is turbulence, the $k-\omega$ SST turbulence model is employed for the gas flow \cite{kwsst-turbulence-model-menter1993zonal}.

\begin{figure}[htbp]
	\centering
	\subfigure{
		\includegraphics[height=6.5cm]{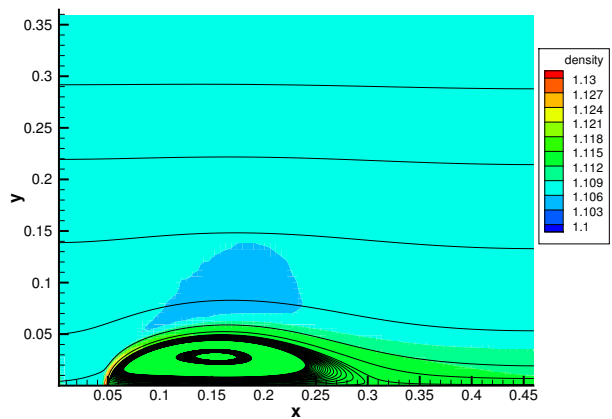}
	}			
	\caption{The density and streamline of gas phase at $t=0.1s$ calculated by the GKS with k-$\omega$ SST turbulence model.}
	\label{Particle jet crossflow gas}		
\end{figure}

\begin{figure}[htbp]
	\centering
	\subfigure{
		\includegraphics[height=5.5cm]{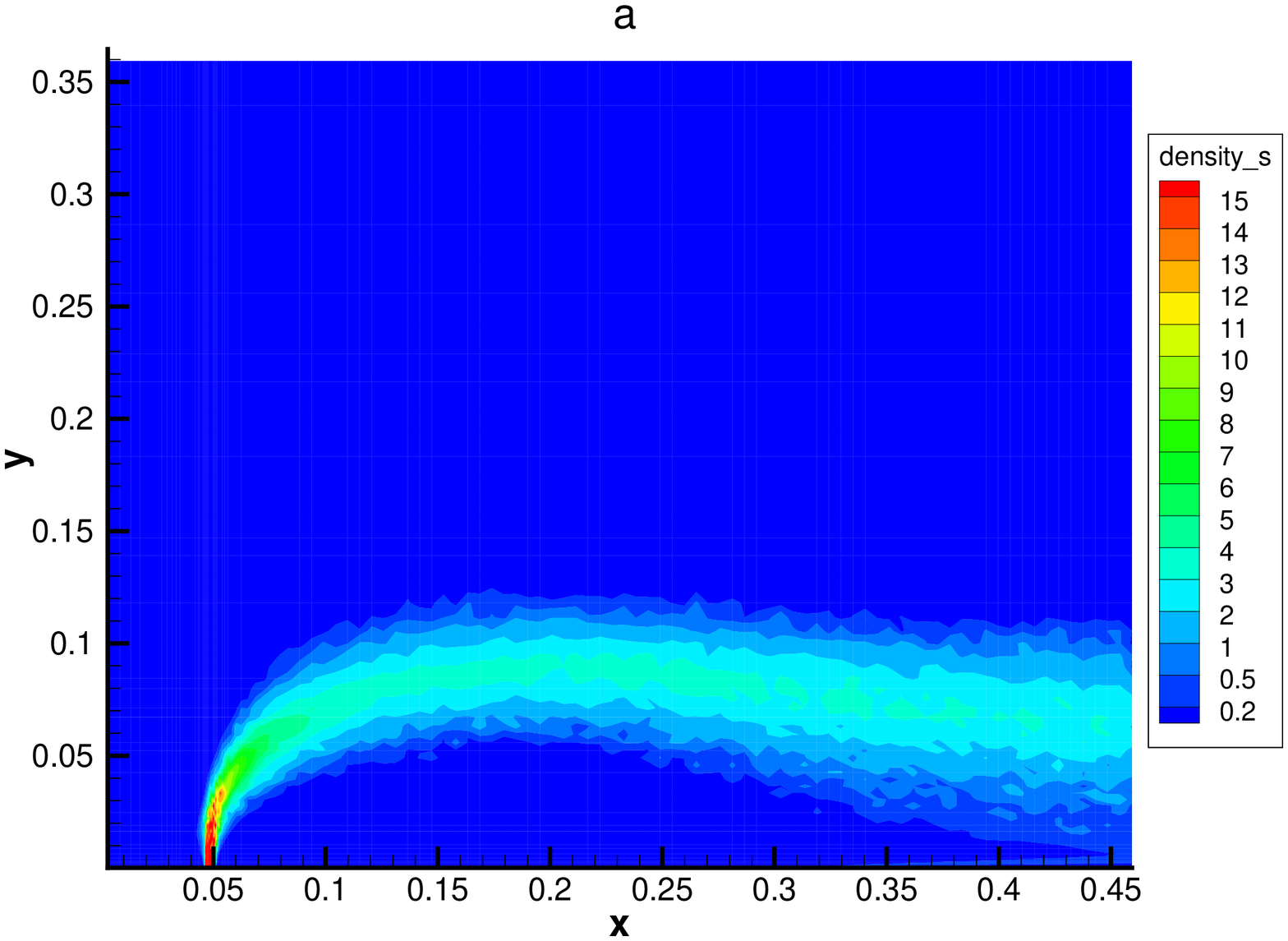}
	}
	\quad
	\subfigure{
		\includegraphics[height=5.5cm]{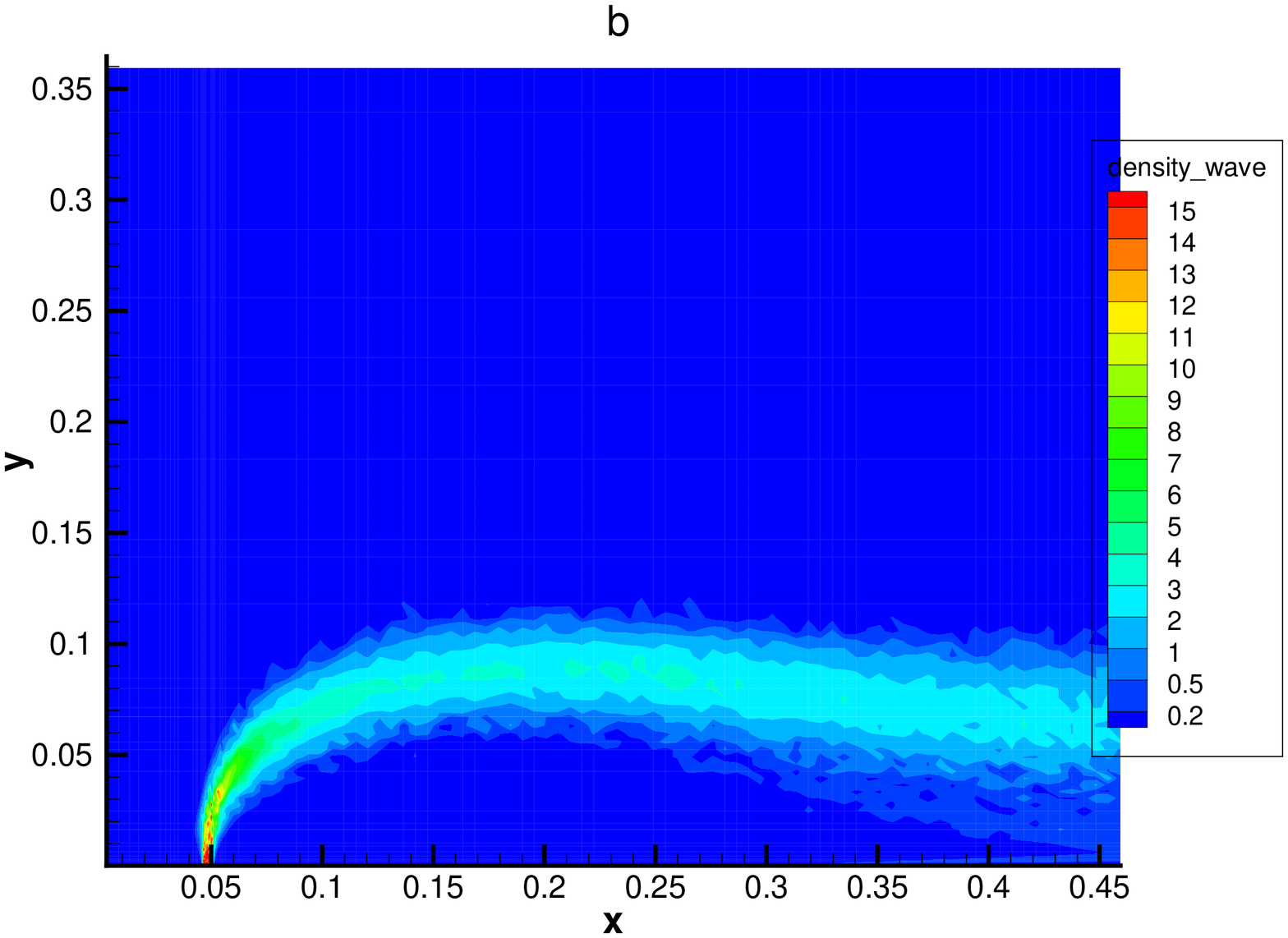}
	}

	\subfigure{
		\includegraphics[height=5.5cm]{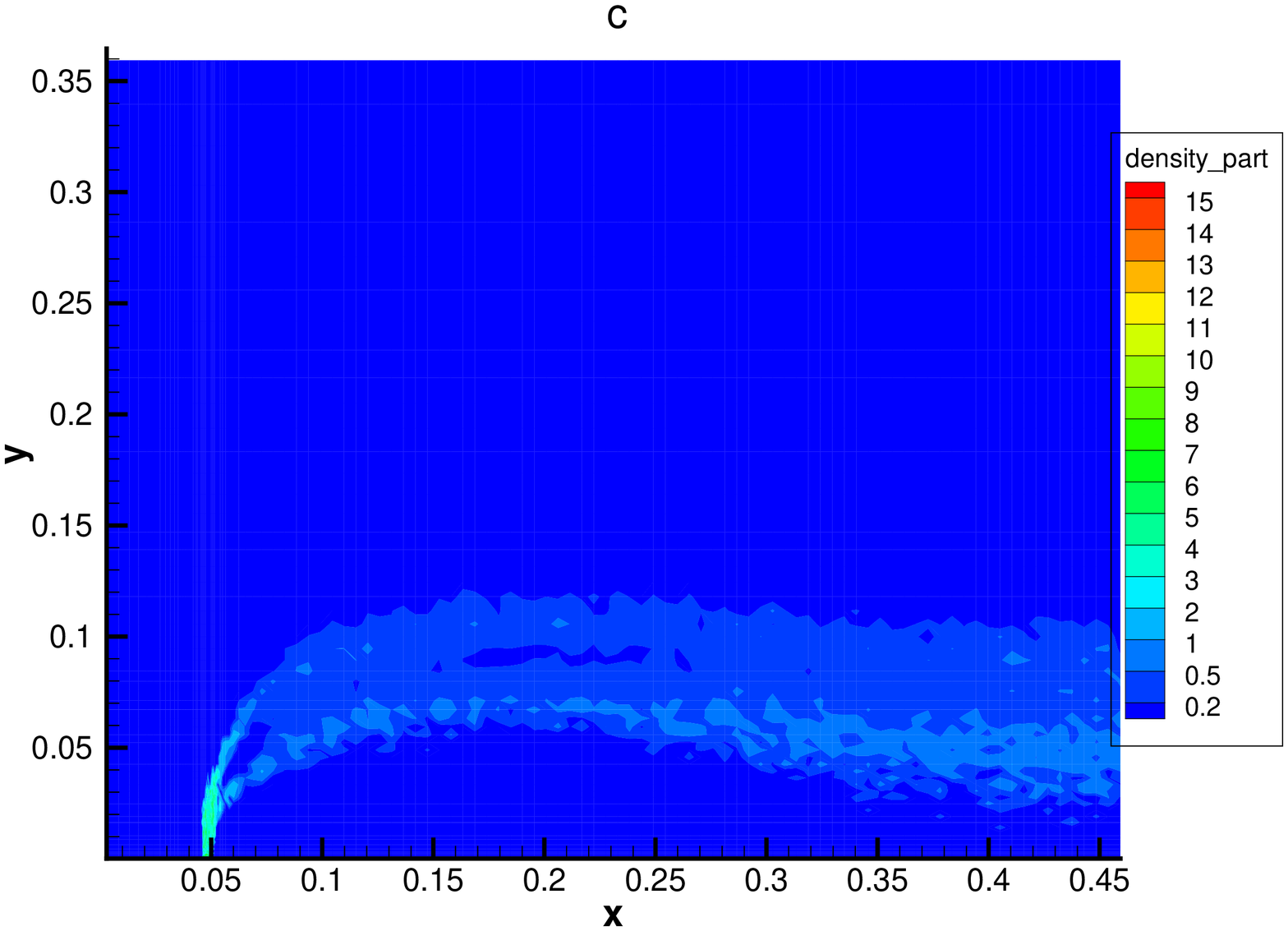}
	}
	\quad
	\subfigure{
		\includegraphics[height=5.5cm]{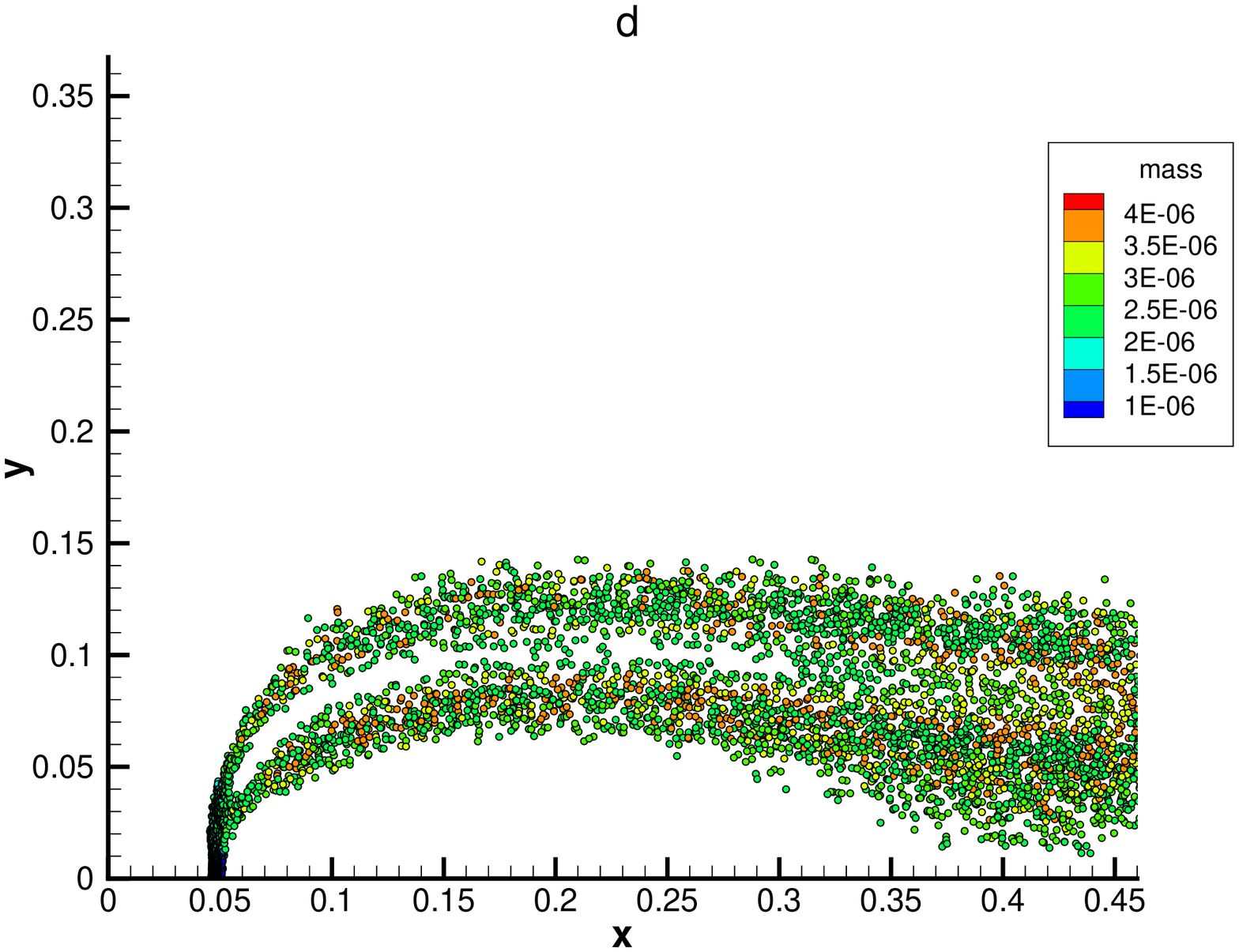}
	}				
	\caption{The apparent density of solid particle phase at $t=0.1 s$ with $St=0.12$ calculated the UGKWP. (a) total apparent density of particle phase; (b) apparent density calculated by wave; (c) apparent density calculated by particle; (d) the scatter plot of sampled particles.}
	\label{Particle jet crossflow St 0dot12}		
\end{figure}

\begin{figure}[htbp]
	\centering
	\subfigure{
		\includegraphics[height=5.5cm]{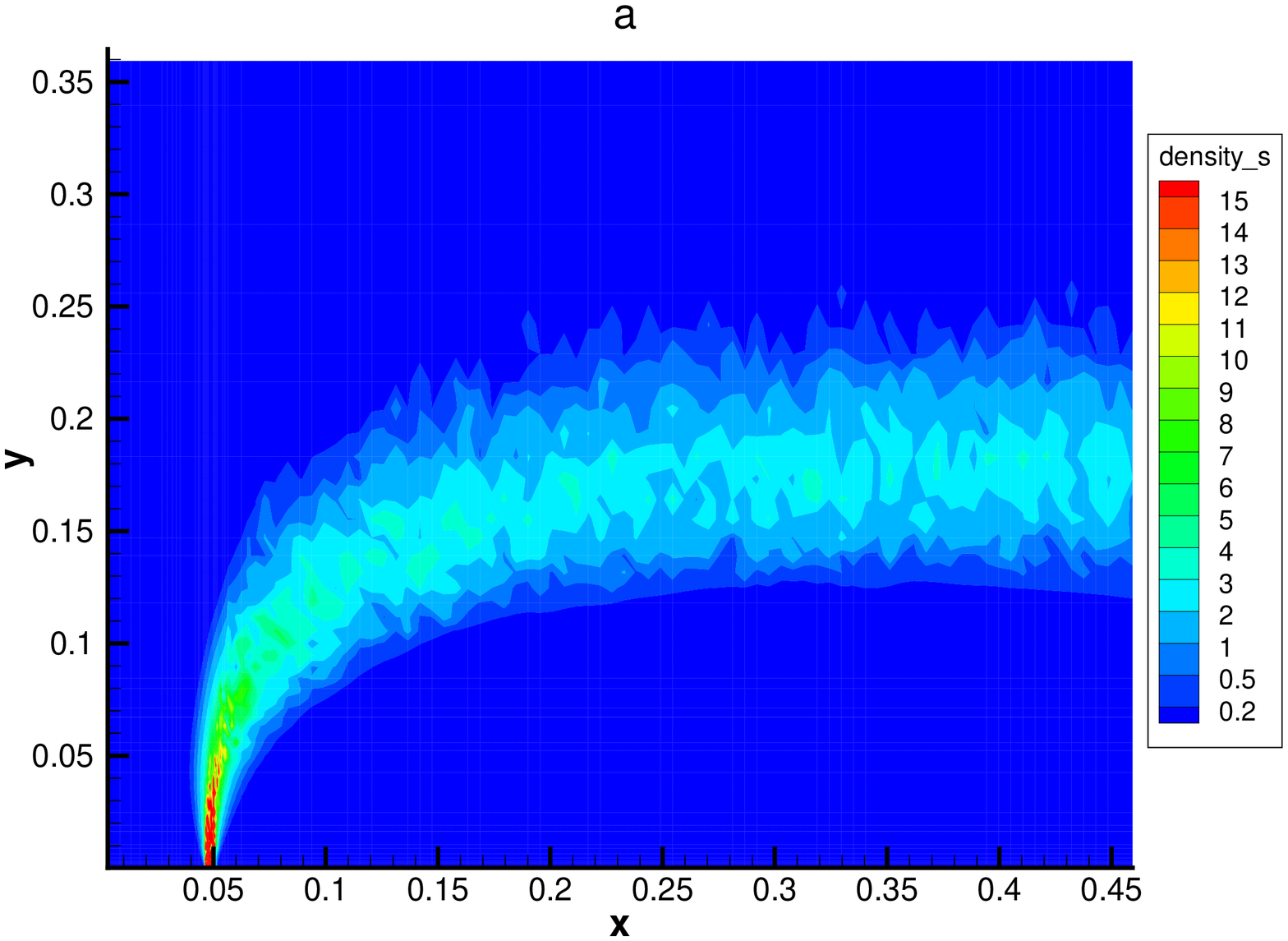}
	}
	\quad
	\subfigure{
		\includegraphics[height=5.5cm]{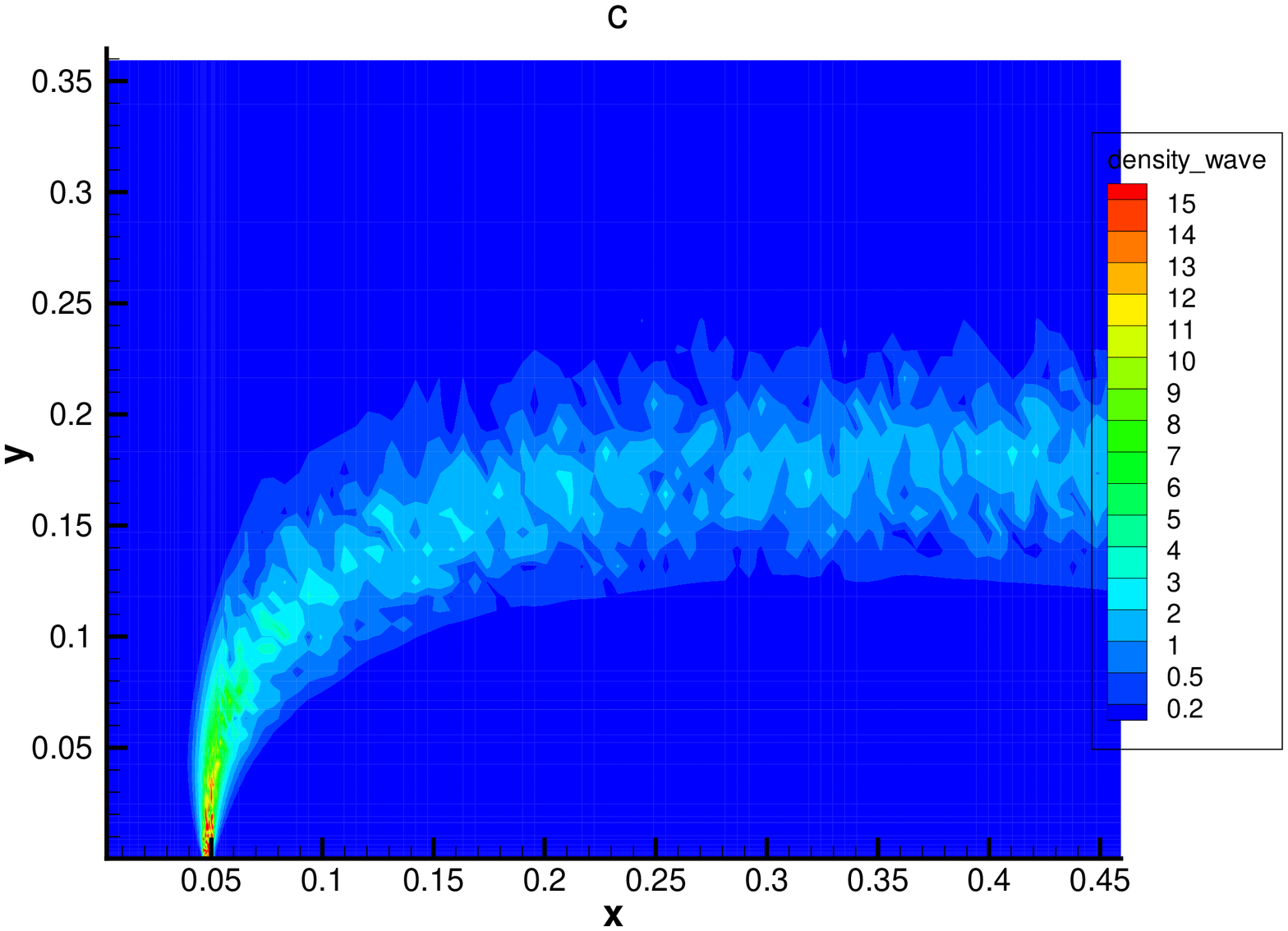}
	}
	
	\subfigure{
		\includegraphics[height=5.5cm]{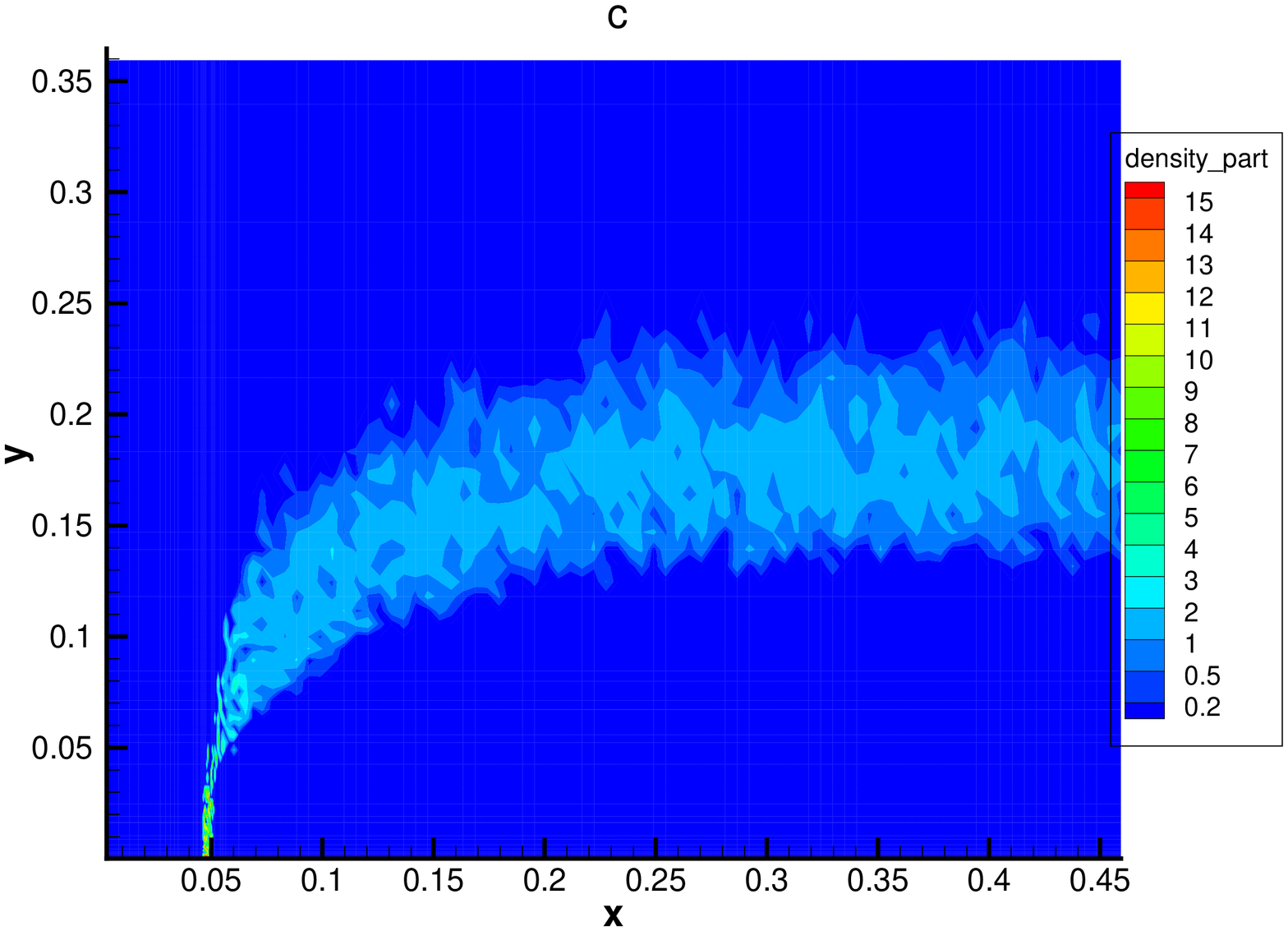}
	}
	\quad
	\subfigure{
		\includegraphics[height=5.5cm]{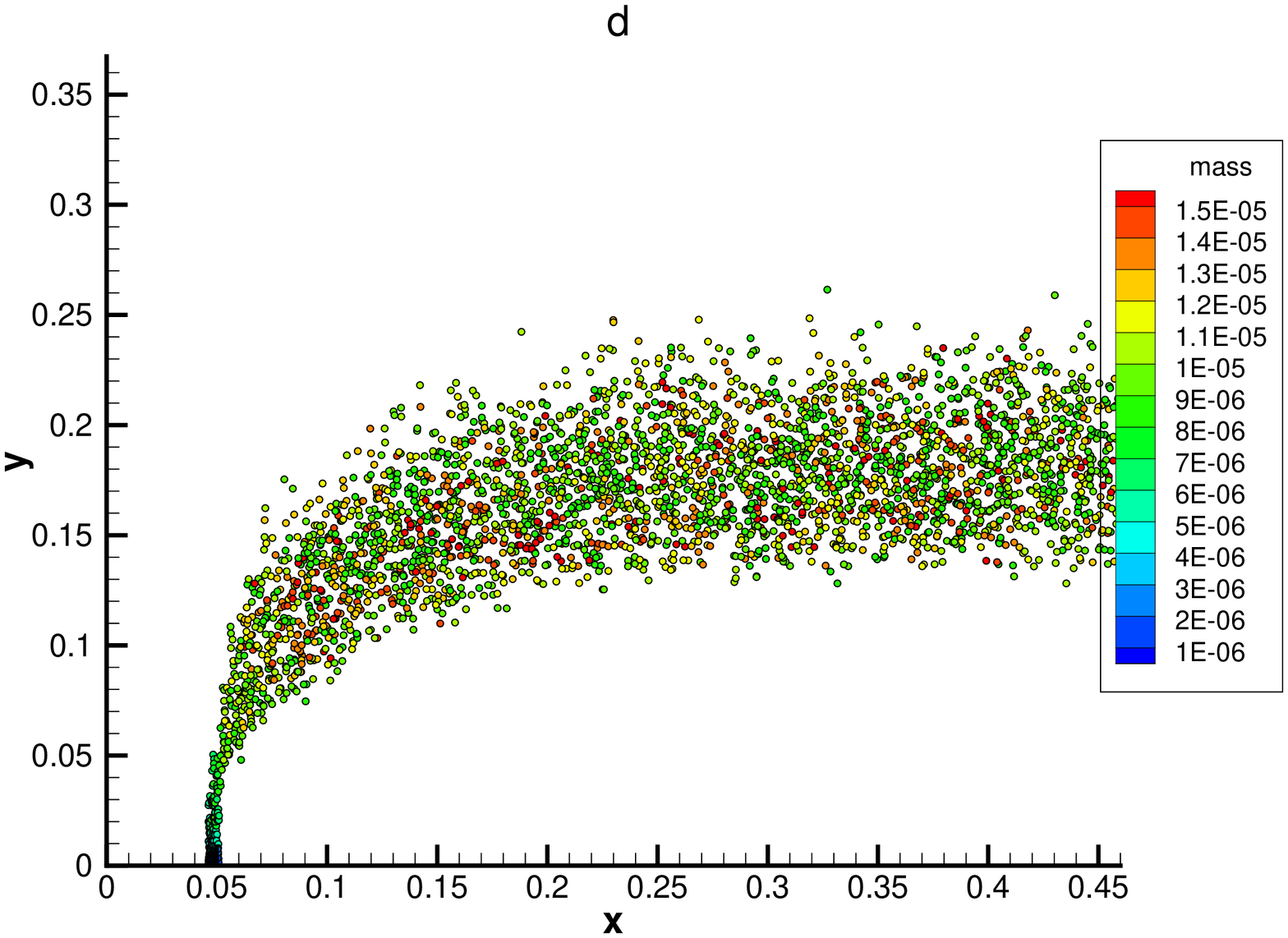}
	}				
	\caption{The apparent density of solid particle phase at $t=0.1 s$ with $St=1.2$ calculated by the UGKWP, (a) total apparent density of particle phase; (b) apparent density calculated by wave; (c) apparent density calculated by particle; (d) the scatter plot of sampled particles.}
	\label{Particle jet crossflow St 1dot2}		
\end{figure}

In this study, the initial cross flow and the gas from the jet will form a flow field which is calculated by the GKS with turbulence modeling.
Then, the movement of the solid particles in the jet will be controlled by their interaction with the gas flow field and their inner solid particle collision.
The influence on the gas phase from the solid particle is ignored. In other words, this test is mainly about the evolution of solid particles under external gas field. Note that the collisions between solid particles will also have dynamic effect on the solid particle evolution.
The density and streamline of gas flow at $t=0.1s$ are presented in Figure \ref{Particle jet crossflow gas}, where a large vortex is formed due to the interaction between the upward gas jet and main flow.
 In order to compensate the lost dispersion effect due to the under-resolved vortex in the turbulence model in the gas phase simulation,
 the turbulence dispersion force is added in the particle phase, where the model proposed by Lahey is employed \cite{Gasparticle-jet-turbulence-dispersion-lahey1993phase},
\begin{equation*}
F_T = -C_T \rho_g k_g \nabla \epsilon_s,
\end{equation*}
where $C_T$ is taken as 0.1, and $k_g$ is the turbulence kinetic energy of gas phase.
  The apparent density of particle phase at $t=0.1 s$ with $St=0.12$ is shown in Figure \ref{Particle jet crossflow St 0dot12}.
As stated earlier, the UGKWP represents the particle phase by both wave and particle, where their separate contributions are shown in Figure \ref{Particle jet crossflow St 0dot12}(b) and Figure \ref{Particle jet crossflow St 0dot12}(c). Figure \ref{Particle jet crossflow St 0dot12}(a) shows the total apparent density of solid particle phase, which is the sum of the above two parts.
The results show that wave and particle are distributed adaptively. In the core region of particle phase, there are more collisions between solid particles and the particle phase has a high percentage representation automatically by the wave in the UGKWP.
Furthermore, Figure \ref{Particle jet crossflow St 0dot12}(d) shows the sampled particles in the simulation, which appear more on the edges of particle phase with less particle-particle collision and these particles have the corresponding density shown in Figure \ref{Particle jet crossflow St 0dot12}(c). The solid particles in the jet change their velocities and move to the right due to their interaction with the gas flow field developed by the external cross-flow and injected gas. Similar results have been found in the previous research \cite{Gasparticle-jet-crossflow-Stefanl2015state}. The dispersion effect is caused by the complex interaction between solid particles with the cross-flow field as well as the particle-particle collisions.
With the increase of particle diameter $d_s$, a large Stokes number $St=1.2$ can be achieved, which is about $10$ times the previous Stokes number. With the Knudsen number $Kn_s=3.4\times10^{-2}$ for the particles in the jet, the simulation results are shown in Figure \ref{Particle jet crossflow St 1dot2}. Since a large $St$ means a weak interaction between the solid particle and the gas phase, the solid particles in the jet can move
 to a higher vertical position, and are pushed to the  right by the gas flow. Furthermore, in comparison with Figure \ref{Particle jet crossflow St 0dot12}, the wave part in this case takes a lower percentage in the particle representation due to large Knudsen number.

\subsection{Particle bed fluidization in shock tube}
\begin{figure}[htbp]
	\centering
	\subfigure{
		\includegraphics[height=6.5cm]{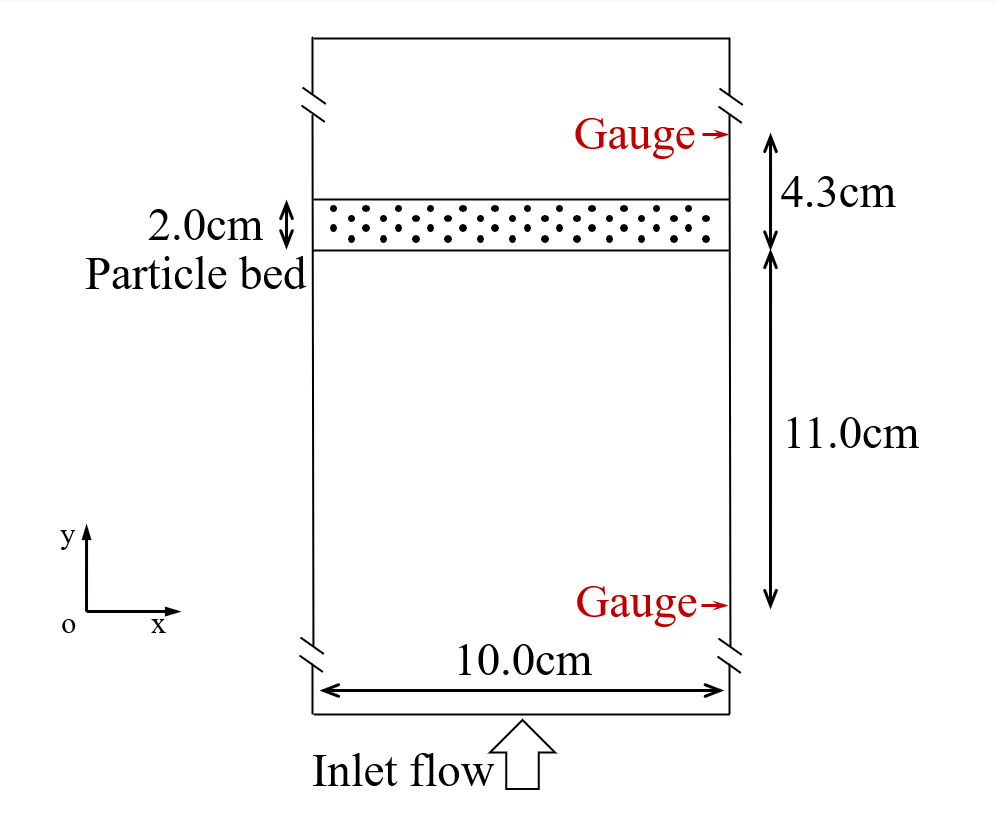}
	}		
	\caption{The sketch of the fluidization of a particle bed induced by shock conducted by Rogue.}
	\label{Rogue problem sketch}		
\end{figure}

\begin{figure}[htbp]
	\centering
	\subfigure{
		\includegraphics[height=6.5cm]{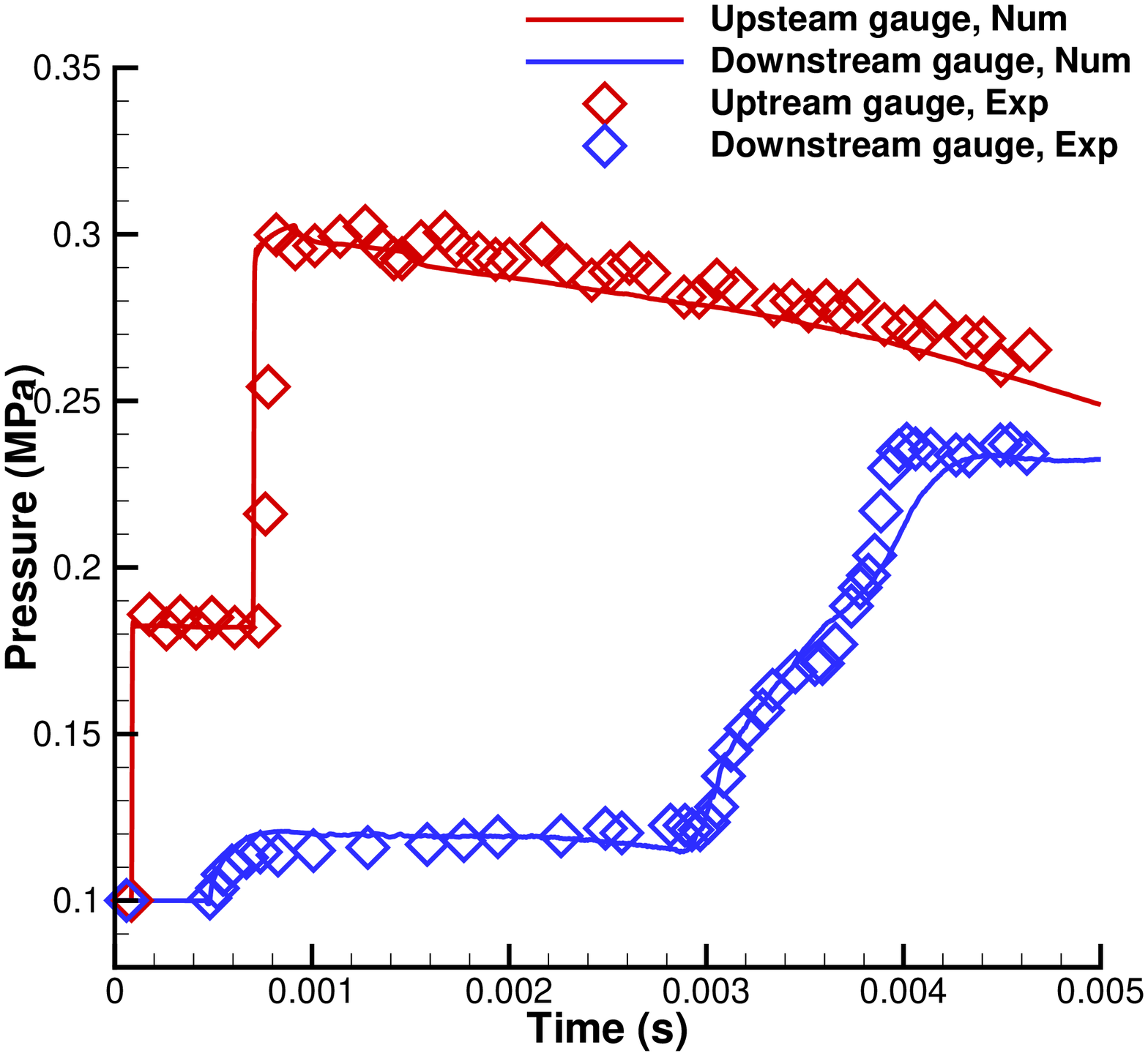}
	}
    \quad
    \subfigure{
    	\includegraphics[height=6.5cm]{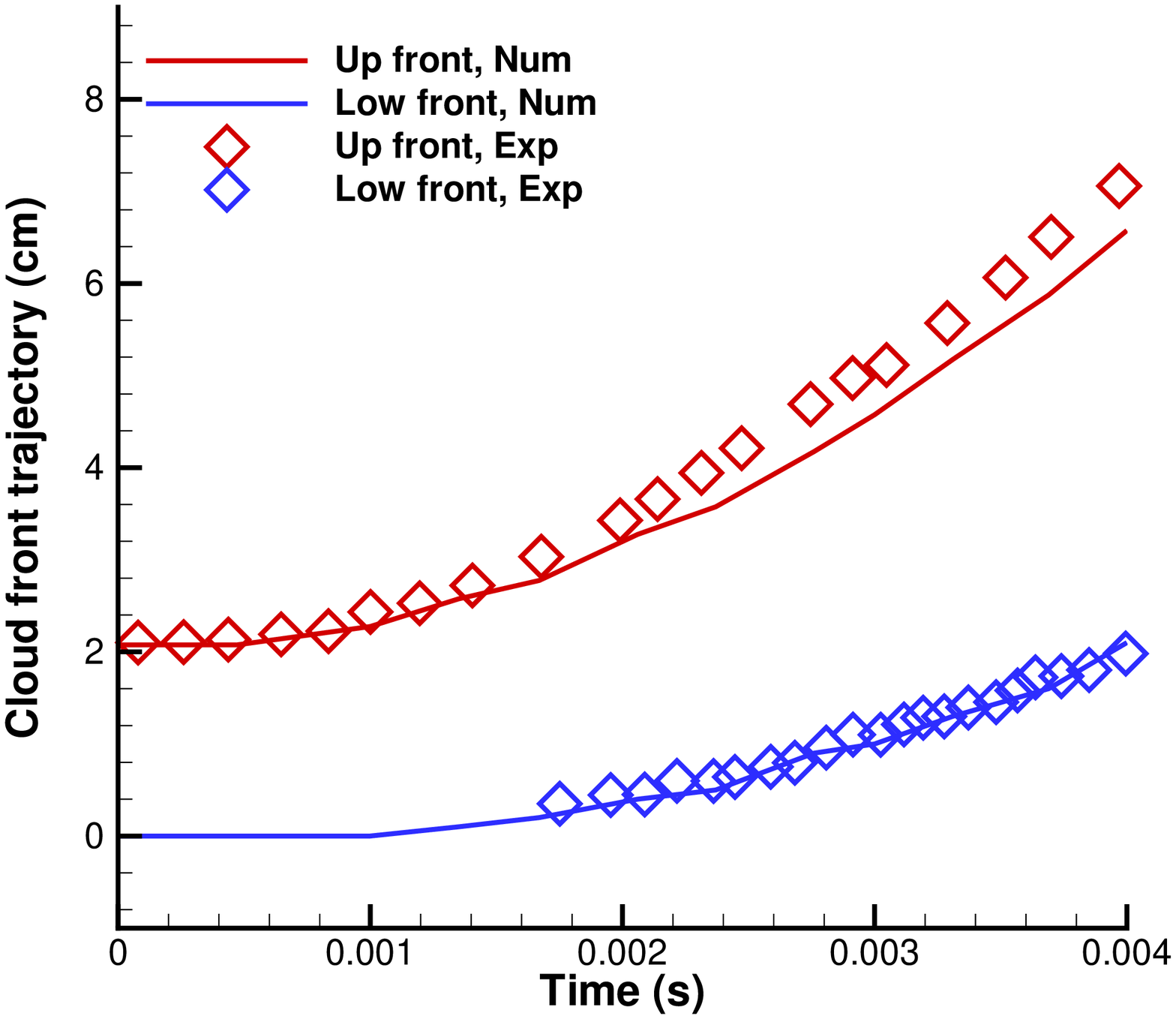}
    }		
	\caption{The comparison of numerical results by the UGKWP with experiment data for particle bed fluidization problem.
  The left figure shows the instantaneous pressure at two gauge positions and the right figure is about the trajectories of upper and lower frontd of solid particle cloud.}
	\label{Rogue problem pressure and front}		
\end{figure}

The fluidization of a particle bed induced by a shock is a challenge problem to verify the reliability of multi-phase numerical simulation \cite{Rogueproblem-Rogue1998fluidization,Rogueproblem-Saurel1999multiphase,Rogueproblem-Houim2016multiphase}. This is a complex physical problem since it involves the movement of solid particles, the interaction of solid particles with shock, the reflection and transmission of a shock.
In this paper, the experiment conducted by Rogue is simulated by the UGKWP \cite{Rogueproblem-Rogue1998fluidization}. Figure \ref{Rogue problem sketch} shows a sketch of this experiment. Initially, a particle bed is located at $15 cm$ in the vertical direction. This particle bed is comprised of a group of glass solid particles with a diameter of $1.5 mm$, and the thickness of this particle bed is $2 cm$. Besides, the density of solid particles is $2500 kg/m^3$, and the volume fraction $\epsilon_{s}$ of solid particle bed is 0.65. The initial density and pressure of the gas in the tube are $1.2 kg/m^3$ and $10^5 Pa$. A shock of Mach number $1.3$ is created by the high-pressure gas with vertical velocity $V_0=151 m/s$ at the inlet boundary. Two pressure gauges are set at the upstream and the downstream, such as $11.0 cm$ below the particle bed and $4.3 cm$ above the particle bed, to monitor the instantaneous pressure value. The computational domain $L\times H$ is $1 cm\times90 cm$ covered by $10\times900$ uniform cells. Considering that the turbulence intensity of the gas will increase after the interaction with a particle bed, the turbulence energy is regarded as internal energy and thus the internal degree of freedom of gas phase will be modeled by $k(t)=k_0+1.5\times(t/t_{end})^3$. In this case, the reference time $t_{ref}$ is defined as the ratio of the whole tube length $H$ to the inflow gas velocity $V_0$. The Stokes number is approximately $St=0.62$ and the Knudsen number of particle phase is around $Kn_s=2\times10^{-3}$. The comparison of the numerical result by the UGKWP with experiment data is shown in Figure \ref{Rogue problem pressure and front}. The instantaneous pressure values at the upstream and downstream gauges show that both the reflected shock and transmitted shock are calculated correctly. In addition, the upper and lower fronts of the solid particle cloud also agrees well with the experiment data.


\section{Conclusion}
The gas-particle two-phase system is complicated. The flow physics of the particles can manifest the dynamics in different regimes.  With the variation of particle Knudsen number, the particle movement can be modeled from the continuum fluid dynamics to the highly non-equilibrium particle free transport, which can be only recovered by a multiscale method. 
In this paper, a combination of GKS and UGKWP methods for the disperse dilute gas-particle system has been developed for the study of multiscale 
flow physics. The GKS, as an efficient kinetic theory-based Navier-Stokes solver, is used for the computation of gas flow. On the other hand, the particle flow in different flow regimes is simulated by the UGKWP, which dynamically distributes the particle and wave to model the particle free transport and equilibrium wave propagation. The UGKWP becomes an efficient and accurate method for the capturing the solid particle's multiscale dynamics. At the same time, the momentum and energy exchanges between the gas and particle are included in GKS-UGKWP through the coupled evolution of the system. The advantage for using UGKWP for the solid particle evolution is that both particle transport and wave propagation behavior in different regime can be integrated under the same formulation and the most efficient approach will automatically be picked up locally by the scheme.
The proposed method can recover the Eulerian-Eulerian formulation, such as TFM, in the  highly collisional regime for the gas-particle system and the Eulerian-Lagrangian model, such as MP-PIC, for the collisionless regime of the solid particle. 
 Many numerical tests have been conducted to validate the proposed method. In the wind-sand shock tube problem, the solution of the new method is consistent with TFM and MP-PIC in continuum and collisionless limits. Also, the non-equilibrium particle flow phenomena, such as PTC and particle wall reflecting, are captured. The vortex-induced particle segregation in the Taylor-Green flow is also investigated under different particle  Knudsen number. The dispersion of particle flow is observed and quantitatively evaluated in the interaction case between the upward particle-laden jet and cross-flow. The shock-induced particle-bed fluidization problem is calculated and the results agree well with the experimental measurements.

\section*{Acknowledgments}

The current research is supported by National Numerical Windtunnel project, National Science Foundation of China (11772281, 91852114), and 
Department of Science and Technology of Guangdong Province (Grant No.2020B1212030001).

\bibliographystyle{plain}%
\bibliography{jixingbib1}

\begin{thebibliography}{10}

\bibitem{Gasparticle-PIC-andrews1996multiphase}
Michael~J Andrews and Peter~J O'Rourke.
\newblock The multiphase particle-in-cell (mp-pic) method for dense particulate
  flows.
\newblock {\em International Journal of Multiphase Flow}, 22(2):379--402, 1996.

\bibitem{Gasparticle-BN-model-baer1986two}
MR~Baer and JW~Nunziato.
\newblock A two-phase mixture theory for the deflagration-to-detonation
  transition (ddt) in reactive granular materials.
\newblock {\em International journal of multiphase flow}, 12(6):861--889, 1986.

\bibitem{Gasparticle-review-balachandar2010turbulent}
S~Balachandar and John~K Eaton.
\newblock Turbulent dispersed multiphase flow.
\newblock {\em Annual review of fluid mechanics}, 42:111--133, 2010.

\bibitem{Gasparticle-DSMC-bird1976molecular}
Graeme~Austin Bird.
\newblock Molecular gas dynamics.
\newblock {\em NASA STI/Recon Technical Report A}, 76:40225, 1976.

\bibitem{GKS-turbulence-CIT-Cao2019}
Guiyu Cao, Liang Pan, and Kun Xu.
\newblock Three dimensional high-order gas-kinetic scheme for supersonic
  isotropic turbulence \uppercase\expandafter{\romannumeral1}: criterion for
  direct numerical simulation.
\newblock {\em Computers \& Fluids}, 192(104273), 2019.

\bibitem{GKS-turbulence-implicitHGKS-Cao2019}
Guiyu Cao, Hongmin Su, Jinxiu Xu, and Kun Xu.
\newblock Implicit high-order gas kinetic scheme for turbulence simulation.
\newblock {\em Aerospace Science and Technology}, 92:958--971, 2019.

\bibitem{CE-expansion}
Sydney Chapman and Thomas~George Cowling.
\newblock {\em The mathematical theory of non-uniform gases: an account of the
  kinetic theory of viscosity, thermal conduction and diffusion in gases}.
\newblock Cambridge university press, 1970.

\bibitem{WP-3D-chen2020three}
Yipei Chen, Yajun Zhu, and Kun Xu.
\newblock A three-dimensional unified gas-kinetic wave-particle solver for flow
  computation in all regimes.
\newblock {\em Physics of Fluids}, 32(9):096108, 2020.

\bibitem{Gasparticle-dusty-hybrid-finite-volume-particle-chertock2017hybrid}
Alina Chertock, Shumo Cui, and Alexander Kurganov.
\newblock Hybrid finite-volume-particle method for dusty gas flows.
\newblock {\em The SMAI journal of computational mathematics}, 3:139--180,
  2017.

\bibitem{Gasparticle-book-crowe2011multiphase}
Clayton~T Crowe, John~D Schwarzkopf, Martin Sommerfeld, and Yutaka Tsuji.
\newblock {\em Multiphase flows with droplets and particles}.
\newblock CRC press, 2011.

\bibitem{Gasparticle-Taylor-Green-vortex-Stephane2007evaluation}
Stephane De~Chaisemartin, Fr{\'e}d{\'e}rique Laurent, Marc Massot, and Julien
  Reveillon.
\newblock Evaluation of eulerian multi-fluid versus lagrangian methods for
  ejection of polydisperse evaporating sprays by vortices.
\newblock 2007.

\bibitem{Gasparticle-MOM-Fox-desjardins2008quadrature}
Olivier Desjardins, Rodney~O Fox, and Philippe Villedieu.
\newblock A quadrature-based moment method for dilute fluid-particle flows.
\newblock {\em Journal of Computational Physics}, 227(4):2514--2539, 2008.

\bibitem{Gasparticle-Taylor-Green-vortex-Moment-method-Fox2008quadrature}
Olivier Desjardins, Rodney~O Fox, and Philippe Villedieu.
\newblock A quadrature-based moment method for dilute fluid-particle flows.
\newblock {\em Journal of Computational Physics}, 227(4):2514--2539, 2008.

\bibitem{Gasparticle-KTGF-ding1990bubbling}
Jianmin Ding and Dimitri Gidaspow.
\newblock A bubbling fluidization model using kinetic theory of granular flow.
\newblock {\em AIChE journal}, 36(4):523--538, 1990.

\bibitem{Gasparticle-MOM-Fox-fox2008quadrature}
Rodney~O Fox.
\newblock A quadrature-based third-order moment method for dilute gas-particle
  flows.
\newblock {\em Journal of Computational Physics}, 227(12):6313--6350, 2008.

\bibitem{Gasparticle-PTC-freret2008turbulent}
L~Fr{\'e}ret, F~Laurent, S~de~Chaisemartin, D~Kah, RO~Fox, P~Vedula,
  J~Reveillon, O~Thomine, and M~Massot.
\newblock Turbulent combustion of polydisperse evaporating sprays with droplet
  crossing: Eulerian modeling of collisions at finite knudsen and validation.
\newblock In {\em Proceedings of the Summer Program}, pages 277--288. Citeseer,
  2008.

\bibitem{Gasparticle-drag-Kliatchko-fuks1955mechanics}
Nikolaj~A Fuks.
\newblock The mechanics of aerosols.
\newblock Technical report, Chemical Warfare Labs Army Chemical Center MD,
  1955.

\bibitem{Gasparticle-review-Ge2017discrete}
Wei Ge, Limin Wang, Ji~Xu, Feiguo Chen, Guangzheng Zhou, Liqiang Lu, Qi~Chang,
  and Jinghai Li.
\newblock Discrete simulation of granular and particle-fluid flows: from
  fundamental study to engineering application.
\newblock {\em Reviews in Chemical Engineering}, 33(6):551--623, 2017.

\bibitem{Gasparticle-DEM-review-guo-curtis-2015discrete}
Yu~Guo and Jennifer~Sinclair Curtis.
\newblock Discrete element method simulations for complex granular flows.
\newblock {\em Annual Review of Fluid Mechanics}, 47:21--46, 2015.

\bibitem{Gasparticle-jet-crossflow-han1992numerical}
Kee~Soo Han and Myung~Kyoon Chung.
\newblock Numerical simulation of a two-phase gas-particle jet in a crossflow.
\newblock {\em Aerosol science and technology}, 17(2):59--68, 1992.

\bibitem{Gasparticle-coarse-grained-DEM-hilton2014comparison}
James~E Hilton and Paul~W Cleary.
\newblock Comparison of non-cohesive resolved and coarse grain dem models for
  gas flow through particle beds.
\newblock {\em Applied Mathematical Modelling}, 38(17-18):4197--4214, 2014.

\bibitem{Rogueproblem-Houim2016multiphase}
Ryan~W Houim and Elaine~S Oran.
\newblock A multiphase model for compressible granular-gaseous flows:
  formulation and initial tests.
\newblock {\em Journal of fluid mechanics}, 789:166, 2016.

\bibitem{Gasparticle-KTGF-jenkins1983theory}
James~T Jenkins, Stuart~B Savage, et~al.
\newblock Theory for the rapid flow of identical, smooth, nearly elastic,
  spherical particles.
\newblock {\em Journal of fluid mechanics}, 130(1):187--202, 1983.

\bibitem{CompactGKS-ji2018-structured}
Xing Ji, Liang Pan, Wei Shyy, and Kun Xu.
\newblock {A compact fourth-order gas-kinetic scheme for the {Euler and
  Navier-Stokes} equations}.
\newblock {\em Journal of Computational Physics}, 372:446 -- 472, 2018.

\bibitem{CompactGKS-ji2020-unstructured}
Xing Ji, Fengxiang Zhao, Wei Shyy, and Kun Xu.
\newblock {A HWENO Reconstruction Based High-order Compact Gas-kinetic Scheme
  on Unstructured Mesh}.
\newblock {\em Journal of Computational Physics}, 109367, 2020.

\bibitem{Gasparticle-jet-turbulence-dispersion-lahey1993phase}
RT~Lahey~Jr, M~Lopez De~Bertodano, and OC~Jones~Jr.
\newblock Phase distribution in complex geometry conduits.
\newblock {\em Nuclear Engineering and Design}, 141(1-2):177--201, 1993.

\bibitem{GKS-hypersonic-LiQibing2005}
Qibing Li, Song Fu, and Kun Xu.
\newblock Application of gas-kinetic scheme with kinetic boundary conditions in
  hypersonic flow.
\newblock {\em AIAA journal}, 43(10):2170--2176, 2005.

\bibitem{UGKS-photon-li2018unified}
Weiming Li, Chang Liu, Yajun Zhu, Jiwei Zhang, and Kun Xu.
\newblock Unified gas-kinetic wave-particle methods iii: Multiscale photon
  transport.
\newblock {\em Journal of Computational Physics}, 408:109280, 2020.

\bibitem{WP-three-photon-transport-li2020unified}
Weiming Li, Chang Liu, Yajun Zhu, Jiwei Zhang, and Kun Xu.
\newblock Unified gas-kinetic wave-particle methods iii: Multiscale photon
  transport.
\newblock {\em Journal of Computational Physics}, 408:109280, 2020.

\bibitem{UGKS-gas-particle-liu2019unified}
Chang Liu, Zhao Wang, and Kun Xu.
\newblock A unified gas-kinetic scheme for continuum and rarefied flows vi:
  Dilute disperse gas-particle multiphase system.
\newblock {\em Journal of Computational Physics}, 386:264--295, 2019.

\bibitem{UGKS-plasma-liu2017unified}
Chang Liu and Kun Xu.
\newblock A unified gas kinetic scheme for continuum and rarefied flows v:
  multiscale and multi-component plasma transport.
\newblock {\em Communications in Computational Physics}, 22(5):1175--1223,
  2017.

\bibitem{UGKS-microflow-linearized-ke-liu2020unified}
Chang Liu and Kun Xu.
\newblock A unified gas-kinetic scheme for micro flow simulation based on
  linearized kinetic equation.
\newblock {\em Advances in Aerodynamics}, 2(1):1--22, 2020.

\bibitem{WP-four-liu2020unified}
Chang Liu and Kun Xu.
\newblock Unified gas-kinetic wave-particle method iv: Multi-species gas
  mixture and plasma transport.
\newblock {\em Advances in Aerodynamics}, 3:9, 2021.

\bibitem{WP-first-liu2020unified}
Chang Liu, Yajun Zhu, and Kun Xu.
\newblock Unified gas-kinetic wave-particle methods i: Continuum and rarefied
  gas flow.
\newblock {\em Journal of Computational Physics}, 401:108977, 2020.

\bibitem{Gasparticle-DNS-gewei-liu2017meso}
Xiaowen Liu, Limin Wang, and Wei Ge.
\newblock Meso-scale statistical properties of gas--solid flow—a direct
  numerical simulation (dns) study.
\newblock {\em AIChE Journal}, 63(1):3--14, 2017.

\bibitem{Gasparticle-DNS-DEM-DFM-comparison-lu-gewei-2017assessing}
Liqiang Lu, Xiaowen Liu, Tingwen Li, Limin Wang, Wei Ge, and Sofiane Benyahia.
\newblock Assessing the capability of continuum and discrete particle methods
  to simulate gas-solids flow using dns predictions as a benchmark.
\newblock {\em Powder Technology}, 321:301--309, 2017.

\bibitem{Gasparticle-coarse-grained-lu2017extension}
Liqiang Lu, Aaron Morris, Tingwen Li, and Sofiane Benyahia.
\newblock Extension of a coarse grained particle method to simulate heat
  transfer in fluidized beds.
\newblock {\em International Journal of Heat and Mass Transfer}, 111:723--735,
  2017.

\bibitem{Gasparticle-coarse-grained-EMMS-DPM-gewei-lu2016computer}
Liqiang Lu, Ji~Xu, Wei Ge, Guoxian Gao, Yong Jiang, Mingcan Zhao, Xinhua Liu,
  and Jinghai Li.
\newblock Computer virtual experiment on fluidized beds using a coarse-grained
  discrete particle method—emms-dpm.
\newblock {\em Chemical Engineering Science}, 155:314--337, 2016.

\bibitem{Gasparticle-KTGF-lun1984kinetic}
CKK Lun, S~Br Savage, DJ~Jeffrey, and N~Chepurniy.
\newblock Kinetic theories for granular flow: inelastic particles in couette
  flow and slightly inelastic particles in a general flowfield.
\newblock {\em Journal of fluid mechanics}, 140:223--256, 1984.

\bibitem{Gasparticle-DNS-immersed-boundary-luokun-luo2019improved}
Kun Luo, Zhuo Wang, Junhua Tan, and Jianren Fan.
\newblock An improved direct-forcing immersed boundary method with inward
  retraction of lagrangian points for simulation of particle-laden flows.
\newblock {\em Journal of Computational Physics}, 376:210--227, 2019.

\bibitem{Gasparticle-momentmethod-Fox2013computational}
Daniele~L Marchisio and Rodney~O Fox.
\newblock {\em Computational models for polydisperse particulate and multiphase
  systems}.
\newblock Cambridge University Press, 2013.

\bibitem{kwsst-turbulence-model-menter1993zonal}
Florianr Menter.
\newblock Zonal two equation kw turbulence models for aerodynamic flows.
\newblock In {\em 23rd fluid dynamics, plasmadynamics, and lasers conference},
  page 2906, 1993.

\bibitem{Gasparticle-PIC-rourke-2012inclusion}
Peter~J O'Rourke and Dale~M Snider.
\newblock Inclusion of collisional return-to-isotropy in the mp-pic method.
\newblock {\em Chemical engineering science}, 80:39--54, 2012.

\bibitem{GKS-multicomponent-pan2017}
Liang Pan, Junxia Cheng, Shuanghu Wang, and Kun Xu.
\newblock A two-stage fourth-order gas-kinetic scheme for compressible
  multicomponent flows.
\newblock {\em Communications in Computational Physics}, 22(4):1123--1149,
  2017.

\bibitem{Gasparticle-jet-crossflow-park2021particle}
Jooyeon Park and Hyungmin Park.
\newblock Particle dispersion induced by vortical interactions in a
  particle-laden upward jet with a partial crossflow.
\newblock {\em Journal of Fluid Mechanics}, 915, 2021.

\bibitem{Gasparticle-MOM-Fox-passalacqua2010fully}
A~Passalacqua, RO~Fox, R~Garg, and S~Subramaniam.
\newblock A fully coupled quadrature-based moment method for dilute to
  moderately dilute fluid--particle flows.
\newblock {\em Chemical Engineering Science}, 65(7):2267--2283, 2010.

\bibitem{Gasparticle-dusty-flow-pelanti2006high}
Marica Pelanti and Randall~J LeVeque.
\newblock High-resolution finite volume methods for dusty gas jets and plumes.
\newblock {\em SIAM Journal on Scientific Computing}, 28(4):1335--1360, 2006.

\bibitem{Gasparticle-jet-crossflow-Stefanl2015state}
Stefan Radl, Begona~C Gonzales, Christoph Goniva, and Stefan Pirker.
\newblock State of the art in mapping schemes for dilute and dense
  euler-lagrange simulations.
\newblock 2015.

\bibitem{Rogueproblem-Rogue1998fluidization}
X~Rogue, G~Rodriguez, JF~Haas, and R~Saurel.
\newblock Experimental and numerical investigation of the shock-induced
  fluidization of a particles bed.
\newblock {\em Shock Waves}, 8(1):29--45, 1998.

\bibitem{Gasparticle-dusty-saito2002numerical}
T~Saito.
\newblock Numerical analysis of dusty-gas flows.
\newblock {\em Journal of computational physics}, 176(1):129--144, 2002.

\bibitem{Gasparticle-Abgrall-saurel1999multiphase}
Richard Saurel and R{\'e}mi Abgrall.
\newblock A multiphase godunov method for compressible multifluid and
  multiphase flows.
\newblock {\em Journal of Computational Physics}, 150(2):425--467, 1999.

\bibitem{Rogueproblem-Saurel1999multiphase}
Richard Saurel and R{\'e}mi Abgrall.
\newblock A multiphase godunov method for compressible multifluid and
  multiphase flows.
\newblock {\em Journal of Computational Physics}, 150(2):425--467, 1999.

\bibitem{Gasparticle-PIC-snider2001incompressible}
Dale~M Snider.
\newblock An incompressible three-dimensional multiphase particle-in-cell model
  for dense particle flows.
\newblock {\em Journal of computational physics}, 170(2):523--549, 2001.

\bibitem{UGKS-radiative-sun2015asymptotic}
Wenjun Sun, Song Jiang, and Kun Xu.
\newblock An asymptotic preserving unified gas kinetic scheme for gray
  radiative transfer equations.
\newblock {\em Journal of Computational Physics}, 285:265--279, 2015.

\bibitem{Gasparticle-DUGKS-immersed-boundary-guozhaoli-tao2018combined}
Shi Tao, Haolong Zhang, Zhaoli Guo, and Lian-Ping Wang.
\newblock A combined immersed boundary and discrete unified gas kinetic scheme
  for particle--fluid flows.
\newblock {\em Journal of Computational Physics}, 375:498--518, 2018.

\bibitem{Gasparticle-PIC-DEM-tian2020compressible}
Baolin Tian, Junsheng Zeng, Baoqing Meng, Qian Chen, Xiaohu Guo, and Kun Xue.
\newblock Compressible multiphase particle-in-cell method (cmp-pic) for full
  pattern flows of gas-particle system.
\newblock {\em Journal of Computational Physics}, 418:109602, 2020.

\bibitem{UGKS-gasparticle-wangzhao-wang2019unified}
Zhao Wang and Hong Yan.
\newblock Unified gas-kinetic scheme for the monodisperse gas-particle flow and
  its application in the shock-driven multiphase instability.
\newblock {\em International Journal of Multiphase Flow}, 119:95--107, 2019.

\bibitem{KP-gasparticle-wangzhao-wang2020unified}
Zhao Wang and Hong Yan.
\newblock Unified gas-kinetic particle method for dilute granular flow and its
  application in a solid jet.
\newblock {\em Acta Mechanica Sinica}, 36(1):22--34, 2020.

\bibitem{GKS-multicomponent-xu1997}
Kun Xu.
\newblock {BGK-based} scheme for multicomponent flow calculations.
\newblock {\em Journal of Computational Physics}, 134(1):122--133, 1997.

\bibitem{GKS-lecture}
Kun Xu.
\newblock Gas-kinetic schemes for unsteady compressible flow simulations.
\newblock {\em Lecture series-van Kareman Institute for fluid dynamics},
  3:C1--C202, 1998.

\bibitem{GKS-2001}
Kun Xu.
\newblock A gas-kinetic {BGK} scheme for the {Navier--Stokes} equations and its
  connection with artificial dissipation and {Godunov} method.
\newblock {\em Journal of Computational Physics}, 171(1):289--335, 2001.

\bibitem{UGKS-xu2010unified}
Kun Xu and Juan-Chen Huang.
\newblock A unified gas-kinetic scheme for continuum and rarefied flows.
\newblock {\em Journal of Computational Physics}, 229(20):7747--7764, 2010.

\bibitem{WP-five-diatomic-molecular-xu2020unified}
Xiaocong Xu, Yipei Chen, Chang Liu, Zhihui Li, and Kun Xu.
\newblock Unified gas-kinetic wave-particle methods v: diatomic molecular flow.
\newblock {\em Journal of Computational Physics}, 442:110496, 2021.

\bibitem{WP-sample-xu2021modeling}
Xiaocong Xu, Yipei Chen, and Kun Xu.
\newblock Modeling and computation for non-equilibrium gas dynamics: Beyond
  single relaxation time kinetic models.
\newblock {\em Physics of Fluids}, 33(1):011703, 2021.

\bibitem{Gasparticle-KTGF-yu2007numerical}
Liang Yu, Jing Lu, Xiangping Zhang, and Suojiang Zhang.
\newblock Numerical simulation of the bubbling fluidized bed coal gasification
  by the kinetic theory of granular flow (ktgf).
\newblock {\em Fuel}, 86(5-6):722--734, 2007.

\bibitem{GKS-acoustic-zhao2019}
Fengxiang Zhao, Xing Ji, Wei Shyy, and Kun Xu.
\newblock An acoustic and shock wave capturing compact high-order gas-kinetic
  scheme with spectral-like resolution.
\newblock {\em International Journal of Computational Fluid Dynamics}, 2019.

\bibitem{CompactGKS-zhao2019-8th-order}
Fengxiang Zhao, Xing Ji, Wei Shyy, and Kun Xu.
\newblock Compact higher-order gas-kinetic schemes with spectral-like
  resolution for compressible flow simulations.
\newblock {\em Advances in Aerodynamics}, 1(1):13, 2019.

\bibitem{WP-second-zhu-unstructured-mesh-zhu2019unified}
Yajun Zhu, Chang Liu, Chengwen Zhong, and Kun Xu.
\newblock Unified gas-kinetic wave-particle methods. ii. multiscale simulation
  on unstructured mesh.
\newblock {\em Physics of Fluids}, 31(6):067105, 2019.

\bibitem{UGKS-first-decade-zhu2021first}
Yajun Zhu and Kun Xu.
\newblock The first decade of unified gas kinetic scheme.
\newblock {\em arXiv preprint arXiv:2102.01261}, 2021.

\end{thebibliography}
\end{document}